\documentclass[10pt]{article}
\usepackage{amsmath}
\usepackage{latexsym}
\usepackage{amssymb}
\usepackage{amsfonts}
\usepackage{amsthm}
\title{NONLINEAR CONFORMAL \\ ELECTROMAGNETISM AND GRAVITATION}
\author{J.-F. Pommaret \\ CERMICS, Ecole des Ponts ParisTech, France \\
jean-francois.pommaret@wanadoo.fr \\
(http://cermics.enpc.fr/$\sim$pommaret/home.html)}
\date{  }
\begin{document}
\maketitle

\vspace{2cm}

\noindent
{\bf ABSTRACT}   \\

\noindent
In 1909 the brothers E. and F. Cosserat discovered a new nonlinear group theoretical approach to elasticity (EL), with the only experimental need to measure the EL constants. In a modern language, their idea has been to use the nonlinear Spencer sequence instead of the nonlinear Janet sequence for the Lie groupoid defining the group of rigid motions of space. Following H. Weyl, our purpose is to compute for the first time the nonlinear Spencer sequence for the Lie groupoid defining the conformal group of space-time in order to provide the physical foundations of both electromagnetism (EM) and gravitation, with the only experimental need to measure the EM constant in vacuum and the gravitational constant. With a manifold of dimension $n$, the difficulty is to deal with the $n$ nonlinear transformations that have been called "elations" by E. Cartan in 1922. Using the fact that dimension $n=4$ has very specific properties for the computation of the Spencer cohomology, we prove that there is no conceptual difference between the Cosserat EL field or induction equations and the Maxwell EM field or induction equations. As a byproduct, the well known field/matter couplings (piezzoelectricity, photoelasticity, ...) can be described abstractly, with the only experimental need to measure the corresponding coupling constants. In the sudy of gravitation, the dimension $n=4$ also allows to have a conformal factor defined everywhere but at the central attractive mass and the inversion law of the subgroupoid made by strict second order jets transforms attraction into repulsion. \\

\vspace{2cm}

\noindent
{\bf KEY WORDS}  \\
\noindent
Nonlinear differential sequences; Linear differential sequences; Lie groupoids; Lie algebroids; \\Conformal geometry; Spencer 
cohomology; Maxwell equations; Cosserat equations.  \\

\newpage

\noindent
{\bf 1) INTRODUCTION}  \\

Let us start this paper with a personal but meaningful story that has oriented my research during the last fourty years or so, since the french " {\it Grandes Ecoles}  " created their own research laboratories. Being a fresh permanent researcher of Ecole Nationale des Ponts et Chauss\'{e}es in Paris, the author of this paper has been asked to become the scientific adviser of a young student in order to introduce him to research. As General Relativity was far too much difficult for somebody without any specific mathematical knowledge while remembering his own experience at the same age, he asked the student to collect about $50$ books of Special Relativity and classify them along the way each writer was avoiding the use of the conformal group of space-time implied by the Michelson and Morley experiment, only caring about the Poincar\'{e} or Lorentz subgroups. After six months, the student (like any reader) arrived at the fact that most books were almost copying each other and could be nevertheless classified into three categories:  \\

\noindent
$\bullet$ $30$ books, including the original $1905$ paper ([9],[23]) by Einstein,  were at once, as a working assumption, deciding to restrict their study to a linear group reducing to the Galil\'{e}e group when the speed of light was going to infinity.  It must be noticed that people did believe that Einstein had not been influenced in $1905$ by the Michelson and Morley experiment of $1887$ till the discovery of hand written notes taken during lectures given by Einstein in Chicago ($1921$) and Kyoto ($1922$).  \\

\noindent
$\bullet$ $15$ books were trying to " {\it prove } " that the conformal factor was indeed reduced to a constant equal to $1$ when space-time  was supposed to be homogeneous and isotropic.  \\

\noindent
$\bullet$ $5$ books {\it only} were claiming that the conformal factor could eventually depend on the property of space-time, adding however that, if there was no surrounding electromagnetism or gravitation, the situation should be reduced to the preceding one but nothing was said otherwise.  \\

The sudent was so disgusted by such a state of affair that he decided to give up on research and to become a normal civil engineer. As a byproduct, if group theory must be used, the underlying group of transformations of space-time {\it must} be related to the propagation of light {\it by itself} rather than by considering tricky signals between observers, thus {\it must} have to do with the biggest group of invariance of Maxwell equations ([22],[54]). However, at the time we got the solution of this problem with the publication of ([27) in $1988$ (See [46] for recent results), a deep confusion was going on which is still not acknowledged though it can be explained in a few lines ([14]). Using standard notations of differential geometry, if the $2$-form $F\in {\wedge}^2T^*$ describing the EM field is satifying the {\it first set} of Maxwell equations, it amounts to say that it is closed, that is killed by the exterior derivative $d:{\wedge}^2T^* \rightarrow {\wedge}^3T^*$. The EM field can be thus (locally) parametrized by the EM potential $1$-form $A\in T^*$ with $dA=F$ where $d:T^* \rightarrow {\wedge}^2T^*$ is again the exterior derivative, because $d^2=d\circ d=0$. Now, if $E$ is a vector bundle over a manifold $X$ of dimension $n$, then we may define its {\it adjoint} vector bundle $ad(E)={\wedge}^nT^*\otimes E^*$ where $E^*$ is obtained from $E$ by inverting the transition rules, like $T^*$ is obtained from $T=T(X)$ and such a construction can be extended to linear partial differential operators between (sections of) vector bundles. When $n=4$, it follows that the {\it second set} of Maxwell equations for the EM induction is just described by $ad(d):{\wedge}^4T^* \otimes {\wedge}^2T \rightarrow {\wedge}^4T^*\otimes T$, {\it independently of any Minkowski constitutive relation between field and induction}. Using {\it Hodge duality} with respect to the volume form $dx=dx^1\wedge ... \wedge dx^4$, this operator is isomorphic to $d: {\wedge}^2T^* \rightarrow {\wedge}^3T^*$. It follows that {\it both the first set and second set of Maxwell equations are invariant by any diffeomorphism} and that the conformal group of space-time is the biggest group of transformations preserving the Minkowski constitutive relations {\it in vacuum} where the speed of light is trully $c$ as a universal constant. It was thus natural to believe that the mathematical structure of electromagnetism and gravitation had only to do with such a group having:  
\[ 4 \,\,translations + 6  \,\,rotations +1 \,\, dilatation +  4 \,\,elations = 15 \,\, parameters \] 
the main difficulty being to deal with these later non-linear tranformations. Of course, such a challenge could not be solved without the help of the non-linear theory of partial differential equations and Lie pseudogroups combined with homological algebra, that is before $1995$ {\it at least} ([28]).   \\

{\it From a purely physical point of view}, these new nonlinear methods have been introduced for the first time in 1909 by the brothers E. and F. Cosserat for studying the mathematical foundations of EL ([1],[7],[8],[18],[29-30],[50]). We have presented their link with the {\it nonlinear Spencer differential sequences} existing in the formal theory of Lie pseudogroups at the end of our book " Differential Galois Theory " published in 1983 ([26]). Similarly, the conformal methods have been introduced by H. Weyl in 1918 for revisiting the mathematical foundations of EM ([54]). We have presented their link with the above approach  through a {\it unique} differential sequence only depending on the structure of the {\it conformal group} in our book " Lie Pseudogroups and Mechanics " published in 1988 ([27]). However, the Cosserat brothers were only dealing with {\it translations} and {\it rotations} while Weyl was only dealing with {\it dilatation} and {\it elations}. Also, as an additional condition not fulfilled by the classical Einstein-Fokker-Nordstr\"{o}m theory ([10]), if the conformal factor has to do with gravitation, it {\it must} be defined everywhere but at the central attractive mass as we already said.  \\

{\it From a purely mathematical point of view}, the concept of a finite length {\it differential sequence}, now called {\it Janet sequence}, has been first described as a footnote by M. Janet in 1920 ([16]). Then, the work of D. C. Spencer in 1970 has been the first attempt to use the formal theory of systems of partial differential equations that he developped himself in order to study the formal theory of Lie pseudogroups ([12-13],[19],[49]). However, the {\it nonlinear Spencer sequences for Lie pseudogroups}, though never used in physics, largely supersede the "{\it Cartan structure equations} " introduced by E.Cartan in 1905 ([4-6],[17]) and are different from the "{\it Vessiot structure equations} " introduced by E. Vessiot in 1903 ([51-52]) for the same purpose but still not known today after more than a century because they have never been acknowledged by Cartan himself or even by his successors.   \\

The purpose of the present difficult paper is to apply these new methods for studying the common nonlinear conformal origin of gravitation 
{\it and} electromagnetism, {\it in a purely mathematical way}, by constructing explicitly the corresponding nonlinear Spencer sequence. All the physical consequences will be presented in another paper.   \\

\noindent
{\bf 2) GROUPOIDS AND ALGEBROIDS} \\ 
 
Let us now turn to the clever way proposed by Vessiot in 1903 ([51]) and 1904 ([52]). Our purpose is only to sketch the main results that we have obtained in many books ([25-28], we do not know other references) and to illustrate them by a series of specific examples, asking the reader to imagine any link with what has been said. We break the study into $9$ successive steps.  \\

\noindent
1) If ${\cal{E}}=X\times X$, we shall denote by ${\Pi}_q={\Pi}_q(X,X)$ the open subfibered manifold of $J_q(X\times X)$ defined independently of the coordinate system by $det(y^k_i)\neq 0$ with {\it source projection} ${\alpha}_q:{\Pi}_q\rightarrow X:(x,y_q)\rightarrow (x)$ and {\it target projection} ${\beta}_q:{\Pi}_q\rightarrow X:(x,y_q)\rightarrow (y)$. We shall sometimes introduce a copy $Y$ of $X$ with local coordinates $(y)$ in order to avoid any confusion between the source and the target manifolds. In order to construct another nonlinear sequence, we need a few basic definitions on {\it Lie groupoids} and {\it Lie algebroids} that will become substitutes for Lie groups and Lie algebras. The first idea is to use the chain rule for derivatives $j_q(g\circ f)=j_q(g)\circ j_q(f)$ whenever $f,g\in aut(X)$ can be composed and to replace both $j_q(f)$ and $j_q(g)$ respectively by $f_q$ and $g_q$ in order to obtain the new section $g_q\circ f_q$. This kind of "composition" law can be written in a pointwise symbolic way by introducing another copy $Z$ of $X$ with local coordinates $(z)$ as follows:\\
\[ {\gamma}_q:{\Pi}_q(Y,Z){\times}_Y{\Pi}_q(X,Y)\rightarrow {\Pi}_q(X,Z):((y,z,\frac{\partial z}{\partial y},...),(x,y,\frac{\partial y}{\partial x},...)\rightarrow (x,z,\frac{\partial z}{\partial y}\frac{\partial y}{\partial x},...)      \]
We may also define $j_q(f)^{-1}=j_q(f^{-1})$ and obtain similarly an "inversion" law.\\

\noindent
{\bf DEFINITION 2.1}: A fibered submanifold ${\cal{R}}_q\subset {\Pi}_q$ is called a {\it system of finite Lie equations} or a {\it Lie groupoid} of order $q$ if we have an induced {\it source projection} ${\alpha}_q:{\cal{R}}_q\rightarrow X$, {\it target projection} ${\beta}_q:{\cal{R}}_q\rightarrow X$, {\it composition} ${\gamma}_q:{\cal{R}}_q{\times}_X{\cal{R}}_q\rightarrow {\cal{R}}_q$, {\it inversion} ${\iota}_q:{\cal{R}}_q\rightarrow {\cal{R}}_q$ and {\it identity} $j_q(id)=id_q:X\rightarrow {\cal{R}}_q$. In the sequel we shall only consider {\it transitive} Lie groupoids such that the map $({\alpha}_q,{\beta}_q):{\cal{R}}_q\rightarrow X\times X $ is an epimorphism and we shall denote by ${\cal{R}}^0_q=id^{-1}({\cal{R}}_q)$ the {\it isotropy Lie group bundle} of ${\cal{R}}_q$. Also, one can prove that the new system ${\rho}_r({\cal{R}}_q)={\cal{R}}_{q+r}$ obtained by differentiating $r$ times all the defining equations of ${\cal{R}}_q$ is a Lie groupoid of order $q+r$. \\
Let us start with a Lie pseudogroup $\Gamma\subset aut(X)$ defined by a system ${\cal{R}}_q\subset {\Pi}_q$ of order $q$. Roughly speaking, if $f,g\in \Gamma\Rightarrow g\circ f, f^{-1} \in \Gamma)$ but such a definition is totally meaningless in actual practice as it cannot be checked most of the 
time. In all the sequel we shall suppose that the system is involutive ([25-28],[30]) and that $\Gamma$ is {\it transitive} that is $\forall x,y\in X, \exists f\in \Gamma, y=f(x)$ or, equivalently, the map $({\alpha}_q,{\beta}_q):{\cal{R}}_q\rightarrow X\times X:(x,y_q)\rightarrow (x,y)$ is surjective.\\

\noindent
2) The Lie algebra $\Theta\subset T$ of infinitesimal transformations is then obtained by linearization, setting $y=x+t\xi(x)+...$ and passing to the limit $t\rightarrow 0$ in order to obtain the linear involutive system $R_q= id^{-1}_q(V({\cal{R}}_q))\subset J_q(T)$ by reciprocal image with $\Theta=\{\xi \in T{\mid}j_q(\xi) \in R_q\}$. We define the {\it isotropy Lie algebra bundle} $R^0_q\subset J^0_q(T)$ by the short exact sequence $0 \rightarrow R^0_q \rightarrow R_q \stackrel{{\pi}^q_0}{\longrightarrow} T \rightarrow 0$. \\

\noindent
3) Passing from source to target, we may {\it prolong} the vertical infinitesimal transformations $\eta={\eta}^k(y)\frac{\partial}{\partial y^k}$ to the jet coordinates up to order $q$ in order to obtain:\\
\[   {\eta}^k(y)\frac{\partial}{\partial y^k}+(\frac{\partial {\eta}^k}{\partial y^r}y^r_i) \frac{\partial}{\partial y^k_i}+(\frac{{\partial}^2{\eta}^k}{\partial y^r\partial y^s}y^r_iy^s_j+\frac{\partial {\eta}^k}{\partial y^r}y^r_{ij})\frac{\partial}{\partial y^k_{ij}}+...    \]
where we have replaced $j_q(f)(x)$ by $y_q$, each component beeing the "formal" derivative of the previous one .\\

\noindent
4) As $[\Theta,\Theta]\subset \Theta$, we may use the Frobenius theorem in order to find a generating fundamental set of {\it differential invariants} $\{{\Phi}^{\tau}(y_q)\}$ up to order $q$ which are such that ${\Phi}^{\tau}({\bar{y}}_q)={\Phi}^{\tau}(y_q)$ by using the chain rule for derivatives whenever $\bar{y}=g(y)\in \Gamma$ acting now on $Y$.  Specializing the ${\Phi}^{\tau}$ at $id_q(x)$ we obtain the {\it Lie form} ${\Phi}^{\tau}(y_q)={\omega}^{\tau}(x)$ of ${\cal{R}}_q$.\\
Of course, in actual practice {\it one must use sections of} $R_q$ {\it instead of solutions} and we now prove why the use of the Spencer operator becomes crucial for such a purpose. Indeed, using the {\it algebraic bracket} $\{ j_{q+1}(\xi),j_{q+1}(\eta)\}=j_q([\xi,\eta]), \forall \xi,\eta\in T$, we may  obtain by bilinearity a {\it differential bracket} on $J_q(T)$ extending the bracket on $T$:\\
\[   [{\xi}_q,{\eta}_q]=\{{\xi}_{q+1},{\eta}_{q+1}\}+i(\xi)D{\eta}_{q+1}-i(\eta)D{\xi}_{q+1}, \forall {\xi}_q,{\eta}_q\in J_q(T) \]
which does not depend on the respective lifts ${\xi}_{q+1}$ and ${\eta}_{q+1}$ of ${\xi}_q$ and ${\eta}_q$ in $J_{q+1}(T)$. This bracket on sections satisfies the Jacobi identity and we set ([25-28]):\\

\noindent
{\bf DEFINITION 2.2}: We say that a vector subbundle $R_q\subset J_q(T)$ is a {\it system of infinitesimal Lie equations} or a {\it Lie algebroid} if $[R_q,R_q]\subset R_q$, that is to say ${\xi}_q,{\eta}_q\in R_q \Rightarrow [{\xi}_q,{\eta}_q]\in R_q$. Such a definition can be tested by means of computer algebra. We shall also say that $R_q$ is {\it transitive} if we have the short exact sequence $0\rightarrow R^0_q \rightarrow R_q \stackrel{{\pi}^q_0}{ \rightarrow} T  \rightarrow 0$. In that case, a {\it splitting} of this sequence, namely a map ${\chi}_q:T \rightarrow R_q$ such that $ {\pi}^q_0\circ {\chi}_q=id_T$ or equivalently a section ${\chi}_q\in T^*\otimes R_q$ over $id_T\in T^*\otimes T$, is called a $R_q$-{\it connection} and its {\it curvature} ${\kappa}_q\in {\wedge}^2T^*\otimes R^0_q$ is defined by ${\kappa}_q(\xi,\eta)=[{\chi}_q(\xi),{\chi}_q(\eta)]-{\chi}_q([\xi,\eta]), \forall \xi,\eta \in T$.\\

\noindent
{\bf PROPOSITION 2.3}: If $[R_q,R_q]\subset R_q$, then $[R_{q+r},R_{q+r}]\subset R_{q+r}, \forall r\geq 0$.  \\

\noindent
{\it Proof}: When $r=1$, we have ${\rho}_1(R_q)=R_{q+1}=\{ {\xi}_{q+1}\in J_{q+1}(T)\mid {\xi}_q\in R_q, D{\xi}_{q+1}\in T^*\otimes R_q\}$ and we just need to use the following formulas showing how $D$ acts on the various brackets (See [25],[28] and [40] for more details):  \\
\[  i(\zeta)D\{{\xi}_{q+1},{\eta}_{q+1}\}=\{i(\zeta)D{\xi}_{q+1},{\eta}_q\}+\{{\xi}_q,i(\zeta)D{\eta}_{q+1}\} ,\hspace {4mm} \forall \zeta \in T \]  
\[ \begin{array}{rcl}
 i(\zeta)D[{\xi}_{q+1},{\eta}_{q+1}]  & =  & [i(\zeta)D{\xi}_{q+1},{\eta}_q]+[{\xi}_q,i(\zeta)D{\eta}_{q+1}]  \\
     &   &+i(L({\eta}_1)\zeta)D{\xi}_{q+1}-i(L({\xi}_1)\zeta)D{\eta}_{q+1}  
\end{array} \]
because the right member of the second formula is a section of $R_q$ whenever ${\xi}_{q+1},{\eta}_{q+1}\in R_{q+1}$. The first formula may be used when $R_q$ is formally integrable. \\
\hspace*{10cm}   Q.E.D.  \\

\noindent
{\bf EXAMPLE 2.4}: With $n=1, q=3, X=\mathbb{R}$ and evident notations, the components of $[{\xi}_3,{\eta}_3]$ at order zero, one, two and three are defined by the totally unusual successive formulas:\\
\[    [\xi,\eta]=\xi{\partial}_x\eta-\eta{\partial}_x\xi     \]
\[    ([{\xi}_1,{\eta}_1])_x=\xi{\partial}_x{\eta}_x-\eta{\partial}_x{\xi}_x    \]
\[    ([{\xi}_2,{\eta}_2])_{xx}={\xi}_x{\eta}_{xx}-{\eta}_x{\xi}_{xx}+\xi{\partial}_x{\eta}_{xx}-\eta{\partial}_x{\xi}_{xx}   \]
\[    ([{\xi}_3,{\eta}_3])_{xxx}=2{\xi}_x{\eta}_{xxx}-2{\eta}_x{\xi}_{xxx}+\xi {\partial}_x{\eta}_{xxx}-\eta{\partial}_x{\xi}_{xxx}   \]
For affine transformations, ${\xi}_{xx}=0,{\eta}_{xx}=0\Rightarrow ([{\xi}_2,{\eta}_2])_{xx}=0$ and thus $[R_2,R_2]\subset R_2$.\\
For projective transformations, ${\xi}_{xxx}=0,{\eta}_{xxx}=0 \Rightarrow ([{\xi}_3,{\eta}_3])_{xxx}=0$ and thus $[R_3,R_3]\subset R_3$.  \\

\noindent
{\bf THEOREM 2.5}: ({\it prolongation/projection} (PP) {\it procedure}) If an arbitrary system $R_q\subseteq J_q(E)$ is given, one can {\it effectively} find two integers $r,s \geq 0$ such that the system $R^{(s)}_{q+r}$ is formally integrable or even involutive. \\

\noindent
{\bf COROLLARY 2.6}: The bracket is compatible with the PP procedure:    \\ 
\[   [R_q,R_q] \subset R_q \hspace{5mm}  \Rightarrow \hspace{5mm} [R^{(s)}_{q+r},R^{(s)}_{q+r} ] \subset  R^{(s)}_{q+r}, \forall r,s \geq 0   \]

\noindent
{\bf EXAMPLE 2.7}: With $n=m=2$ and $q=1$, let us consider the Lie pseudodogroup $\Gamma \subset aut(X)$ of finite transformations $y=f(x)$ such that $y^2dy^1=x^2dx^1=\omega =(x^2,0)\in T^*$. Setting $y=x+t {\xi}(x)+...$ and linearizing, we get the Lie operator ${\cal{D}}\xi={\cal{L}}(\xi)\omega $ where ${\cal{L}}$ is the Lie derivative because it is well known that $[ {\cal{L}}(\xi),{\cal{L}}(\eta)]={\cal{L}}(\xi)\circ {\cal{L}}(\eta) - {\cal{L}}(\eta) \circ {\cal{L}}(\xi)={\cal{L}}([\xi,\eta])$ in the operator sense. The system $R_1\subset J_1(T)$ of linear infinitesimal Lie equations is:\\
\[  x^2{\partial}_1{\xi}^1 + {\xi}^2=0, \hspace{1cm}  {\partial}_2{\xi}^1=0   \]
Replacing $j_1(\xi)$ by a section ${\xi}_1 \in J_1(T)$, we have:  \\
\[   {\xi}^1_1= - \frac{1}{x^2}  {\xi}^2, \hspace{1cm}  {\xi}^1_2=0   \]
Let us choose the two sections:  \\
\[  {\xi}_1= \{ {\xi}^1=0, {\xi}^2= - x^2, {\xi}^1_1=1, {\xi}^1_2=0, {\xi}^2_1=0, {\xi}^2_2=0\}\in R_1  \] 
\[    {\eta}_1= \{ {\eta}^1=x^2, {\eta}^2=0, {\eta}^1_1=0, {\eta}^1_2=- x¬2, {\eta}^2_1=0, {\eta}^2_2=1 \} \in R_1    \]
We let the reader check that $[{\xi}_1,{\eta}_1] \in R_1$. However, we have the strict inclusion $R^{(1)}_1 \subset R_1$ defined by the additional equation ${\xi}^1_1 + {\xi}^2_2=0$ and thus ${\xi}_1, {\eta}_1 \notin R^{(1)}_1$ though we have indeed $[ R^{(1)}_1, R^{(1)}_1] \subset R^{(1)}_1$, a result not evident at all because the sections ${\xi}_1$ and ${\eta}_1$ have {\it nothing to do} with solutions. The reader may proceed in the same way with $ x^2dx^1 -x^1dx^2$ and compare.  \\

\noindent
5) The main discovery of Vessiot, as early as in $1903$ and thus fifty years in advance, has been to notice that the prolongation at order $q$  of any horizontal vector field $\xi={\xi}^i(x)\frac{\partial}{\partial x^i}$ commutes with the prolongation at order $q$ of any vertical vector field $\eta={\eta}^k(y)\frac{\partial}{\partial y^k}$, exchanging therefore the differential invariants. Keeping in mind the well known property of the Jacobian determinant while passing to the finite point of view, any (local) transformation $y=f(x)$ can be lifted to a (local) transformation of the differential invariants between themselves of the form $u\rightarrow \lambda(u,j_q(f)(x))$ allowing to introduce a {\it natural bundle} $\cal{F}$ over $X$ by patching changes of coordinates $\bar{x}=\varphi(x), \bar{u}=\lambda(u,j_q(\varphi)(x))$. A section $\omega$ of $\cal{F}$ is called a {\it geometric object} or {\it structure} on $X$ and transforms like ${\bar{\omega}}(f(x))=\lambda(\omega(x),j_q(f)(x))$ or simply $\bar{\omega}=j_q(f)(\omega)$. This is a way to generalize vectors and tensors ($q=1$) or even connections ($q=2$). As a byproduct we have $\Gamma=\{f\in aut(X){\mid} {\Phi}_{\omega}(j_q(f))=j_q(f)^{-1}(\omega)=\omega\}$ as a new way to write out the Lie form and we may say that $\Gamma$ {\it preserves} $\omega$. We also obtain ${\cal{R}}_q=\{f_q\in {\Pi}_q{\mid} f_q^{-1}(\omega)=\omega\}$. Coming back to the infinitesimal point of view and setting $f_t=exp(t\xi)\in aut(X), \forall \xi\in T$, we may define the {\it ordinary Lie derivative} with value in $F={\omega}^{-1}(V({\cal{F}}))$ by introducing the {\it vertical bundle} of ${\cal{F}}$ as a vector bundle over 
${\cal{F}}$ and the formula :\\
\[   {\cal{D}}\xi={\cal{L}}(\xi)\omega=\frac{d}{dt}j_q(f_t)^{-1}(\omega){\mid}_{t=0} \Rightarrow \Theta=\{\xi\in T{\mid}{\cal{L}}(\xi)\omega=0\}      \]
while we have $x\rightarrow x+t\xi(x)+...\Rightarrow u^{\tau}\rightarrow u^{\tau}+t{\partial}_{\mu}{\xi}^kL^{\tau\mu}_k(u)+...$ where $\mu=({\mu}_1,...,{\mu}_n)$ is a multi-index as a way to write down the system $R_q\subset J_q(T)$ of infinitesimal Lie equations in the {\it Medolaghi form}:\\
\[     {\Omega}^{\tau}\equiv ({\cal{L}}(\xi)\omega)^{\tau}\equiv -L^{\tau\mu}_k(\omega(x)){\partial}_{\mu}{\xi}^k+{\xi}^r{\partial}_r{\omega}^{\tau}(x)=0    \]

\noindent
{\bf EXAMPLE 2.8}: With $n=1$, let us consider the Lie group of projective transformations $y=(ax+b)/(cx+d)$ as a lie pseudogroup. Differentiating three times in order to eliminate the parameters, we obtain the third order {\it Schwarzian} OD equation and its linearization over $y=x$:  \\
\[  {\cal{R}}_3 \subset {\Pi}_3      \hspace{3cm}  \Psi  \equiv  \frac{y_{xxx}}{y_x} - \frac{3}{2}( \frac{y_{xx}}{y_x})^2=0   \]
\[  R_3 \subset J_3(T)    \hspace{3cm}   {\xi}_{xxx}=0    \]
Accordingly, the prolongation $\sharp ({\eta}_3)$of any ${\eta}_3 \in J_3(T(Y))$ over $Y$ such that ${\eta}_{yyy}=0$ becomes:  \
\[  {\eta}(y) \frac{\partial}{\partial y} + {\eta}_y(y) ( y_x \frac{\partial}{\partial y_x} + y_{xx} \frac{\partial}{\partial y_{xx}}+ y_{xxx}\frac{\partial}{\partial y_{xxx}}) + {\eta}_{yy}(y)( (y_x)^2 \frac{\partial}{\partial y_{xx}} + 3 y_xy_{xx} \frac{\partial}{\partial y_{xxx}})  \]
It follows that $\Psi$ is a generating third order differential invariant and ${\cal{R}}_3$ is in Lie form. \\
Now, we have: 
\[   \bar{x}=\varphi(x) \Rightarrow  y_x=y_{\bar{x}}{\partial}_x\varphi, y_{xx}= y_{\bar{x}\bar{x}}({\partial}_x\varphi)^2 + y_{\bar{x}}{\partial}_{xx}\varphi, y_{xxx}= y_{\bar{x}\bar{x}\bar{x}}({\partial}_x\varphi)^3 + 3 y_{\bar{x}\bar{x}}{\partial}_x\varphi {\partial}_{xx}\varphi + y_{\bar{x}}{\partial}_{xxx}\varphi  \]
and the natural bundle ${\cal{F}}$ is thus defined by the transition rules:  \\
\[  \bar{x}=\varphi (x),\,\,\,   u= \bar{u}({\partial}_x\varphi)^2 + (  \frac{{\partial}_{xxx}\varphi}{{\partial}_x\varphi} - 
\frac{3}{2}( \frac{{\partial}_{xx}\varphi}{{\partial} _x\varphi})^2)  \]
The general Lie form of ${\cal{R}}_3$  is:   \\
\[  \frac{y_{xxx}}{y_x} -\frac{3}{2} (\frac{y_{xx}}{y_x})^2 + \omega(y)(y_x)^2=\omega(x)  \]
We obtain $R_3\subset J_3(T)$ in Medolaghi form as follows:   \\
\[    \Omega \equiv {\cal{L}}(\xi)\omega \equiv {\partial}_{xxx}\xi +2 \, {\omega (x) {\partial}_x\xi + \xi \partial}_x\omega(x) =0  \]
Using a section ${\xi}_3\in J_3(T)$, we finally get the formal Lie derivative:  \\
\[ \Omega \equiv L({\xi}_3)\omega \equiv {\xi}_{xxx} + 2 \, \omega (x) {\xi}_x + \xi {\partial}_x\omega (x) =0  \]
and let the reader ckeck directly that $[L({\xi}_3),L({\eta}_3)]= L ([{\xi}_3,{\eta}_3]), \forall {\xi}_3,{\eta}_3 \in J_3(T)$, a result absolutely not evident at first sight ([47]).   \\

\noindent
7) By analogy with "special" and "general" relativity, we shall call the given section {\it special} and any other arbitrary section {\it general}. The problem is now to study the formal properties of the linear system just obtained with coefficients only depending on $j_1(\omega)$, exactly like L.P. Eisenhart did for ${\cal{F}}=S_2T^*$ when finding the constant Riemann curvature condition for a metric $\omega$ with 
$det(\omega)\neq 0$ ([11],[28], Example 10, p 246 to 256). Indeed, if any expression involving $\omega$ and its derivatives is a scalar object, it must reduce to a constant because $\Gamma$ is assumed to be transitive and thus cannot be defined by any zero order equation. Now one can prove that the CC for $\bar{\omega}$, thus for $\omega$ too, only depend on the $\Phi$ and take the quasi-linear symbolic form $v\equiv I(u_1)\equiv A(u)u_x+B(u)=0$, allowing to define an affine subfibered manifold ${\cal{B}}_1\subset J_1({\cal{F}})$ over $\cal{F}$. 
Now, if one has two sections $\omega$ and $\bar{\omega}$ of $\cal{F}$, the {\it equivalence problem} is to look for $f\in aut(X)$ such that $j_q(f)^{-1}(\omega)=\bar{\omega}$. When the two sections satisfy the same CC, the problem is sometimes locally possible (Lie groups of transformations, Darboux problem in analytical mechanics,...) but sometimes not ([23], p 333).\\

\noindent
8) Instead of the CC for the equivalence problem, let us look for the {\it integrability conditions} (IC) for the system of infinitesimal Lie equations and suppose that, for the given section, all the equations of order $q+r$ are obtained by differentiating $r$ times only the equations of order $q$, then it was claimed by Vessiot ([50] with no proof, see [28], p 207-211) that such a property is held if and only if there is an equivariant section 
$c:{\cal{F}}\rightarrow {\cal{F}}_1:(x,u)\rightarrow (x,u,v=c(u))$ where ${\cal{F}}_1=J_1({\cal{F}})/{\cal{B}}_1$ is a natural vector bundle over $\cal{F}$ with local coordinates $(x,u,v)$. Moreover, any such equivariant section depends on a finite number of constants $c$ called {\it structure constants} and the IC for the {\it Vessiot structure equations} $I(u_1)=c(u)$ are of a polynomial form $J(c)=0$.\\

\noindent
{\bf EXAMPLE 2.9}: Comig back to Example $2.7$ first considered by Vessiot as early as in $1903$ ([51]), the geometric object $\omega=(\alpha,\beta)\in T^*{\otimes}_X{\wedge}^2T^*$ must satisfy the Vessiot structure equation $d\alpha=c \, \beta$ with a single Vessiot structure constant $c= - 1$ in the situation considered where $\alpha=x^2dx^1$ and $\beta = dx^1 \wedge dx^2$ (See ([40]) for other examples and applications). As a byproduct, there is no conceptual difference between such a constant and the constant appearing in the constant Riemannian curvature condition of Eisenhart ([11]).  \\

\noindent
9) Finally, when $Y$ is no longer a copy of $X$, a system ${\cal{A}}_q\subset J_q(X\times Y)$ is said to be an {\it automorphic system} for a Lie pseudogroup $\Gamma\subset aut(Y)$ if, whenever $ y=f(x)$ and $\bar{y}=\bar{f}(x)$ are two solutions, then there exists one and only one transformation $\bar{y}=g(y)\in \Gamma$ such that $\bar{f}=g\circ f$. Explicit tests for checking such a property formally have been given in ([26],[42]) and can be implemented on computer in the differential algebraic framework.\\

\noindent
{\bf 3)  NONLINEAR SEQUENCES}\\

Contrary to what happens in the study of Lie pseudogroups and in particular in the study of the algebraic ones that can be found in mathematical physics, {\it nonlinear operators do not in general admit} CC, unless they are defined by differential polynomials, as can be seen by considering the two following examples with $m=1, n=2, q=2$. With standard notations from differential algebra, if we are dealing with a ground differential field $K$, like $\mathbb{Q}$ in the next examples, we denote by $K\{y\}$ the ring (which is even an integral domain) of differential polynomials in $y$ with coefficients in $K$ and by $K<y>= Q(K\{y\})$ the corresponding quotient field of differential rational functions in $y$. Then, if $u,v\in K<y>$, we have the two towers $K \subset K<u> \subset K<y>$ and $K \subset K<v> \subset K<y>$ of extensions, thus the tower $K \subset K<u,v> \subset K<y>$. Accordingly, the differential extension $K<u,v> /K$ is a finitely generated differential extension. If we consider $u$ and $v$ as new indeterminates, then $K<u>$ and $K<v>$  are both differential transcendental extensions of $K$ and the kernel of the  canonical differential morphism 
$K\{u\} {\otimes }_K K\{v\} \rightarrow K<y>$ is a prime differential ideal in the differential integral domain $K<u> {\otimes}_K K<v>$, a way to describe by residue the smallest differential field containing $K<u>$ and $K<v>$ in $K<y>$. Of course, the true difficulty is to find out such a prime differential ideal.  \\

\noindent
{\bf EXAMPLE 3.1}: First of all, let us consider the following nonlinear system in $y$ with second member $(u,v)$:   \\
\[ P \equiv  y_{22}-\frac{1}{3}(y_{11})^3=u,\,\, Q\equiv  y_{12}-\frac{1}{2}(y_{11})^2=v \Rightarrow  y_{11}= \frac{u_1 - v_2}{v_1}  \]
The differential ideal $\mathfrak{a}$ generated by $P$ and $Q$ in $\mathbb{Q}\{y\}$ is prime because $d_2Q + d_1P - y_{11}d_1Q=0$ and thus $\mathbb{Q} \{y\}/\{P,Q\}\simeq \mathbb{Q}[ y,y_1,y_2,y_{11}, y_{111}, ...]$ is an integral domain. \\
We may consider the following nonlinear involutive system with two equations:   \\ 
\[  \left \{  \begin{array}{cclc}
 y_{22}-\frac{1}{3}(y_{11})^3 & = & 0  \\
 y_{12}-\frac{1}{2}(y_{11})^2 & = & 0 
 \end{array} \right.  \fbox{ $ \begin{array}{cc}
 1 & 2 \\
 1 & \bullet 
 \end{array} $}  \]
We have also the linear inhomogeneous finite type second order system with three equations:  \\
\[   \left \{  \begin{array}{ccl}
y_{22} & =  & u + \frac{1}{3}( \frac{u_1 - v_2}{v_1})^3   \\
y_{12} & = & v+ \frac{1}{2}(\frac{u_1 - v_2}{v_1})^2  \\
y_{11} & =  & \frac{u_1- v_2}{v_1}
\end{array} \right.   \fbox{  $  \begin{array} {cc}
1 & 2   \\
1 & \bullet  \\
1 &  \bullet  
\end{array}  $  }  \] 
Though we have {\it a priori} two CC, we let the reader prove, as a delicate exercise, that there is only the single nonlinear second order CC obtained from the bottom dot.:  \\
\[  \fbox{$ d_2 (\frac{u_1 - v_2}{v_1}) - d_1 (v + \frac{1}{2} ( \frac{u_1 - v_2}{v_1})^2) =0 $ }  \]

\noindent
{\bf EXAMPLE 3.2}: On the contrary, if we consider the following new nonlinear system:  \\
\[ P \equiv  y_{22}-\frac{1}{2}(y_{11})^2=u,\,\, Q\equiv  y_{12}-y_{11}=v \Rightarrow  (y_{11}-1)y_{111}= v_2 + v_1 - u_1 = w  \]
we obtain successively:  \\
\[  d_2 Q +  d_1 Q - d_1P\equiv (y_{11} - 1)y_{111}  \]
\[  y_{111} ( d_{12} Q +  d_{11}Q - d_{11} P) - y_{1111}( d_2 Q + d_1 Q - d_1 P)= (y_{111})^3  \]
The symbol at order 3 is thus not a vector bundle and no direct study as above can be used because the differential ideal generated by $(P,Q)$ is not perfect as it contains $(y_{111})^3$ without containing $y_{111}$ (See [26] and [41] for more details). The following nonlinear system is {\it not} involutive:  \\
\[  \left \{  \begin{array}{lcc}
 y_{22}-\frac{1}{2}(y_{11})^2 & = & 0  \\
 y_{12}- y_{11} & = & 0 
 \end{array} \right.  \fbox{ $ \begin{array}{cc}
 1 & 2 \\
 1 & \bullet 
 \end{array} $}  \]
We have the following four generic nonlinear additional finite type third order equations:  \\
\[  \left \{   \begin{array}{lcl}
 y_{222} -  y_{11}(v_1 + \frac{w}{y_{11}-1})&=& u_2  \\
 y_{122} - y_{11}\frac{w}{y_{11} -1} & = & u_1  \\
 y_{112} - \frac{w}{y_{11} - 1} & = & v_1   \\
 y_{111} - \frac{w}{y_{11}- 1}  & =  & 0
\end{array} \right.   \fbox{  $  \begin{array} {cc}
1 & 2   \\
1 & \bullet  \\
1 &  \bullet  \\
1 & \bullet
\end{array}  $  }  \]  
Though we have now {\it a priori} three CC and thus three additional equations because the system is not involutive, setting $y_{11}- 1=z \Rightarrow y_{112}=z_2,y_{111}=z_1$, there is only the single additional nonlinear second  order equation:   \\
 \[  \fbox{ $v_{11}z^2 + (w_1 - w_2) z + v_1 w =0 $}   \] 
Differentiating once and using the relation $zz_1=w$, we get:  \\
\[  \fbox { $ v_{111} z^3 + (w_{11}- w_{12}) z^2 + (v_1w_1 + 3 v_{11}w) z + (w_1 - w_2)w= 0 $ } \]
a result leading to a tricky resultant providing a third order differential polynomial in $(u,v)$.  \\

However, the kernel of a linear operator ${\cal{D}}:E\rightarrow F$ is always taken with respet to the zero section of $F$, while it must be taken with respect to a prescribed section by a {\it double arrow} for a nonlinear operator. Keeping in mind the linear Janet sequence and the examples of Vessiot structure equations already presented, one obtains:  \\

\noindent
{\bf THEOREM 3.3}: There exists a {\it nonlinear Janet sequence} associated with the Lie form of an involutive system of finite Lie equations:   \\
\[  \begin{array}{rcccl}
  & {\Phi}\circ j_q &   &  I\circ j_1  &   \\
  0\rightarrow \Gamma \rightarrow aut(X) &\rightrightarrows &  {\cal{F}}  &\rightrightarrows   &  {\cal{F}}_1 \\
     & \omega\circ\alpha  &  &  0 &
  \end{array}  \]
where the kernel of the first operator $f\rightarrow {\Phi}\circ j_q(f)={\Phi}(j_q(f))=j_q(f)^{-1}(\omega)$ is taken with respect to the section $\omega$ of $\cal{F}$ while the kernel of the second operator $\omega\rightarrow I(j_1(\omega))\equiv A(\omega){\partial}_x\omega+B(\omega)$ is taken with respect to the zero section of the vector bundle ${\cal{F}}_1$ over ${\cal{F}}$.\\

\noindent
{\bf COROLLARY 3.4}: By linearization at the identity, one obtains the involutive {\it Lie operator} ${\cal{D}}:T\rightarrow F_0:\xi\rightarrow {\cal{L}}(\xi)\omega $ with kernel $\Theta=\{\xi\in T{\mid}{\cal{L}}(\xi)\omega=0\}\subset T$ satisfying $[\Theta,\Theta]\subset \Theta$ and the corresponding {\it linear Janet sequence}:   \\
\[   0 \rightarrow  \Theta  \rightarrow T  \stackrel{{\cal{D}}}{\longrightarrow} F_0 \stackrel{{\cal{D}}_1}{\longrightarrow}  F_1       \]
where $F_0=F={\omega}^{-1}(V({\cal{F}}))$ and $F_1={\omega}^{-1}({\cal{F}}_1)$.\\

Now we notice that $T$ is a natural vector bundle of order $1$ and $J_q(T)$ is thus a natural vector bundle of order $q+1$. Looking at the way a vector field and its derivatives are transformed under any $f\in aut(X)$ while replacing $j_q(f)$ by $f_q$, we obtain:\\
\[  {\eta}^k(f(x))=f^k_r(x){\xi}^r(x) \Rightarrow {\eta}^k_u(f(x))f^u_i(x)=f^k_r(x){\xi}^r_i(x)+f^k_{ri}(x){\xi}^r(x)\]
and so on, a result leading to:\\

\noindent
{\bf LEMMA 3.5}: $J_q(T)$ is {\it associated} with ${\Pi}_{q+1}={\Pi}_{q+1}(X,X)$ that is we can obtain a new section ${\eta}_q=f_{q+1}({\xi}_q)$ from any section ${\xi}_q \in J_q(T)$ and any section $f_{q+1}\in {\Pi}_{q+1}$ by the formula:\\
\[ d_{\mu}{\eta}^k\equiv {\eta}^k_rf^r_{\mu}+ ...=f^k_r{\xi}^r_{\mu}+  ...  +f^k_{\mu+1_r}{\xi}^r , \forall 0\leq {\mid}\mu {\mid}\leq q\]
where the left member belongs to $V({\Pi}_q)$. Similarly $R_q\subset J_q(T)$ is associated with ${\cal{R}}_{q+1}\subset {\Pi}_{q+1}$.\\

More generally, looking now for transformations "close" to the identity, that is setting $y=x+t\xi(x)+...$ when $t\ll 1$ is a small constant parameter and passing to the limit $t\rightarrow 0$, we may linearize any (nonlinear) {\it system of finite Lie equations} in order to obtain a (linear) {\it system of infinitesimal Lie equations} $R_q\subset J_q(T)$ for vector fields. Such a system has the property that, if $\xi,\eta$ are two solutions, then $[\xi,\eta]$ is also a solution. Accordingly, the set $\Theta\subset T$ of its solutions satisfies $[\Theta,\Theta]\subset \Theta$ and can therefore be considered as the Lie algebra of $\Gamma$. \\

More generally, the next definition will extend the {\it classical Lie derivative}:    \\
\[  {\cal{L}}(\xi)\omega =(i(\xi)d +di(\xi))\omega=\frac{d}{dt}j_q(exp\hspace{2mm} t\xi)^{-1}(\omega){\mid}_{t=0}.  \]

\noindent
{\bf DEFINITION 3.6}: We say that a vector bundle $F$ is {\it associated} with $R_q$ if there exists a first order differential operator $L({\xi}_q):F \rightarrow  F$ called {\it formal Lie derivative} and such that:  \\
1)  $L({\xi}_q+{\eta}_q)=L({\xi}_q)+L({\eta}_q)     \hspace{2cm} \forall {\xi}_q,{\eta}_q\in R_q$.  \\
2)  $L(f{\xi}_q)=fL({\xi}_q)  \hspace{2cm}   \forall {\xi}_q\in R_q, \forall f\in C^{\infty}(X)$.   \\
3)  $[L({\xi}_q),L({\eta}_q)]=L({\xi}_q)\circ L({\eta}_q)-L({\eta}_q)\circ L({\xi}_q)=L([{\xi}_q,{\eta}_q])  \hspace{5mm}  \forall {\xi}_q,{\eta}_q\in R_q$.   \\
4)  $ L({\xi}_q)(f\eta)=fL({\xi}_q)\eta + ({\xi}.f)\eta \hspace{15mm} \forall {\xi}_q\in R_q,\forall f\in C^{\infty}(X), \forall \eta \in F$. \\

\noindent
{\bf LEMMA 3.7}: If $E$ and $F$ are associated with $R_q$, we may set on $E\otimes F$:Ê\\
\[    L({\xi}_q)(\eta \otimes \zeta)=L({\xi}_q)\eta\otimes \zeta+\eta\otimes L({\xi}_q)\zeta \hspace{2cm} \forall {\xi}_q\in R_q,\forall \eta \in E, \forall \zeta \in F   \] 

If $\Theta \subset T$ denotes the solutions of $R_q$, then we may set ${\cal{L}}(\xi)=L(j_q(\xi)), \forall \xi\in \Theta$ but no explicit computation can be done when $\Theta$ is infinite dimensional. However, we have:    \\

\noindent
{\bf PROPOSITION 3.8}: $J_q(T)$ is associated with $J_{q+1}(T)$ if we define:  \\
\[     L({\xi}_{q+1}){\eta}_q=\{{\xi}_{q+1},{\eta}_{q+1}\}+i(\xi)D{\eta}_{q+1}=[{\xi}_q,{\eta}_q]+i(\eta)D{\xi}_{q+1}        \]
and thus $R_q$ is associated with $R_{q+1}$.  \\

\noindent
{\it Proof}: It is easy to check the properties 1, 2, 4 and it only remains to prove property 3 as follows.\\
\[  \begin{array}{rcl}
[L({\xi}_{q+1}),L({\eta}_{q+1})]{\zeta}_q & = & L({\xi}_{q+1})(\{{\eta}_{q+1},{\zeta}_{q+1}\}+i(\eta)D{\zeta}_{q+1}) \\
 &   &-L({\eta}_{q+1})(\{{\xi}_{q+1},{\zeta}_{q+1}\}+i(\xi)D{\zeta}_{q+1})   \\
       &  =  & \{{\xi}_{q+1},\{{\eta}_{q+2},{\zeta}_{q+2}\}\}  -\{{\eta}_{q+1},\{{\xi}_{q+2},{\zeta}_{q+2}\}\}  \\
   &   & +\{{\xi}_{q+1},i(\eta)D{\zeta}_{q+2}\}-\{{\eta}_{q+1},i(\xi)D{\zeta}_{q+2}\}   \\
   &   & +i(\xi)D\{{\eta}_{q+2},{\zeta}_{q+2}\}-i(\eta)D\{{\xi}_{q+2},{\zeta}_{q+2}\}   \\
   &   & +i(\xi)D(i(\eta)D{\zeta}_{q+2})-i(\eta)D(i(\xi)D{\zeta}_{q+2})    \\
   &= & \{\{{\xi}_{q+2},{\eta}_{q+2}\},{\zeta}_{q+1}\}+\{i(\xi)D{\eta}_{q+2},{\zeta}_{q+1}\} \\ 
   &  &  - \{i(\eta)D{\xi}_{q+2},{\zeta}_{q+1}\} + i([\xi,\eta])D{\zeta}_{q+1}   \\
   &= & \{[{\xi}_{q+1},{\eta}_{q+1}],{\zeta}_{q+1}\}+i([\xi,\eta])D{\zeta}_{q+1}
   \end{array}   \]
by using successively the Jacobi identity for the algebraic bracket and the last proposition.Ê\\
\hspace*{10cm}   Q.E.D.   \\

\noindent
{\bf EXAMPLE 3.9}: $T$ and $T^*$ both with any tensor bundle are associated with $J_1(T)$. For $T$ we may define $L({\xi}_1)\eta=[\xi,\eta]+i(\eta)D{\xi}_1=\{{\xi}_1,j_1(\eta)\}$. We have ${\xi}^r{\partial}_r{\eta}^k-{\eta}^s{\partial}_s{\xi}^k+{\eta}^s({\partial}_s{\xi}^k-{\xi}^k_s)=-{\eta}^s{\xi}^k_s+{\xi}^r{\partial}_r{\eta}^k$ and the four properties of the formal Lie derivative can be checked directly. Of course, we find back ${\cal{L}}(\xi)\eta=[\xi,\eta], \forall \xi,\eta \in T$. We let the reader treat similarly the case of $T^*$. \\

\noindent
{\bf PROPOSITION 3.10}: There is a {\it first nonlinear Spencer sequence}:    \\
\[ 0\longrightarrow aut(X) \stackrel{j_{q+1}}{\longrightarrow} {\Pi}_{q+1}(X,X)\stackrel{\bar{D}}{\longrightarrow}T^*\otimes J_q(T)\stackrel{{\bar{D}}'}{\longrightarrow} {\wedge}^2T^*\otimes J_{q-1}(T)  \]
with $\bar{D}f_{q+1}\equiv f_{q+1}^{-1}\circ j_1(f_q)-id_{q+1}={\chi}_q \Rightarrow {\bar{D}}'{\chi}_q(\xi,\eta)\equiv D{\chi}_q(\xi,\eta)-\{{\chi}_q(\xi),{\chi}_q(\eta)\}=0 $. Moreover, setting ${\chi}_0=A-id\in T^*\otimes T$, this sequence is locally exact if $det(A)\neq 0$.\\

\noindent
{\it Proof}: There is a canonical inclusion ${\Pi}_{q+1}\subset J_1({\Pi}_q)$ defined by $y^k_{\mu,i}=y^k_{\mu+1_i}$ and the composition $f^{-1}_{q+1}\circ j_1(f_q)$ is a well defined section of $J_1({\Pi}_q)$ over the section $f^{-1}_q\circ f_q=id_q$ of ${\Pi}_q$ like $id_{q+1}$. The difference ${\chi}_q=f^{-1}_{q+1}\circ j_1(f_q)-id_{q+1}$ is thus a section of $T^*\otimes V({\Pi}_q)$ over $id_q$ and we have already noticed that 
$ id^{-1}_q(V({\Pi}_q))=J_q(T)$. For $q=1$ we get with $g_1=f^{-1}_1$:\\
\[ {\chi}^k_{,i}=g^k_l{\partial}_if^l-{\delta}^k_i=A^k_i-{\delta}^k_i,\hspace{5mm} {\chi}^k_{j,i}=g^k_l({\partial}_if^l_j-A^r_if^l_{rj})  \]
We also obtain from Lemma 3.5 the useful formula $ f^k_r{\chi}^r_{\mu,i}+...+f^k_{\mu+1_r}{\chi}^r_{,i}={\partial}_if^k_{\mu}-f^k_{\mu+1_i}$ allowing to determine ${\chi}_q$ inductively.\\
We refer to ([28], p 215-216) for the inductive proof of the local exactness, providing the only formulas that will be used later on and can be checked directly by the reader: \\
\[  {\partial}_i{\chi}^k_{,j}-{\partial}_j{\chi}^k_{,i}-{\chi}^k_{i,j}+{\chi}^k_{j,i}-({\chi}^r_{,i}{\chi}^k_{r,j}-{\chi}^r_{,j}{\chi}^k_{r,i})=0  
\eqno{(1)}     \]
\[  {\partial}_i{\chi}^k_{l,j}-{\partial}_j{\chi}^k_{l,i}-{\chi}^k_{li,j}+{\chi}^k_{lj,i}-({\chi}^r_{,i}{\chi}^k_{lr,j}+{\chi}^r_{l,i}{\chi}^k_{r,j}-{\chi}^r_{l,j}{\chi}^k_{r,i}-{\chi}^r_{,j}{\chi}^k_{lr,i})=0   \eqno{(2)}      \]
\[  {\partial}_i{\chi}^k_{lr,j} - {\partial}_j{\chi}^k_{lr,i}  - {\chi}^k_{lri,j} + 
{\chi}^k_{lrj,i}\]
\[ -  ({\chi}^s_{,i}{\chi}^k_{lrs,j} + {\chi}^s_{r,i} {\chi}^k_{ls,j} + {\chi}^s_{l,i}{\chi}^k_{rs,j} + {\chi}^s_{lr,i} {\chi}^k_{s,j} - {\chi}^s_{,j} {\chi}^k_{lrs,i} - {\chi}^s_{r,j}{\chi}^k_{ls,i} - {\chi}^s_{l,j}{\chi}^k_{rs,i} - {\chi}^s_{lr,j}{\chi}^k_{s,i}) = 0    \eqno{(3)} \]

There is no need for double-arrows in this framework as the kernels are taken with respect to the zero section of the vector bundles involved. We finally notice that the main difference with the gauge sequence is that {\it all the indices range from} $1$ {\it to} $n$ and that the condition $det(A)\neq 0$ amounts to $\Delta=det({\partial}_if^k)\neq 0$ because $det(f^k_i)\neq 0$ by assumption. \\
\hspace*{12cm}   Q.E.D.  \\

\noindent
{\bf COROLLARY 3.11}: There is a {\it restricted first nonlinear Spencer sequence}:\\
\[ 0\longrightarrow \Gamma \stackrel{j_{q+1}}{\longrightarrow} {\cal{R}}_{q+1} \stackrel{\bar{D}}{\longrightarrow} T^*\otimes R_q\stackrel{{\bar{D}}'}{\longrightarrow}{\wedge}^2T^*\otimes J_{q-1}(T)  \]

\noindent
{\bf DEFINITION 3.12}: A {\it splitting} of the short exact sequence $0\rightarrow R^0_q\rightarrow R_q\stackrel{{\pi}^q_0}{\rightarrow} T \rightarrow 0$ is a map ${\chi}'_q:T\rightarrow R_q$ such that ${\pi}^q_0\circ {\chi}'_q=id_T$ or equivalently a section of $T^*\otimes R_q$ over $id_T\in T^*\otimes T$ and is called a $R_q$-{\it connection}. Its {\it curvature} ${\kappa}'_q\in {\wedge}^2T^*\otimes R^0_q$ is defined by ${\kappa}'_q(\xi,\eta)=[{\chi}'_q(\xi),{\chi}'_q(\eta)]-{\chi}'_q([\xi,\eta])$. We notice that ${\chi}'_q=-{\chi}_q$ is a connection with ${\bar{D}}'{\chi}'_q={\kappa}'_q$ if and only if $A=0$. In particular $({\delta}^k_i,-{\gamma}^k_{ij})$ is the only existing symmetric connection for the Killing system.         \\

\noindent
{\bf REMARK 3.13}: Rewriting the previous local formulas with $A$ instead of ${\chi}_0$ we get:  \\
 \[ {\partial}_iA^k_j-{\partial}_jA^k_i-A^r_i{\chi}^k_{r,j}+A^r_j{\chi}^k_{r,i}=0  \eqno{(1^*)} \]
\[ {\partial}_i{\chi}^k_{l,j}-{\partial}_j{\chi}^k_{l,i}-{\chi}^r_{l,i}{\chi}^k_{r,j}+{\chi}^r_{l,j}{\chi}^k_{r,i}-A^r_i{\chi}^k_{lr,j}+A^r_j{\chi}^k_{lr,i}=0  \eqno{(2^*)}   \]
\[  {\partial}_i{\chi}^k_{lr,j} - {\partial}_j{\chi}^k_{lr,i} - A^sj{\chi}^k_{lri,j} + A^s_i{\chi}^k_{lrj,i}\]
\[ -  ( {\chi}^s_{r,i} {\chi}^k_{ls,j} + {\chi}^s_{l,i}{\chi}^k_{rs,j} + {\chi}^s_{lr,i} {\chi}^k_{s,j} - {\chi}^s_{r,j}{\chi}^k_{ls,i} - {\chi}^s_{l,j}{\chi}^k_{rs,i} - {\chi}^s_{lr,j}{\chi}^k_{s,i}) = 0  \eqno{(3^*)}   \]

When $q=1, g_2=0$ and though surprising it may look like, we find back {\it exactly} all the formulas presented by E. and F. Cosserat in ([8], p 123 and [16]). Even more strikingly, {\it in the case of a Riemann structure, the last two terms disappear but the quadratic terms are left while, in the case of screw and complex structures, the quadratic terms disappear but the last two terms are left}. We finally notice that ${\chi}'_q= - {\chi}_q$ is a $R_q$-connection if and only if $A=0$, {\it a result contradicting the use of connections in physics}. However, when $A=0$, we have ${\chi}'_0(\xi)=\xi$ and thus:  \\
\[  \begin{array}{rcl}
 {\bar{D}}'{\chi}_{q+1} & = & (D{\chi}_{q+1})(\xi,\eta)-([{\chi}_q(\xi),{\chi}_q(\eta)] + i(\xi)D({\chi}_{q+1}(\eta)) - i(\eta)D({\chi}_{q+1}(\xi))) \\
    &   =   & - [{\chi}_q(\xi),{\chi}_q(\eta)]  - {\chi}_q([\xi,\eta])  \\
    &  =    &   - ([{\chi}'_q(\xi),{\chi}'_q(\eta)]  - {\chi}'_q([\xi,\eta]))  \\

    &  =  & - {\kappa}'_q(\xi,\eta)
    \end{array}   \]
does not depend on the lift of ${\chi}_q$.  \\
\\

\noindent
{\bf COROLLARY 3.14}: When $det(A)\neq 0$ there is a {\it second nonlinear Spencer sequence} stabilized at order $q$:\\
\[ 0 \longrightarrow {aut(X)} \stackrel{j_q}{\longrightarrow} {\Pi}_q \stackrel{{\bar{D}}_1}{\longrightarrow} C_1(T) \stackrel{{\bar{D}}_2}{\longrightarrow} C_2(T)  \]
where ${\bar{D}}_1$ and ${\bar{D}}_2$ are involutive and a {\it  restricted second nonlinear Spencer sequence}:ÊÊ\\
\[ 0\longrightarrow \Gamma \stackrel{j_q}{\longrightarrow} {\cal{R}}_q \stackrel{{\bar{D}}_1}{\longrightarrow} C_1 \stackrel{{\bar{D}}_2}{\longrightarrow} C_2   \]
such that ${\bar{D}}_1$ and ${\bar{D}}_2$ are involutive whenever ${\cal{R}}_q$ is involutive.\\

\noindent
{\it Proof}: With ${\mid}\mu{\mid}=q$ we have ${\chi}^k_{\mu,i}=-g^k_lA^r_if^l_{\mu+1_r}+\, terms(order \leq q)$. Setting ${\chi}^k_{\mu,i}=A^r_i{\tau}^k_{\mu,r}$, we obtain ${\tau}^k_{\mu,r}= - g^k_lf^l_{\mu+1_r}+terms(order\leq q)$ and ${\bar{D}}:{\Pi}_{q+1}\rightarrow T^*\otimes J_q(T)$ restricts to ${\bar{D}}_1:{\Pi}_q\rightarrow C_1(T)$. \\
Finally, setting $ A^{-1}=B=id-{\tau}_0$, we obtain successively:\\
\[  {\partial}_i{\chi}^k_{\mu,j}-{\partial}_j{\chi}^k_{\mu,i}+terms({\chi}_q) -(A^r_i{\chi}^k_{\mu+1_r,j}-A^r_j{\chi}^k_{\mu+1_r,i})=0ÊÊ\]
\[  B^i_rB^j_s({\partial}_i{\chi}^k_{\mu,j}-{\partial}_j{\chi}^k_{\mu,i})+terms ({\chi}_q)-({\tau}^k_{\mu+1_r,s}-{\tau}^k_{\mu+1_s,r})=0  \]
We obtain therefore $D{\tau}_{q+1}+terms({\tau}_q)=0$ and ${\bar{D}}':T^*\otimes J_q(T)\rightarrow {\wedge}^2T^*\otimes J_{q-1}(T)$ restricts to ${\bar{D}}_2:C_1(T)\rightarrow C_2(T)$. \\
In the case of Lie groups of transformations, the symbol of the involutive system $R_q$ {\it must} be $g_q=0$ providing an isomorphism ${\cal{R}}_{q+1}\simeq {\cal{R}}_q\Rightarrow R_{q+1}\simeq R_q$ and we have therefore $C_r={\wedge}^rT^*\otimes R_q$ for $ r=1,...,n$ like in the linear Spencer sequence.   \\
\hspace*{12cm}   Q.E.D.  \\

\noindent
{\bf REMARK 3.15}: In the case of the (local) action of a Lie group $G$ on $X$, we may consider the graph of this action, that is the morphism $X\times G \rightarrow X\times X: (x,a) \rightarrow (x,y=f(x,a)) $. If $q$ is large enough, then there is an isomorphism $X\times G \rightarrow {\cal{R}}_q \subset {\Pi}_q: (x,a) \rightarrow j_q(f)(x,a)$ obtained by eliminating the parameters and $C_r ={\wedge}^rT^*\otimes R_q$. If $\{ {\theta}_{\tau}\}$ with $1\leq  \tau \leq dim(G)$ is a basis of infinitesimal generators of this action, there is a morphism of Lie algebroids over $X$, namely $X\times {\cal{G}} \rightarrow R_q: {\lambda}^{\tau}(x) \rightarrow {\lambda}^{\tau}(x)j_q({\theta}_{\tau})$ when $q$ is large enough and the linear Spencer sequence $R_q \stackrel{D_1}{\longrightarrow} T^*\otimes R_q \stackrel{D_2}{\longrightarrow} {\wedge}^2T^*\otimes R_q \stackrel{D_3}{\longrightarrow} ... $ is locally exact because it is locally isomoprphic to the tensor product by ${\cal{G}}$ of the Poincar\'{e} sequence ${\wedge}^0T^* \stackrel{d}{\longrightarrow} {\wedge}^1T^* \stackrel{d}{\longrightarrow}   {\wedge}^2T^* \stackrel{d}{\longrightarrow} ...$ where $d$ is the exterior derivative ([28]).  \\
We may also consider  similarly $dy=dax=daa^{-1}y$ and $dx=dbb^{-1}dx=- a^{-1}dax$, depending on the choice of the independent variable among the {\it source} $x$ or the {\it target} $y$. \\

Surprisingly, in the case of Lie pseudogroups or Lie groupoids, the situation is quite different. We recall the  way to introduce a groupoid structure on ${\Pi}_{q,1}\subset J_1({\Pi}_q)$ from the groupoid structure on ${\Pi}_q$ when $\Delta=det({\partial}_if^k(x)\neq 0$, that is how to define $j_1(h_q)=j_1(g_q\circ f_q)=j_1(g_q)\circ j_1(f_q)$. We get successively with $y=f(x)$:  \\
\[  h(x)=(g\circ f)(x)= g(f(x)) \Rightarrow  \frac{\partial h^r}{\partial x^i}=\frac{\partial g^r}{\partial y^k}\frac{\partial f^k}{\partial x^i} \Rightarrow h^r_i(x)=g^r_k(f(x)f^k_i (x)              \]

\[  \frac{\partial h^r_i}{\partial x^j}= \frac{\partial g^r_k}{\partial y^l}f^k_i\frac{\partial f^l}{\partial x^j} + 
g^r_k\frac{\partial f^k_i}{\partial x^j}  \Rightarrow h^r_{ij}(x)=g^r_{kl}(f(x))f^k_i(x)f^l_j(x) + g^r_k(f(x)) f^k_{ij}(x) \]
\[ \frac{\partial h^r_{ij}}{\partial x^s}= \frac{\partial g^r_{ki}}{\partial y^u}f^k_if^l_j\frac{\partial f^u}{\partial x^s} + g^r_{kl}( \frac{\partial f^k_i}{\partial x^s} f^l_j + f^k_i \frac{\partial f^l_j}{\partial x^s}) +
\frac{\partial g^r_k}{\partial y^u} \frac{\partial f^u}{\partial x^s} f^k_{ij} + 
g^r_k \frac{\partial f^k_{ij}}{\partial x^s} \]
\[  \Rightarrow h^r_{ijs}= g^r_{klu} f^k_if^l_j f^u_s + g^r_{kl}(f^k_{is}f^l_j + f^k_if^l_{js}) + g^r_{ku} f^u_sf^k_{ij} + 
g^r_k f^k_{ijs}  \]
and so on with more  and more involved formulas.  \\
Now, if we want to obtain objects over the {\it source} $x$ according to the non-linear Spencer sequence, we have only two possibilities in actual practice, namely:  \\
\[     {\chi}_q= f^{-1}_{q+1}\circ j_1(f_q) - id_{q+1}  \in T^*\otimes J_q(T) \,\,\,  \leftrightarrow \,\,\,  
      {\bar{\chi}}_q= j_1(f_q)^{-1}\circ f_{q+1} - id_{q+1} \in T^*\otimes J_q(T)        \]
As we have already considered the first, we have now only to study the second. In $J_1({\Pi}_q)$, we have:  \\
\[ {\chi}_q + id_{q+1}=(A^k_r, {\chi}^k_{i,r}, {\chi}^k_{ij,r}, ... ) \,\, and  \,\, {\bar{\chi}}_q+id_{q+1}=({\bar{A}}^k_r, {\bar{\chi}}^k_{i,r}, {\bar{\chi}}^k_{ij,r}, ... ) \,\, over \,\, (x,x,\delta, 0, ... )  \]

\noindent
{\bf LEMMA 3.16 }: ${\bar{\chi}}_q$ is a quasi-linear rational function of ${\chi}_q$, $\forall q\geq 0$. With more details, when $q=0$, we have ${\bar{{\chi}}}_0=\bar{A} - id$ and ${\chi}_0=A - id$ with $\bar{A}=A^{-1}=B$ and when $q\geq 1$, we have ${\bar{\chi}}_q \circ A = - {\chi}_q$, that is to say ${\bar{\chi}}_q= - {\tau}_q$.   \\

\noindent
{\it Proof}: In the groupoid framework, we have:  \\
\[   ({\bar{\chi}}_q + id_{q+1}) \circ ({\chi}_q + id_{q+1})= id_{q+1} \in J_1({\Pi}_q)  \]
Doing the substitutions:  \\
\[  \frac{\partial g^r}{\partial y^k} \rightarrow {\bar{A}}^r_k, \,\,\, \frac{\partial g^r_k}{\partial y^l} \rightarrow {\bar{\chi}}^r_{k,l}, \,\,\, \frac{\partial g^r_{kl}}{\partial y^u} \rightarrow {\bar{\chi}}^r_{kl,u} \]
\[  \frac{\partial f^k}{\partial x^i} \rightarrow A ^k_i, \,\,\, \frac{\partial f^k_i}{\partial x^j} \rightarrow {\chi}^k_{i,j},\,\,\,  \frac{\partial f^k_{ij}}{\partial x^s} \rightarrow {\chi}^k_{ij,s}  \]
while using the fact that $f^k_i={\delta}^k_i, f^k_{ij}=0, ...$ and $g^r_k={\delta}^r_k, g^r_{kl}=0, .. $, we obtain at once:  \\
\[ {\bar{A}}^r_kA^k_i={\delta}^r_i, {\bar{\chi}}^r_{k,l} A^l_j+ {\chi}^k_{i,j}=0, {\bar{\chi}}^r_{ij,u}A^u_s +{\chi}^r _{ij,s}=0, ...  \]
Proceeding by induction, we finally obtain:  \\
\[  {\bar{\chi}}^k_{\mu,r} A^r_s + {\chi}^k_{\mu, i}=0  \]
that is to say ${\bar{\chi}}^k_{\mu,i} + {\tau}^k_{\mu,i}=0$ because $ \Delta \neq 0 \Rightarrow det(A)\neq 0$, thus ${\bar{\chi}}_q\circ A = -  {\chi}_q$ or, equivalently, ${\bar{\chi}}_q= - {\tau}_q$.   \\
\hspace*{12cm}    Q.E.D.   \\

\noindent
{\bf REMARK 3.17}: The passage from ${\chi}_q$ to ${\tau}_q$ is {\it exactly} the one done by E. and F. Cosserat in ([8], p 190), even though it is based on a {\it subtle misunderstanding} that we shall correct later on.\\   

\noindent
{\bf REMARK 3.18}: According to the previous results, the "{\it field} " must be a section of the natural bundle ${\cal{F}}$ of geometric objects if we use the nonlinear Janet sequence or a section of the first Spencer bundle $C_1$ if we use the nonlinear Spencer sequence. The aim of this paper is to prove that the second choice is by far more convenient for mathematical physics.  \\   \\

\noindent
{\bf 4) VARIATIONAL CALCULUS }   \\   
  
It remains to graft a variational procedure adapted to the previous results. Contrary to what happens in analytical mechanics or elasticity for example, {\it the main idea is to vary sections but not points}. Hence, we may introduce the variation $\delta f^k(x)={\eta}^k(f(x))$ and set ${\eta}^k(f(x))=
{\xi}^i{\partial}_if^k(x)(x)$ along the "{\it vertical machinery} " but {\it notations like} $\delta x^i={\xi}^i$ {\it or} $\delta y^k={\eta}^k$ {\it have no meaning at all}. \\

As a major result first discovered in specific cases by the brothers Cosserat in 1909 and by Weyl in 1916, we shall prove and apply the following 
key result:  \\

\noindent
THE PROCEDURE ONLY DEPENDS ON THE LINEAR SPENCER OPERATOR AND ITS FORMAL ADJOINT. \\

In order to prove this result, if $f_{q+1},g_{q+1},h_{q+1} \in {\Pi}_{q+1}$ can be composed in such a way that $g'_{q+1}=g_{q+1}\circ f_{q+1} = f_{q+1}\circ h_{q+1}$, we get:\\
\[ \begin{array}{rl}
{\bar{D}}g'_{q+1}&=f^{-1}_{q+1}\circ g^{-1}_{q+1}\circ j_1(g_q)\circ j_1(f_q)-id_{q+1}    =   f^{-1}_{q+1}\circ {\bar{D}}g_{q+1}\circ j_1(f_q)+{\bar{D}}f_{q+1} \\
     &= h^{-1}_{q+1}\circ f^{-1}_{q+1}\circ j_1(f_q)\circ j_1(h_q) - id_{q+1}     =    h^{-1}_{q+1} \circ {\bar{D}}f_{q+1} \circ j_1(h_q) + \bar{D} h_{q+1}
 \end{array}    \]
Using the local exactness of the first nonlinear Spencer sequence or ([23], p 219), we may state:  \\
 
\noindent
{\bf LEMMA 4.1}: For any section $f_{q+1}\in {\cal{R}}_{q+1}$, the {\it finite gauge transformation}:\\
\[   {\chi}_q \in T^*\otimes R_q  \longrightarrow  {\chi}'_q= f^{-1}_{q+1}\circ {\chi}_q\circ j_1(f_q)+{\bar{D}}f_{q+1} \in T^* \otimes R_q \]
exchanges the solutions of the {\it field equations} ${\bar{D}}'{\chi}_q=0$.  \\

\noindent
Introducing the {\it formal Lie derivative} on $J_q(T)$ by the formulas:\\
\[  L({\xi}_{q+1}){\eta}_q=\{{\xi}_{q+1},{\eta}_{q+1}\}+i(\xi)D{\eta}_{q+1}=[{\xi}_q,{\eta}_q]+i(\eta)D{\xi}_{q+1} \]
\[      (L(j_1({\xi}_{q+1})){\chi}_q)(\zeta)= L({\xi}_{q+1})({\chi}_q(\zeta))-{\chi}_q([\xi,\zeta])   \]

\noindent
{\bf LEMMA 4.2}: Passing to the limit {\it over the source} with $h_{q+1}=id_{q+1}+t {\xi}_{q+1}+ ... $ for $t\rightarrow 0$, we get an  {\it infinitesimal gauge transformation} leading to the {\it infinitesimal variation}:  \\
\[        \delta {\chi}_q= D{\xi}_{q+1}+ L(j_1({\xi}_{q+1})){\chi}_q    \eqno{(3)}   \]
which {\it does not depend on the parametrization} of ${\chi}_q$. Setting ${\bar{\xi}}_{q+1}={\xi}_{q+1} + {\chi}_{q+1}(\xi)$, we get:   \\
\[   \delta{\chi}_q=D{\bar{\xi}}_{q+1}-\{{\chi}_{q+1},{\bar{\xi}}_{q+1}\}   \eqno{(3^*)}    \]\\

\noindent
{\bf LEMMA 4.3}: Passing to the limit {\it over the target} with ${\chi}_q=\bar{D}f_{q+1}$ and $g_{q+1}=id_{q+1}+ t {\eta}_{q+1}+ ... $, we get the other {\it infinitesimal variation} where $D{\eta}_{q+1}$ is {\it over the target}:     \\
\[         \delta {\chi}_q= f^{-1}_{q+1}\circ D{\eta}_{q+1}\circ j_1(f_q)    \eqno{(4)}    \]
which {\it depends on the parametrization} of ${\chi}_q$. \\

\noindent
{\bf EXAMPLE 4.4}: We obtain for $q=1$:  \\
\[ \begin{array}{ll}
\delta{\chi}^k_{,i}& =({\partial}_i{\xi}^k-{\xi}^k_i)+({\xi}^r{\partial}_r{\chi}^k_{,i}+{\chi}^k_{,r}{\partial}_i{\xi}^r-{\chi}^r_{,i}{\xi}^k_r)   \\
  &=({\partial}_i{\bar{\xi}}^k-{\bar{\xi}}^k_i)+({\chi}^k_{r,i}{\bar{\xi}}^r-{\chi}^r_{,i}{\bar{\xi}}^k_r)     \\
\delta {\chi}^k_{j,i}&=({\partial}_i{\xi}^k_j-{\xi}^k_{ij})+({\xi}^r{\partial}_r{\chi}^k_{j,i}+{\chi}^k_{j,r}{\partial}_i{\xi}^r+{\chi}^k_{r,i}{\xi}^r_j-{\chi}^r_{j,i}{\xi}^k_r-{\chi}^r_{,i}{\xi}^k_{jr})   \\
  & =({\partial}_i{\bar{\xi}}^k_j-{\bar{\xi}}^k_{ij})+({\chi}^k_{rj,i}{\bar{\xi}}^r+{\chi}^k_{r,i}{\bar{\xi}}^r_j-{\chi}^r_{j,i}{\bar{\xi}}^k_r-{\chi}^r_{,i}{\bar{\xi}}^k_{jr})   
  \end{array}  \]
  Introducing the inverse matrix  $B=A^{-1}$, we obtain therefore equivalently:  \\
  \[ \delta A^k_i= {\xi}^r{\partial}_rA^k_i + A^k_r{\partial}_i{\xi}^r - A^r_i{\xi}^k_r \,\,\,  \Leftrightarrow \,\,\,  \delta B^i_k={\xi}^r{\partial}_rB^i_k -B^r_k{\partial}_r{\xi}^i + B^i_r{\xi}^r_k    \]
both with:
\[  \delta {\chi}^k_{j,i}=({\partial}_i{\xi}^k_j - A^r_i{\xi}^k_{jr})+({\xi}^r{\partial}_r{\chi}^k_{j,i}+{\chi}^k_{j,r}{\partial}_i{\xi}^r+{\chi}^k_{r,i}{\xi}^r_j-{\chi}^r_{j,i}{\xi}^k_r)      \]
For the Killing system $R_1\subset J_1(T)$ with $g_2=0$, these variations are {\it exactly} the ones that can be found in ([8], (50)+(49), p 124 with a printing mistake corrected on p 128) when replacing a $3\times 3$ skewsymmetric matrix by the corresponding vector. {\it The last unavoidable Proposition is thus essential in order to bring back the nonlinear framework of finite elasticity to the linear framewok of infinitesimal elasticity that only depends on the linear Spencer operator}.\\
For the conformal Killing system ${\hat{R}}_1\subset J_1(T)$ (see next section) we obtain:    \\
\[ \delta {\chi}^r _{r,i} = ({\partial}_i{\xi}^r_r-{\xi}^r_{ri})+({\xi}^r{\partial}_r {\chi}^s_{s,i}+{\chi}^s_{s,r}{\partial}_i{\xi}^r - {\chi}^s_{,i}{\xi}^r_{rs}) \]
but ${\chi}^r_{r,i}(x)dx^i$ is {\it far} from being a $1$-form. However, $({\chi}^k_{j,i} + {\gamma}^k_{js}{\chi}^s_{,i})\in T^*\otimes T^* \otimes T$ and thus $({\alpha}_i={\chi}^r_{r,i} +{\gamma}^r_{rs}{\chi}^s_{,i})\in T^*$ is a pure $1$-form if we replace $({\chi}^r_{r,i}, {\chi}^r_{,i})$ by $({\alpha}_i,0)$. Hence, $\alpha(\zeta)$ is a scalar for any $\zeta \in T$ and we have $L({\xi}_1)(\alpha(\zeta))- \alpha ([\xi,\zeta])=({\alpha}_r{\partial}_i{\xi}^r + {\xi}^r{\partial}_r {\alpha}_i){\zeta}^i$. As we shall see in section $V.A$, we have $(L({\xi}_2)\gamma)^k_{ij}={\xi}^k_{ij}$ for any section ${\xi}_2 \in J_2(T)$ and we obtain therefore successively:   \\
\[  \delta {\alpha}_i=({\partial}_i{\xi}^r_r - {\xi}^r_{ri}) + ( {\alpha}_r{\partial}_i{\xi}^r + 
{\xi}^r{\partial}_r{\alpha}_i)  \]
\[  {\varphi}_{ij}={\partial}_i{\alpha}_j - {\partial}_j{\alpha}_i \,\, \Rightarrow \,\, \delta {\varphi}_{ij}= 
({\partial}_j{\xi}^r_{ri} - {\partial}_i{\xi}^r_{rj}) + ({\varphi}_{rj}{\partial}_i{\xi}^r + {\varphi}_{ir}{\partial}_j{\xi}^r 
+ {\xi}^r {\partial}_r {\varphi}_{ij})  \]

These are {\it exactly} the variations obtained by Weyl ([54], (76), p 289) who was assuming implicitly $A=0$ when setting ${\bar{\xi}}^r_r=0\Leftrightarrow {\xi}^r_r=-{\alpha}_i{\xi}^i$ by introducing a connection. Accordingly, ${\xi}^r_{ri}$ is the variation of the EM potential itself, that is the $\delta A_i$ of engineers used in order to exhibit the Maxwell equations from a variational principle ([54], $\S$ 26) but the introduction of the Spencer operator is new in this framework.\\

The explicit general formulas of the two lemma cannot be found somewhere else (The reader may compare them to the ones obtained in [19] by means of the so-called " diagonal " method that cannot be applied to the study of explicit examples). The following unusual difficult proposition generalizes well known variational techniques used in continuum mechanics and will be crucially used for applications:  \\

\noindent
{\bf PROPOSITION 4.5}: The same variation is obtained whenever ${\eta}_q=f_{q+1}({\xi}_q+{\chi}_q(\xi))$ with ${\chi}_q=\bar{D}f_{q+1}$, a transformation only depending on $j_1(f_q)$ and invertible if and only if $det(A)\neq 0$.\\

\noindent
{\it Proof}: First of all, setting ${\bar{\xi}}_q={\xi}_q + {\chi}_q(\xi)$, we get $\bar{\xi}=A(\xi)$ for $q=0$, a transformation which is invertible if and only if $det(A)\neq 0$. In the nonlinear framework, we have to keep in mind that there is no need to vary the object $\omega$ which is given but only the need to vary the section $f_{q+1}$ as we already saw, using ${\eta}_q\in R_q(Y)$ {\it over the target} or ${\xi}_q\in R_q$ {\it over the source}. With ${\eta}_q=f_{q+1}({\xi}_q)$, we obtain for example: \\
\[ \begin{array}{rcccl}
  \delta f^k & = & {\eta}^k & = & f^k_r{\xi}^r \\
   \delta f^k_i & = & {\eta}^k_uf^u_i & = & f^k_r{\xi}^r_i+f^k_{ri}{\xi}^r \\
  \delta f^k_{ij} & = & {\eta}^k_{uv}f^u_if^v_j+{\eta}^k_uf^u_{ij} & = & f^k_r{\xi}^r_{ij} + f^k_{ri}{\xi}^r_j+f^k_{rj}{\xi}^r_i+f^k_{rij}{\xi}^r 
  \end{array}  \]
and so on. Introducing the formal derivatives $d_i$ for $i=1,...,n$, we have:  \\
\[  \delta f^k_{\mu}={\zeta}^k_{\mu}(f_q,{\eta}_q)= d_{\mu}{\eta}^k={\eta}^k_uf^u_{\mu} + ... = f^k_r{\xi}^r_{\mu} + ... + f^k_{\mu +1_r} {\xi}^r       \]
We shall denote by $\sharp({\eta}_q)={\zeta}^k_{\mu}(y_q,{\eta}_q)\frac{\partial}{\partial y^k_{\mu}}\in V({\cal{R}}_q) $ with ${\zeta}^k={\eta}^k$ the corresponding vertical vector field, namely:    \\
\[   \sharp({\eta}_q)= 0\frac{\partial}{\partial x^i}+{\eta}^k(y)\frac{\partial}{\partial y^k}+({\eta}^k_u(y)y^u_i)\frac{\partial}{\partial y^k_i}+({\eta}^k_{uv}(y)y^u_iy^v_j+{\eta}^k_u(y)y^u_{ij} )\frac{\partial}{\partial y^k_{ij}}+ ...  \]
However, the standard prolongation of an infinitesimal change of source coordinates described by the horizontal vector field $\xi$, obtained by replacing all the derivatives of $\xi$ by a section ${\xi}_q \in R_q$ over $\xi \in T$, is the vector field: \\
\[   \flat({\xi}_q)={\xi}^i(x)\frac{\partial}{\partial x^i}+ 0\frac{\partial}{\partial y^k} - (y^k_r{\xi}^r_i(x))\frac{\partial}{\partial y^k_i}-(y^k_r{\xi}^r_{ij}(x)+y^k_{rj}{\xi}^r_i(x)+y^k_{ri}{\xi}^r_j(x)) \frac{\partial}{\partial y^k_{ij}}+ ...                                                             \]
It can be proved that $[\flat({\xi})_q,\flat({\xi}'_q]=\flat([{\xi}_q,{\xi}'_q]), \forall {\xi}_q,{\xi}'_q\in R_q$ {\it over the source}, with a similar property for $\sharp(.)$ {\it over the target} ([25]). However, $\flat({\xi}_q)$ {\it is not a vertical vector field and cannot therefore be compared to} $\sharp({\eta}_q)$.The solution of this problem explains a strange comment made by Weyl in ([53], p 289 + (78), p 290) and which became a founding stone of classical gauge theory. Indeed, ${\xi}^r_r$ is {\it not} a scalar because ${\xi}^k_i$ is {\it not} a $2$-tensor. However, when $A=0$, then $-{\chi}_q$ is a $R_q$-connection and ${\bar{\xi}}^r_r={\xi}^r_r+{\chi}^r_{r,i}{\xi}^i$ is a true scalar that may be set equal to zero in order to obtain ${\xi}^r_r=-{\chi}^r_{r,i}{\xi}^i$, a fact explaining why the EM-potential is considered as a connection in quantum mechanics instead of using the second order jets ${\xi}^r_{ri}$ of the conformal system, with a {\it shift by one step in the physical interpretation of the Spencer sequence} (See [27] for more historical details).\\
The main idea is to consider the vertical vector field $T(f_q)(\xi) - \flat({\xi}_q)\in V({\cal{R}}_q)$ whenever $y_q=f_q(x)$. Passing to the limit $t\rightarrow 0$ in the formula $g_q\circ f_q=f_q\circ h_q$, we first get $g\circ f = f\circ h \Rightarrow f(x)+t\eta (f(x)) + ... = f(x + t \xi(x) + ... )$. Using the chain rule for derivatives and substituting jets, we get successively:    \\
\[ \delta f^k(x)={\xi}^r{\partial}_r f^k, \hspace{2mm}  \delta f^k_i={\xi}^r{\partial}_rf^k_i + f^k_r {\xi}^r_i,\hspace{2mm}  \delta f^k_{ij}={\xi}^r{\partial}_rf^k_{ij}+f^k_{rj}{\xi}^r_i + f^k_{ri}{\xi}^r_j + f^k_r{\xi}^r_{ij}           \]
and so on, replacing ${\xi}^rf^k_{\mu + 1_r}$ by ${\xi}^r{\partial}_rf^k_{\mu}$ in ${\eta}_q=f_{q+1}({\xi}_q)$ in order to obtain:  \\
\[   \delta f^k_{\mu} = {\eta}^k_rf^r_{\mu} + ... ={\xi}^i({\partial}_if^k_{\mu}-f^k_{\mu+1_i})+ f^k_{\mu +1_r}{\xi}^r +  ... +f^k_r{\xi}^r_{\mu}   \]
where the right member only depends on $j_1(f_q)$ when $\mid\mu\mid=q$. \\
Finally, we may write the symbolic formula $f_{q+1}({\chi}_q)=j_1(f_q)-f_{q+1}=Df_{q+1}\in T^*\otimes V({\cal{R}}_q)$ in the explicit form:\\
\[        f^k_r{\chi}^r_{\mu,i} + ... +f^k_{\mu +1_r}{\chi}^r_{,i} = {\partial}_if^k_{\mu}-f^k_{\mu +1_i}   \]
Substituting in the previous formula provides ${\eta}_q=f_{q+1}({\xi}_q+ {\chi}_q(\xi))$ and we just need to replace $q$ by $q+1$ in order to achieve the proof. \\
Checking directly the proposition is not evident even when $q=0$ as we have:  \\
\[  (\frac{\partial {\eta}^k}{\partial y^u}-{\eta}^k_u){\partial}_if^u = f^k_r [({\partial}_i{\bar{\xi}}^r-{\bar{\xi}}^r_i) - ({\chi}^s_{,i}{\bar{\xi}}^r_s - {\chi}^r_{s,i}{\bar{\xi}}^s)]    \]
but cannot be done by hand when $q\geq 1$.Ê \\
\hspace*{12cm}  Q.E.D.   \\

For an arbitrary vector bundle $E$ and involutive system $R_q\subseteq J_q(E)$, we may define the $r$-{\it prolongations} ${\rho}_r(R_q)=R_{q+r}=J_r(R_q)\cap J_{q+r}(E)\subset J_r(J_q(E))$ and their respective {\it symbols} $g_{q+r}={\rho}_r(g_q)$ defined from $g_q \subseteq S_qT^*\otimes E$ where $S_qT^*$ is the vector bundle of $q$-symmetric covariant tensors. Using the Spencer $\delta$-map, we now recall the definition of the {\it Spencer bundles}:  \\
\[    C_r={\wedge}^rT^*\otimes R_q / \delta ({\wedge}^{r-1}T^*\otimes g_{q+1}) \subseteq {\wedge}^rT^*\otimes J_q(E) / \delta ({\wedge}^{r-1}T^*\otimes S_{q+1})T^*\otimes E)=C_r(E)  \]
and of the {\it Janet bundles}:   \\
\[   F_r={\wedge}^rT^*\otimes J_q(E)/({\wedge}^rT^* \otimes R_q + \delta {\wedge}^{r-1}T^*\otimes S_{q+1}T^*\otimes E)  \]
When ${\cal{D}}=\Phi \circ j_q$, we may obtain by induction on $r$ the following {\it fundamental diagram} $I$ relating the {\it second linear Spencer sequence} to the {\it linear Janet sequence} with epimorphisms $\Phi={\Phi}_0, ..., {\Phi}_n$:  \\
 \[ \begin{array}{rcccccccccccl}
 &&&&& 0 &&0&&0&  &0&  \\
 &&&&& \downarrow && \downarrow && \downarrow &    & \downarrow &  \\
  & 0& \rightarrow& \Theta &\stackrel{j_q}{\longrightarrow}&C_0 &\stackrel{D_1}{\longrightarrow}& C_1 &\stackrel{D_2}{\longrightarrow} & C_2 &\stackrel{D_3}{\longrightarrow} ... \stackrel{D_n}{\rightarrow}& C_n &\rightarrow 0 \\
  &&&&& \downarrow & & \downarrow & & \downarrow & &\downarrow &     \\
   & 0 & \rightarrow & E & \stackrel{j_q}{\longrightarrow} & C_0(E) & \stackrel{D_1}{\longrightarrow} & C_1(E) &\stackrel{D_2}{\longrightarrow} & C_2(E) &\stackrel{D_3}{\longrightarrow} ... \stackrel{D_n}{\longrightarrow} & C_n(E) &   \rightarrow 0 \\
   & & & \parallel && \hspace{5mm}\downarrow {\Phi}_0 & &\hspace{5mm} \downarrow {\Phi}_1 & & \hspace{5mm}\downarrow {\Phi}_2 &  & \hspace{5mm}\downarrow {\Phi}_n & \\
   0 \rightarrow & \Theta &\rightarrow & E & \stackrel{\cal{D}}{\longrightarrow} & F_0 & \stackrel{{\cal{D}}_1}{\longrightarrow} & F_1 & \stackrel{{\cal{D}}_2}{\longrightarrow} & F_2 & \stackrel{{\cal{D}}_3}{\longrightarrow} ... \stackrel{{\cal{D}}_n}{\longrightarrow} & F_n & \rightarrow  0 \\
   &&&&& \downarrow & & \downarrow & & \downarrow &   &\downarrow &   \\
   &&&&& 0 && 0 && 0 &&0 &  
   \end{array}     \]

Chasing in the above diagram, the Spencer sequence is locally exact at $C_1$ if and only if the Janet sequence is locally exact at $F_0$ because the central sequence is locally exact (See [25],[28],[30] for more details). In the present situation, we shall always have $E=T$. {\it The situation is much more complicate in the nonlinear framewok and we provide details for a later use}.\\

Let $\bar{\omega}$ be a section of $\cal{F}$ satisfying the same CC as $\omega$, namely $I(j_1(\omega))=0$. As ${\cal{F}}$ is a quotient of ${\Pi}_q$, we may find a section $f_q\in {\Pi}_q $ such that:  \\
\[ {\Phi}\circ f_q\equiv f^{-1}_q(\omega)=\bar{\omega} \,\,\,  \Rightarrow  \,\,\,{\rho}_1(\Phi) \circ j_1(f_q)\equiv j_1(f^{-1}_q)(j_1(\omega))=j_1(f^{-1}_q(\omega))=j_1({\bar{\omega}})\]
Similarly, as ${\cal{F}}$ is a natural bundle of order $q$, then $J_1({\cal{F}})$ is a natural bundle of order $q+1$ and we can find a section $f_{q+1}\in {\Pi}_{q+1}$ such that:  \\
\[   {\rho}_1(\Phi)\circ f_{q+1} \equiv f^{-1}_{q+1}(j_1(\omega))=j_1(\bar{\omega})  \]
and we are facing two possible but quite different situations:  \\

\noindent
$\bullet $  Eliminating $\bar{\omega}$, we obtain:  \\  
\[ j_1(f^{-1}_q)(j_1(\omega))=f^{-1}_{q+1}(j_1(\omega)) \Rightarrow (f_{q+1}\circ  j_1(f^{-1}_q))^{-1}(j_1(\omega))-j_1(\omega)=L({\sigma}_q)\omega=0 \]
and thus ${\sigma}_q={\bar{D}}f^{-1}_{q+1}\in T^*\otimes R_q = - f_{q+1}\circ {\chi}_q\circ j_1(f)^{-1}$ {\it over the target} if we set 
${\chi}_q =\bar{D}f_{q+1}=f^{-1}_{q+1} \circ j_1(f_q) - id_{q+1}$ {\it over the source}, even if $f_{q+1} $ may not be a section of ${\cal{R}}_{q+1}$. As ${\sigma}_q$ is killed by ${\bar{D}}'$, we have related cocycles at $\cal{F}$ in the Janet sequence {\it over the source} with cocycles at $T^*\otimes R_q$ or $C_1$ {\it over the target}.\\

\noindent
$\bullet$  Eliminating $\omega$, we obtain successively: \\
\[  \begin{array}{rcl}
(f^{-1}_{q+1}\circ j_1(f_q))(j_1(\bar{\omega})) - j_1(\bar{\omega}) & = & -(f^{-1}_{q+1}\circ j_1(f_q)) [f^{-1}_{q+1}\circ j_1(f_q))^{-1}(j_1(\bar{\omega}) - j_1(\bar{\omega}) ] \\
 &  = & -  (f^{-1}_{q+1}\circ j_1(f_q) )L({\chi}_q)\bar{\omega}
 \end{array}  \]
where we have {\it over the source}: \\
\[ L({\chi}_q)\bar{\omega}=({\bar{\Omega}}^{\tau}_i\equiv - L^{\tau \mu}_k(\bar{\omega}(x)){\chi}^k_{\mu,i} + {\chi}^r_{,i}{\partial}_r{\bar{\omega}}^{\tau}(x)) \in T^*\otimes F_0\]
However, we know that $F_0$ is associated with $R_q$ and is thus not affected by $f^{-1}_{q+1}\circ j_1(f_q)$ which projects onto $f^{-1}_q \circ f_q=id_q$. Hence, only $T^*$ is affected by $f^{-1}_1\circ j_1(f)=A$ in a covariant way and we obtain therefore {\it over the source}:  \\
\[ (f^{-1}_{q+1}\circ j_1(f_q))(j_1(\bar{\omega})) - j_1(\bar{\omega}) = - B L({\chi}_q)\bar{\omega}= - L({\tau}_q)\bar{\omega}=0  \]
where $B=A^{-1}$. It follows that ${\chi}_q \in T^*\otimes R_q(\bar{\omega})$ with ${\bar{D}}'{\chi}_q=0$ in the first non-linear Spencer sequence 
for $R_q(\bar{\omega})\subset J_q(T)$.  \\

We invite the reader to follow all the formulas involved in these technical results on the next examples. Of course, whenever ${\cal{R}}_q$ is formally integrable and $f_{q+1} \in {\cal{R}}_{q+1}$ is a lift of $f_q \in {\cal{R}}_q$, then we have $\bar{\omega} = \omega$ and ${\xi}_q \in T^*\otimes R_q$ because $R_q(\omega)=R_q$.  \\

\noindent
{\bf EXAMPLE 4.6}: In the case of Riemannian structures, we have ${\cal{F}}\in S_2T^*$ because we deal with a non-degenerate metric $\omega=({\omega}_{ij})\in S_2T^*$ with $det(\omega)\neq 0$ and may introduce ${\omega}^{-1}=({\omega}^{ij})\in S_2T$. We have by definition
 $  {\omega}_{kl}(f(x))f^k_i(x)f^l_j(x)={\bar{\omega}}_{ij}(x)$ that we shall simply write $  {\omega}_{kl}(f)f^k_if^l_j={\bar{\omega}}_{ij}(x)$
and obtain therefore:  \\
\[{\omega}_{kl}(f)f^l_j{\partial}_rf^k_i + {\omega}_{kl}(f)f^k_i{\partial}_rf^l_j + \frac{\partial{\omega}_{kl}}{\partial y^u}(f)f^k_if^l_j{\partial}_rf^u - {\partial}_r{\bar{\omega}}_{ij}(x)=0  \]
Our purpose is now to compute the expression:  \\
\[ {\omega}_{kl}(f)f^l_jf^k_{ir}+ {\omega}_{kl}(f)f^k_if^l_{jr} + \frac{\partial{\omega}_{kl}}{\partial y^u}(f)f^k_if^l_jf^u_r - {\partial}_r{\bar{\omega}}_{ij}(x) \neq 0  \]
In order to eliminate the derivatives of $\omega$ over te target we may multiply the first equation by $B$ and substract from the second while using the fact that ${\omega}_{kl}(f)={\bar{\omega}}_{ij}(x)g^i_kg^j_l$ with ${\chi}_0=A-id_T\Rightarrow {\tau}_0= B{\chi}_0=id_T - B$ in order to get:   \\
\[  - ({\bar{\omega}}_{sj}{\tau}^s_{i,r} + {\bar{\omega}}_{is}{\tau}^s_{j,r} + {\tau}^s_{,r}{\partial}_s{\bar{\omega}}_{ij})=
 - (L({\tau}_1)\bar{\omega})_{ij,r}   \]

These results can be extended at once to any tensorial geometric object but the conformal case needs more work and we let the reader treat it as an exercise. He will discover that the standard elimination of a conformal factor is not the best way to use in order to understand the conformal structure which has to do with a tensor density and no longer with a tensor.  \\

\noindent
{\bf REMARK 4.7}: In the non-linear case, the non-linear CC of the system ${\cal{R}}_q$ defined by $\Phi(y_q)=\bar{\omega}(x)$ only depend on the differential invariants and are exactly the ones satisfied by $\omega$ in the sense that they have the same Vessiot structure constants whenever ${\cal{R}}_q$ is formally integrable, in particular involutive as shown in Example $2.7$. Accordingly, we can always find $f_{q+1}$ over $f_q$. In the linear case, the procedure is similar but slightly simpler. Indeed, if ${\cal{D}}:T \rightarrow F_0$ is an involutive Lie operator, we may consider only the initial part of the {\it fundamental diagram I}:   \\
\[   SPENCER   \]
 \[  \begin{array}{rccccccccl}
 &&&&& 0 &&0&&0  \\
 &&&&& \downarrow && \downarrow && \downarrow  \\
  & 0& \rightarrow& \Theta &\stackrel{j_q}{\rightarrow}&C_0 &\stackrel{D_1}{\rightarrow}& C_1 &\stackrel{D_2}{\rightarrow} & C_2  \\
  &&&&& \downarrow & & \downarrow & & \downarrow     \\
   & 0 & \rightarrow & T & \stackrel{j_q}{\rightarrow} & C_0(T) & \stackrel{D_1}{\rightarrow} & C_1(T) &\stackrel{D_2}{\rightarrow} & C_2(T) \\
   & & & \parallel && \hspace{6mm}\downarrow  {\Phi}_0& & \hspace{6mm}\downarrow  {\Phi}_1& & \\
   0 \rightarrow & \Theta &\rightarrow & T & \stackrel{\cal{D}}{\rightarrow} & F_0 & \stackrel{{\cal{D}}_1}{\rightarrow} & F_1 &  &  \\
   &&&&& \downarrow & & \downarrow & &    \\
   &&&&& 0 && 0 &&    
   \end{array}     \]
\[   JANET   \]
\[  \begin{array}{rcccccccl}
 &&&& 0 &&0& \\
 &&&& \downarrow && \downarrow &&  \\
    & & 0 & \rightarrow &g_{q+1} &\stackrel{-\delta}{\rightarrow}& \delta (g_{q+1}) &\rightarrow &  0\\
  &&&& \downarrow & & \downarrow    \\
    0 & \rightarrow & \Theta & \stackrel{j_{q+1}}{\rightarrow} & R_{q+1} & \stackrel{D}{\rightarrow} &T^*\otimes R_q  & \\
    & & \parallel && \downarrow& & \downarrow & & \\
   0 & \rightarrow & \Theta & \stackrel{j_q}{\rightarrow }& R_q & \stackrel{D_1}{\rightarrow} & C_1 &    \\
   &&&& \downarrow & & \downarrow & &  \\
   &&&& 0 && 0 &&  
   \end{array}     \]
and study the linear inhomogeneous involutive system ${\cal{D}}\xi=\Omega$ with $\Omega \in F_0$ and ${\cal{D}}_1\Omega=0$. If we pick up any lift ${\xi}_q\in C_0(T)=J_q(T)$ of $\Omega$ and chase, we notice that $X_1=D_1{\xi}_q\in C_1\subset C_1(T)$ is such that $D_2X_1=0$. In the Example $2.7$, using the involutive system $R'_1=R^{(1)}_1\subset R_1\subset J_1(T)$, we have $m=n=2, q=1$ and the fiber dimensions:  \\
 \[  \begin{array}{rccccccccccl}
 &&&&& 0 &&0&&0 & \\
 &&&&& \downarrow && \downarrow && \downarrow & &\\
  & 0& \rightarrow& \Theta &\stackrel{j_1}{\rightarrow}&3 &\stackrel{D_1}{\rightarrow}& 5 &\stackrel{D_2}{\rightarrow} & 2 &\rightarrow & 0 \\
  &&&&& \downarrow & & \downarrow & & \parallel  & & \\
   & 0 & \rightarrow & 2 & \stackrel{j_1}{\rightarrow} & 6 & \stackrel{D_1}{\rightarrow} & 6 &\stackrel{D_2}{\rightarrow} & 2 &\rightarrow & 0 \\
   & & & \parallel && \hspace{6mm}\downarrow  {\Phi}_0& & \hspace{6mm}\downarrow  {\Phi}_1& & \downarrow  & &\\
   0 \rightarrow & \Theta &\rightarrow & 2 & \stackrel{\cal{D}}{\rightarrow} & 3 & \stackrel{{\cal{D}}_1}{\rightarrow} & 1 & \rightarrow  
   & 0 & & \\
   &&&&& \downarrow & & \downarrow & &  & & \\
   &&&&& 0 && 0 && &  & 
   \end{array}     \]
\[  \begin{array}{rcccccccl}
 &&&& 0 &&0& \\
 &&&& \downarrow && \downarrow &&  \\
    & & 0 & \rightarrow &1 &\stackrel{-\delta}{\rightarrow}& 1 &\rightarrow &  0\\
  &&&& \downarrow & & \downarrow    \\
    0 & \rightarrow & \Theta & \stackrel{j_2}{\rightarrow} & 4 & \stackrel{D}{\rightarrow} &6 & \\
    & & \parallel && \downarrow& & \downarrow & & \\
   0 & \rightarrow & \Theta & \stackrel{j_1}{\rightarrow }& 3 & \stackrel{D_1}{\rightarrow} & 5 &    \\
   &&&& \downarrow & & \downarrow & &  \\
   &&&& 0 && 0 &&  
   \end{array}     \]
It is important to point out the importance of formal integrability and involution in this case. For this, let us start with a $1$-form $\alpha=({\alpha}_1,{\alpha}_2)$, denote its variation by $A=(A_1,A_2)$ and consider only the linear inhomogeneous system ${\cal{D}}\xi={\cal{L}}(\xi)\alpha=A $ with no CC for $A$. If the ground differential field is $K=\mathbb{Q}(x^1,x^2)$ with commuting derivations $(d_1,d_2)$, let us choose $\alpha=x^2dx^1=(x^2,0), A=(x^2,x^1)$. As a lift ${\xi}_1\in J_1(T)$ of $A$, we let the reader check that we may choose in $K$:   \\
\[  {\xi}^1=0, {\xi}^2=0, {\xi}^1_1=1,{\xi}^1_2=\frac{x^1}{x^2}, {\xi}^2_1=0, {\xi}^2_2=0     \]
 Using one prolongation, we have:  \\
\[  d_1A_1\equiv x^2{\xi}^1_{11}+{\xi}^2_1=0, d_2A_1\equiv x^2{\xi}^1_{12}+{\xi}^1_1 +{\xi}^2_2=1, 
d_1A_2\equiv x^2{\xi}^1_{12}=1, d_2A_2\equiv x^2{\xi}^1_{22}+{\xi}^1_2=0  \]
If $\beta= - d\alpha=dx^1\wedge dx^2$, we may denote its variation by $B$ and get at once $B=d_2A_1-d_1A_2\equiv {\xi}^1_1 +{\xi}^2_2=0$. Such a 
result is contradicting our inital choice $1+0=1$ and we cannot therefore find a lift ${\xi}_2$ of $j_1(A)$. Hence, we have to introduce the new geometric object $\omega=(\alpha, \beta)$ with $\Omega=(A,B)$ and CC $d\alpha + \beta = 0$ leading to $d_1A_2 - d_2A_1 + B=0$ while using the previous diagrams. We can therefore lift $\Omega=(A,B)$ to ${\xi}_1\in J_1(T)$ by choosing in $K$:  \\
\[  {\xi}^1=0, {\xi}^2=0, {\xi}^1_1=1, {\xi}^1_2=\frac{x^1}{x^2}, {\xi}^2_1=0,{\xi}^2_2= - 1   \]
However, we have now to add:  \\
\[  d_1B\equiv {\xi}^1_{11}+{\xi}^2_{12}=0, d_2B\equiv {\xi}^1_{12} + {\xi}^2_{22}=0  \]
and lift $j_1(\Omega)$ to ${\xi}_2\in J_2(T)$ over ${\xi}_1\in J_1(T)$ by choosing in $K$:  \\
\[  {\xi}^1_{11}=0, {\xi}^1_{12}=\frac{1}{x^2}, {\xi}^1_{22}= - \frac{x^1}{(x^2)^2}, 
{\xi}^2_{11}=0, {\xi}^2_{12}= 0, {\xi}^2_{22}= - \frac{1}{x^2}  \]
The image of the Spencer operator is $X_1=D{\xi}_2=j_1({\xi}_1) - {\xi}_2$ that is to say:   \\
\[  X^1_{,1}= - 1, X^1_{,2}= - \frac{x^1}{x^2}, X^2_{,1}=0, X^2_{,2}=1, \]
\[  X^1_{1,1}=0, X^1_{2,1}=0, X^1_{1,2}= - \frac{1}{x^2}, X^1_{2,2}=0, 
    X^2_{1,1}=0, X^2_{2,1}=0, X^2_{1,2}=0, X^2_{2,2}=\frac{1}{x^2}  \]
and we check that $X_1 \in T^*\otimes R_1$, namely:  \\
\[  x^2X^1_{1,i}+ X^2_{,i}=0, X^1_{2,i}=0, X^1_{1,i} + X^2_{2,i}=0, \,\,\,  \forall i=1,2   \]
a result which is not evident at first sight and has no meaning in any classical approach because we use {\it sections} and not {\it solutions}.   \\  \\

Now, if $f_{q+1},f'_{q+1}\in {\Pi}_{q+1}$ are such that $f^{-1}_{q+1}(j_1(\omega))=f'^{-1}_{q+1}(j_1(\omega))=j_1({\bar{\omega}})$, it follows that $(f'_{q+1}\circ f^{-1}_{q+1})(j_1(\omega))=j_1(\omega)   \Rightarrow
\exists  g_{q+1}\in {\cal{R}}_{q+1}$ such that $f'_{q+1}=g_{q+1}\circ f_{q+1}$ and the new ${\sigma}'_q=\bar{D}f'^{-1}_{q+1}$ differs from the initial ${\sigma}_q=\bar{D}f^{-1}_{q+1}$ by a gauge transformation.\\
Conversely, let $f_{q+1},f'_{q+1}\in {\Pi}_{q+1}$ be such that ${\sigma}_q=\bar{D}f^{-1}_{q+1}=
\bar{D}f'^{-1}_{q+1}={\sigma}'_q$. It follows that $\bar{D}(f^{-1}_{q+1}\circ f'_{q+1})=0$ and one can find $g\in aut(X)$ such that $f'_{q+1}=f_{q+1}\circ j_{q+1}(g)$ providing ${\bar{\omega}}'=f'^{-1}_q(\omega)=(f_q\circ j_q(g))^{-1}(\omega)=j_q(g)^{-1}(f^{-1}_q(\omega))=j_q(g)^{-1}(\bar{\omega})$.\\

\noindent
{\bf PROPOSITION 4.8}: Natural transformations of $\cal{F}$ {\it over the source} in the nonlinear Janet sequence correspond to gauge transformations of $T^*\otimes R_q$ or $C_1$ {\it over the target} in the nonlinear Spencer sequence. Similarly, the Lie derivative ${\cal{D}}\xi={\cal{L}}(\xi)\omega\in F_0$ in the linear Janet sequence corresponds to the Spencer operator $D{\xi}_{q+1}\in T^*\otimes R_q$ or $D_1{\xi}_q\in C_1$ in the linear Spencer sequence.\\

With a slight abuse of language $\delta f=\eta\circ f\Leftrightarrow \delta f\circ f^{-1}=\eta \Leftrightarrow f^{-1}\circ \delta f=\xi$ when $\eta=T(f)(\xi)$ and we get $j_q(f)^{-1}(\omega)=\bar{\omega} \Rightarrow j_q(f+\delta f)^{-1}(\omega)=\bar{\omega}+\delta \bar{\omega}$ that is $j_q(f^{-1}\circ(f+\delta f))^{-1}(\bar{\omega})=\bar{\omega}+\delta \bar{\omega} \Rightarrow \delta \bar{\omega}={\cal{L}}(\xi)\bar{\omega}$ and $j_q((f+\delta f)\circ f^{-1}\circ f)^{-1}(\omega)=j_q(f)^{-1}(j_q((f+\delta f)\circ f^{-1})^{-1}(\omega)) \Rightarrow \delta \bar{\omega}=j_q(f)^{-1}({\cal{L}}(\eta)\omega)$.\\
Passing to the infinitesimal point of view, we obtain the following generalization of Remark 3.12 which is important for applications (See [2] for details). \\

\noindent
{\bf COROLLARY 4.9}:   $ \bar{\Omega}=\delta \bar{\omega}= L({\xi}_q)\bar{\omega}=f_q^{-1}(L({\eta}_q)\omega) \hspace{4mm}  \Rightarrow  \hspace{4mm}  \delta \bar{\omega}={\cal{L}}(\xi)\bar{\omega}=j_q(f)^{-1}({\cal{L}}(\eta)\omega)$.\\

 Recapitulating the results so far obtained concerning the links existing between the source and the target points of view, we may set in a symbolic way:  \\
 \[  {\delta}f_q  \stackrel{(f_q)}{\longleftrightarrow}{\eta}_q \stackrel{(f_{q+1})}{\longleftrightarrow} {\bar{\xi}}_q \stackrel{({\chi}_q)}{\longleftrightarrow} {\xi}_q   \]
In order to help the reader maturing the corresponding nontrivial formulas, we compute explicitly the case $n=1, q=1,2$ and let the case $n$ arbitrary left to the reader as a difficult exercise that cannot be achieved by hand when $q\geq 3$:  \\

\noindent
{\bf EXAMPLE 4.10}: Using the previous formulas, we have  $\delta f(x)=\eta (f(x)), \delta f_x(x)={\eta}_y(f(x))f_x(x) $ and :  \\
\[   {\eta}_1=f_2( {\bar{\xi}}_1)  \Rightarrow  ( {\eta}(f(x))=f_x(x){\bar{\xi}}(x),\hspace{5mm} {\eta}_y(f(x)f_x(x)=f_x(x){\bar{\xi}}_x(x) + f_{xx}(x){\bar{\xi}}(x) )  \]
The delicate point is that we have successively:   \\
\[ {\chi}_{,x}=\frac{{\partial}_xf}{f_x}-1= A-1, \hspace{5mm} {\chi}_{x,x}=\frac{1}{f_x}({\partial}_xf_x-\frac{{\partial}_xf}{f_x}f_{xx})  \]
\[{\bar{\xi}}=\xi + {\chi}_{,x}(\xi)=\frac{{\partial}_xf}{f_x} \xi =A \xi, \hspace{5mm}{\bar{\xi}}_x={\xi}_x + {\chi}_{x,x}\xi \hspace{5mm} \Rightarrow  
\hspace{5mm}   \eta={\partial}_xf \xi , \hspace{5mm} {\eta}_y = {\xi}_x +  \frac{{\partial}_xf_x}{f_x}  \xi    \]
\[   f_x {\eta}_{yy}={\xi}_{xx}+\frac{f_{xx}}{f_x}{\xi}_x + (\frac{{\partial}_x f_{xx}}{f_x} - \frac{ f_{xx}}{(f_x)^2}{\partial}_x f_x)\xi   \]
When $z=g(y), y=f(x) \Rightarrow z=(g\circ f) (x)=h(x)$, we obtain therefore the simple groupoid composition formulas $h_x(x)=g_y(f(x))f_x(x)$ and thus:  \\
\[ \zeta={\partial}_xh\xi={\partial}_yg \eta={\partial}_yg{\partial}_xf \xi ,\hspace{5mm}
{\zeta}_z={\eta}_y + \frac{{\partial}_yg_y}{g_y}\eta={\xi}_x + (\frac{{\partial}_yg_y}{g_y}{\partial}_xf +\frac{{\partial}_xf_x}{f_x})\xi ={\xi}_x +\frac{{\partial}_xh_x}{h_x} \xi  \]
 Using indices in arbitrary dimension, we get successively:  \\
 \[{\eta}^k=f^k_r{\bar{\xi}}^r, {\eta}^k_uf^u_i=f ^k_r{\bar{\xi}}^r_i + f^k_{ri}{\bar{\xi}}^r\ {\eta}^k\Rightarrow {\eta}^k={\xi}^r {\partial}_rf^k, {\eta}^k_uf^u_i =f^k_s({\xi}^s _i  +   g^s_u({\partial}_rf^u_i - A^t_rf^u_{ti}){\xi}^r)  + f^k_{ti}A^t_r{\xi}^r   \]
 \[  {\eta}^k_u= g^i_uf^k_s {\xi}^s_i + {\xi}^r  g^i_u{\partial}_rf^k_i \Rightarrow  {\eta}^k_k={\xi}^r_r+ {\xi}^rg^i_u {\partial}_rf^u_i     \]
 As a very useful application, we obtain successively:   \\
 \[  \Delta (x) =det ({\partial}_if^k(x))  \Rightarrow \delta \Delta=\Delta \frac{\partial{\eta}^k}{\partial y^k}=\Delta {\partial}_r{\xi}^r + {\xi}^r{\partial}_r\Delta={\partial}_r({\xi}^r\Delta)   \]
 \[   \delta det(A)=det(A)(\frac{\partial {\eta}^k}{\partial y^k}-{\eta}^k_k)=det(A)({\partial}_r{\xi}^r -{\xi}^r_r) +{\xi}^r {\partial}_r det(A)  \]
 where sections of jet bundles are used in an essential way, and the important lemma:  \\
 
 \noindent
 {\bf LEMMA 4.11}:  When the transformation $y=f(x)$ is invertible with inverse $x=g(y)$, we have the {\it fundamental identity} over the source or over the target:  \\
 \[   \frac{\partial}{\partial x^i}(\Delta (x) \frac{\partial g^i}{\partial y^k}(f(x)))\equiv 0, \hspace{2mm} \forall x\in X \,\, \Leftrightarrow  \,\,
  \frac{\partial}{\partial y^k}(\frac{1}{\Delta (g(y))} \frac{\partial f^k}{\partial x^i}(g(y))) \equiv 0 , \hspace{2mm}\forall y\in Y \]
 
 \noindent
 {\bf EXAMPLE 4.12}: We proceed the same way for studying the links existing between ${\chi}_q=\bar{D}f_{q+1}$ {\it over the source,} ${{\chi}_q}^{-1}={\sigma}_q=\bar{D}f_{q+1}^{-1}$ {\it over the target} and the nonlinear Spencer operator. First of all, we notice that:  \\
 \[{\sigma}_q=f_{q+1}\circ j_1({f_q}^{-1}) - id_{q+1}=f_{q+1}\circ (id_{q+1}- {f_{q+1}}^{-1}j_1(f_q)) \circ j_1(f_q)^{-1}= - f_{q+1}\circ {\chi}_q \circ j_1(f_q)^{-1} \]
 and the components of ${\sigma}_q$ thus factor through linear combinations of the components of ${\chi}_q$.
 After tedious computations, we get successively when $m=n=1$:   \\
 \[{\chi}_{,x}=\frac{{\partial}_xf}{f_x} -1=A -1=\frac{1}{f_x}({\partial}_xf - f_x)  \]
 \[  {\chi}_{x,x}=  \frac{1}{f_x}({\partial}_xf_x - \frac{{\partial}_xf}{f_x}f_{xx}) = \frac{1}{f_x}({\partial}_xf_x-f_{xx})- \frac{f_{xx}}{(f_x)^2}
 ({\partial}_xf-f_x)  \]
 \[  \begin{array}{lcl}
 {\chi}_{xx,x}  & =  &  \frac{1}{f_x}({\partial}_xf_{xx}- \frac{{\partial}_xf}{f_x}f_{xxx}) - 2 \frac{f_{xx}}{(f_x)^2}({\partial}_xf_x -\frac{{\partial}_xf}{f_x}f_{xx})    \\
   &  =  & \frac{1}{f_x}({\partial}_xf_{xx} -f_{xxx}) - 2 \frac{f_{xx}}{(f_x)^2}({\partial}_xf_x - f_{xx}) + (2 \frac{(f_{xx})^2}{(f_x)^3}-\frac{f_{xxx}}{(f_x)^2})({\partial}_xf-f_x)   
   \end{array}  \]
These formulas agree with the successive constructive/inductive identities:  \\
\[  \left  \{  \begin{array}{rcl} 
{\chi}_{,x} f_x  &  =  &  {\partial}_xf - f_x  \\
  {\chi}_{x,x}f_x + {\chi}_{,x}f_{xx}  &   =  &  {\partial}_xf_x -f_{xx}   \\
 {\chi}_{xx,x}f_x + 2 {\chi}_{x,x}f_{xx} + {\chi}_{,x}f_{xxx}   &  = &  {\partial}_xf_{xx}- f_{xxx}  
 \end{array}  \right.   \]
showing that ${\chi}_q$ is linearly depending on $Df_{q+1}$ and we finally get:   \\
\[ \left  \{ \begin{array}{rcl}
{\sigma}_{,y}  &  =  &  -({\partial}_xf - f_x)\frac{1}{{\partial}_xf}  =  \frac{f_x}{{\partial}_xf} -1 = - f_x{\chi}_{,x}\frac{1}{{\partial}_xf}  \\
   &   &     \\
{\sigma}_{y,y}  &  = &  - \frac{1}{f_x}({\partial}_xf_x - f_{xx})\frac{1}{{\partial}_xf} = - ({\chi}_{x,x}+ \frac{f_{xx}}{f_x}{\chi}_{,x})\frac{1}{{\partial}_xf}    \\
     &    &     \\
{\sigma}_{yy,y} & = & -((\frac{1}{(f_x)^2}({\partial}_xf_{xx} - f_{xxx}) - \frac{f_{xx}}{(f_x)^3}({\partial}_xf_x-f_{xx})) \frac{1}{{\partial}_xf}\\
  & =  & - (\frac{1}{f_x}{\chi}_{xx,x} + \frac{f_{xx}}{(f_x)^2}{\chi}_{x,x} +(\frac{f_{xxx}}{(f_x)^2} - \frac{(f_{xx})^2}{(f_x)^3}){\chi}_{,x})\frac{1}{{\partial}_xf} 
\end{array}  \right. \]
while using successively the relations $g_yf_x=1, {\partial}_yg{\partial}_xf=1, g_{yy}(f_x)^2+g_yf_{xx}=0$ and so on when $x=g(y)$ is the inverse of $y=f(x)$, in a coherent way with the action of $f_2$ on $J_2(T)$ which is described as follows:  \\
\[  \left  \{  \begin{array}{lcl} 
\eta & = & f_x \xi  \\
{\eta}_y  & = & {\xi}_x + \frac{f_{xx}}{f_x}\xi  \\
{\eta}_{yy} & = & \frac{1}{f_x}  {\xi}_{xx} +\frac{f_{xx}}{(f_x)^2}{\xi}_x + (\frac{f_{xxx}}{(f_x)^2} - \frac{(f_{xx})^2}{(f_x)^3}) \xi
 \end{array}  \right.   \]
Restricting these formulas to the affine case defined by $y_{xx}=0 \Rightarrow {\xi}_{xx}=0$, we have thus $y_{xx}=0, y_{xxx}=0 \Rightarrow f_{xx}=0, f_{xxx}=0$. It follows that $\eta=f_x\xi, {\eta}_y={\xi}_x, {\eta}_{yy}=\frac{1}{f_x}{\xi}_{xx}=0$ on one side and ${\chi}_{xx,x}=0 \Leftrightarrow {\sigma}_{yy,y}=0$ in a coherent way. It is finally important to notice that these results are not evident, even when $m=n=1$, as soon as second order jets are involved. \\

We shall use {\it all} the preceding formulas in the next example showing that, contrary to what happens in elasticity theory where the source is usually identified with the Lagrange variables, in both the Vessiot/Janet and the Cartan/Spencer approaches, {\it the source must be identified with the Euler variables without any possible doubt}.   \\

\noindent
{\bf EXAMPLE 4.13}: With $n=1, q=1,{\cal{F}}=T^*$ and the finite OD Lie equation $\omega(y)y_x=\omega(x)$ with $\omega \in T^*$ and corresponding Lie operator ${\cal{D}}\xi\equiv {\cal{L}}(\xi)\omega=\omega(x){\partial}_x\xi + \xi {\partial}_x\omega(x) $ {\it over the source}, we have:  \\
\[  \omega(f(x))f_x(x)={\bar{\omega}}(x), \hspace{5mm}  \omega(f(x))f_{xx}(x)+ {\partial}_y\omega(f(x))f^2_x(x)={\partial}_x{\bar{\omega}}(x)  \]
Differentiating once the first equation and substracting the second, we obtain therefore:   \\
\[ \omega {\sigma}_{y,y}+{\sigma}_{,y}{\partial}_y\omega \equiv -\omega(1/f_x)({\partial}_xf_x-f_{xx})(1/{\partial}_xf)+((f_x/{\partial}_xf)-1){\partial}_y\omega=0   \] 
whenever $y=f(x)$. Finally, setting $\omega(f(x)){\partial}_xf(x)=\bar{\omega}(x)$, we get {\it over the target}:   \\
\[  \delta \bar{\omega}= \omega(f(x)) \frac{\partial \eta}{\partial y} {\partial}_xf (x)+ {\partial}_xf(x) \frac{\partial \omega}{\partial y}(f(x))\eta ={\partial}_xf {\cal{L}}(\eta) \omega \]
Differentiating $\eta =\xi {\partial}_xf$ in order to obtain $\frac{\partial \eta}{\partial y} ={\partial}_x \xi + \xi ({\partial}_{xx}f/{\partial}_xf)$, we get {\it over the source}: \\ 
\[    \delta \bar{\omega}= \bar{\omega}{\partial}_x\xi+ \xi {\partial}_x{\bar{\omega}}= {\cal{L}}(\xi)\bar{\omega}  \]
We may summarize these results as follows:  \\
\[  \delta \bar{\omega}= {\cal{L}}(\xi)\bar{\omega}  \,\,\,  \stackrel{(j_1(f))}{\longrightarrow}\,\,\,  \delta {\bar{\omega}}= 
{\partial}_xf {\cal{L}}(\eta)\omega    \]
We invite the reader to extend this result to an arbitrary dimension $n\geq 2$.  \\

\noindent
{\bf EXAMPLE 4.14}: The case of an affine stucture needs more work with $n=m=1, q=2$. Indeed, let us consider the action of the affine Lie group of transformations $\bar{y}=ay+b$ with $a,b=cst$ acting on the target $y\in Y$ considered as a copy of the real line $X$. We obtain the prolongations up to order $2$ of the $2$ infinitesimal generators of the action:  \\
\[  a \rightarrow   y\frac{\partial}{\partial y} +  y_x\frac{\partial}{\partial y_x}+  y_{xx}\frac{\partial}{\partial y_{xx}}, \hspace{5mm}    b \rightarrow   \frac{\partial}{\partial y} +  0\frac{\partial}{\partial y_x}+  0\frac{\partial}{\partial y_{xx}} \]
There cannot be any differential invariant of order $1$ and the only generating one of order $2$ can be $\Phi\equiv y_{xx}/y_x$. When $\bar{x}=\varphi(x)$ we get successively $y_x=y_{\bar{x}}{\partial}_x\varphi, y_{xx}=y_{\bar{x}\bar{x}}({\partial}_x\varphi)^2 +y_{\bar{x}}{\partial}_{xx}\varphi$ and $\Phi $ transforms like $u={\partial}_x\varphi \,\,\bar{u} + \frac{{\partial}_{xx}\varphi}{{\partial}_x\varphi}$ a result providing the bundle of geometric objects ${\cal{F}}$ with local coordinates $(x,u)$ and corresponding transition rules. For any section $\gamma$, we get the Vessiot general system ${\cal{R}}_2 \subset {\Pi}_2$ of second order finite Lie equations $\frac{y_{xx}}{y_x} + {\gamma }(y)y_x =\gamma (x)$ which is already in Lie form and relates the jet coordinates $(x,y,y_x,y_{xx})$ of order $2$. The special section is $\gamma =0$ and we may consider the automorphic system $\Phi \equiv \frac{y_{xx}}{y_x}={\bar{\gamma}}(x)$ obtained by introducing {\it any second order section} $f_2(x)=(f(x), f_x(x),f_{xx}(x))$, for example $f_2=j_2(f)$ providing $(f(x), {\partial}_xf(x), {\partial}_{xx}f(x))$. It is not at all evident, {\it even on such an elementary example}, to compute the variation $\bar{\Gamma}=\delta {\bar{\gamma}}$ induced by the previous formulas and to prove that, like any {\it field} quantity, it only depends on $\bar{\gamma}$ on the condition to use {\it only} source quantities. For this, setting $\frac{f_{xx}(x)}{f_x(x)}={\bar{\gamma}}(x)$, varying and substituting, we obtain:  \\
\[ \bar{\Gamma}=\delta {\bar{\gamma}}= \frac{\delta f_{xx}}{f_x} - \frac{f_{xx}}{(f_x)^2}\delta f_x=f_x {\eta}_{yy}={\xi}_{xx} + 
{\bar{\gamma}}{\xi}_x + \xi {\partial}_x{\bar{\gamma}}  \]
Now, linearizing the preceding Lie equation over the identity transformation $y=x$, we get the Medolaghi equation:  \\
\[         L({\xi}_2)\gamma \equiv{\xi}_{xx} + \gamma (x){\xi}_x + \xi {\partial}_x\gamma (x)=0, \forall {\xi}_2\in R_2 \subset J_2(T)   \]
and the striking formula ${\bar{\Gamma}}=\delta {\bar{\gamma}}= L({\xi}_2)\bar{\gamma}$ over the source for an arbitrary ${\xi}_2 \in J_2(T)$. We finally point out the fact that, as we have just shown above and contrary to what the brothers Cosserat had in mind, the first order operators involved in the nonlinear Spencer sequence have {\it strictly nothing to do} with the operators involved in the nonlinear Janet sequence whenever $q\geq 2$. For example, in the present situation, ${\chi}_{,x}= \frac{{\partial}_xf}{f_x} - 1$ has nothing to do with $\Phi\equiv \frac{f_{xx}}{f_x}$. Similarly, using Remark $4.7$ in the linear framework, we have the first order Spencer operator $D_1: (\xi, {\xi}_x) \rightarrow ({\partial}_x\xi - {\xi}_x,{\partial}_x{\xi}_x )$ on one side and the second order Lie operator ${\cal{D}}:\xi \rightarrow {\partial}_{xx}\xi$ on the other side.  \\

The next delicate example proves nevertheless that target quantities may also be used.  \\

\noindent
{\bf EXAMPLE 4.15}: In the last example, depending on the way we use $\bar{\gamma}(x)$ on the source or $\gamma (y)$ on the target, we may consider the two (very different) Medolaghi equations:   \\
\[      {\xi}_{xx} + {\bar{\gamma}}(x) {\xi}_x + \xi {\partial}_x {\bar{\gamma}}(x)=0  \hspace{5mm}  \leftrightarrow  \hspace{5mm} 
         {\eta}_{yy} + \gamma (y) {\eta}_y + \eta {\partial}_y {\gamma} (y)=0   \]
Now, starting from the single OD equation $\frac{f_{xx}}{f_x}=\bar{\gamma}(x)$ in sectional notations, we may successively {\it differentiate} and {\it prolongate} once in order to get:  \\
\[   \frac{{\partial}_x f_{xx}}{f_x} - \frac{f_{xx}}{(f_x)^2} {\partial}_xf_x={\partial}_x{\bar{\gamma}}(x)                                              \hspace{5mm} \leftrightarrow  \hspace{5mm}  
     \frac{f_{xxx}}{f_x} - (\frac{f_{xx}}{f_x})^2= {\partial}_x{\bar{\gamma}}(x)     \]
Substracting the second from the first as a way to eliminate $\bar{\gamma}$, we obtain a linear relation involving only the components of the nonlinear Spencer operator in a coherent way with the theory of nonlinear systems ([30],[41]), namely:  \\
\[  \frac{1}{f_x} ( {\partial}_x f_{xx} - f_{xxx}) - \frac{f_{xx}}{(f_x)^2}({\partial}_xf_x - f_{xx}) =0   \]
At first sight it does not seem possible to know whether we have a linear combination of the components of ${\chi}_2$ or of the components of ${\sigma}_2$. However, if we come back to the original situation $f_q^{-1}(\omega)=\bar{\omega}$, we have eliminated $j_1(\bar{\gamma})$ over the source and we are thus only left with $j_1(\gamma)$ over the target. Hence it can only depend on ${\sigma}_2$ and we find indeed the striking relation:  \\
\[  -  \frac{1}{f_x}    [ \frac{1}{f_x} ( {\partial}_x f_{xx} - f_{xxx}) - \frac{f_{xx}}{(f_x)^2}({\partial}_xf_x - f_{xx}) ] \frac{1}{{\partial}_xf}=
{\sigma}_{yy,y}=0          \] 
provided by the simple second order Medolaghi equation $\gamma=0  \Rightarrow  {\eta}_{yy}=0$ over the target. It is essential to notice that no classical technique can provide these results which are essentially depending on the nonlinear Spencer operator and are thus not known today.  \\

\noindent
{\bf EXAMPLE 4.16}: The above methods can be applied to any explicit example. The reader may treat as an exercise the case of the pseudogroup of isometries of a non degenerate metric which can be found in any textbook of continuum mechanics or elasticity theory, though in a very different framework with methods only valid for tensors. With the previous notations, let $\omega \in S_2T^*$ with $det(\omega)\neq 0$ and consider the following nonlinear system $ {\omega}_{kl}(f(x)){\partial}_if^k(x){\partial}_jf^l(x)={\bar{\omega}}_{ij}(x)$ with $1 \leq i,j,k,l \leq n$. One obtains therefore:  \\
\[  {\delta \bar{\omega}}_{ij}= {\bar{\omega}}_{rj}{\partial}_i{\xi}^r + {\bar{\omega}}_{ir}{\partial}_i{\xi}^u + {\xi}^r{\partial}_r {\bar{\omega}}_{ij}  =  {\partial}_if^k{\partial}_jf^l ({\omega}_{ul}\frac{\partial {\eta}^u}{\partial y^k}+ {\omega}_{ku}\frac{\partial {\eta}^u}{\partial y^l}+ {\eta}^u \frac{\partial {\omega}_{kl}}{\partial y^u}  ) \]
and thus the same recapitulating formulas linking the source and target variations:  \\
\[    \delta \bar{\omega}={\cal{L}}(\xi )\bar{\omega}  \,\,\,  \stackrel {(j_1(f))}{\longrightarrow} \,\,\,  \delta {\bar{\omega}}=({\partial}_if^k{\partial}_jf^l ({\cal{L}}(\eta)\omega)_{kl})  \]

It is also difficult to compute or compare the variational formulas over the source and target in the nonlinear Spencer sequence, even when $m=n=1$ and $q=0, 1$ ([47]).  \\

\noindent
{\bf EXAMPLE 4.17}: Let us prove that the explicit computation of the gauge transformation is at the limit of what can be done with the hand, even when $m=n=1, q=1$. We have successively:
\[{\chi}_{,x}=\frac{{\partial}_xf}{f_x} - 1, \,\, {\chi}_{x,x}= \frac{1}{f_x}({\partial}_xf_x - \frac{{\partial}_xf}{f_x} f_{xx})\] 
\[f'(x)=g(f(x)) \Rightarrow  f'_x=g_yf_x, \,\,\, f'_{xx}=g_{yy}(f_x)^2 + g_yf_{xx}\] 
and thus:  
\[ \begin{array}{lcl}
{\chi}'_{,x} & = & \frac{{\partial}_xf'}{f'_x} - 1 =\frac{{\partial}_yg{\partial}_xf}{g_yf_x} - 1= ({\chi}_{,y} + 1)\frac{{\partial}_xf}{f_x} - 1=
 \frac{{\partial}_xf}{f_x}{\chi}_{,y}+ ( \frac{{\partial}_xf}{f_x}- 1)   \\
 {\chi}'_{x,x} & = &  \frac{1}{f'_x}({\partial}_xf'_x - \frac{{\partial}_xf'}{f'_x} f'_{xx})    \\
                   & =  &   \frac{1}{g_yf_x}( {\partial}_yg_y({\partial}_xf)f_x  + g_y{\partial}_xf_x - \frac{{\partial}_yg{\partial}_xf}{g_yf_x}( g_{yy}(f_x)^2 +g_yf_{xx})) \\
                   &  =  &  \frac{1}{g_y}{\partial}_yg_y {\partial}_xf + \frac{{\partial}_xf_x}{f_x}- \frac{{\partial}_yg{\partial}_xf g_{yy}}{(g_y)^2}
                                    - \frac{{\partial}_yg{\partial}_xff_{xx}}{g_y(f_x)^2}   \\
                   &  =  &  ({\partial}_xf{\chi}_{y,y}- \frac{{\partial}_xff_{xx}}{(f_x)^2} {\chi}_{,y})  + \frac{1}{f_x}( {\partial}_xf_x - \frac{{\partial}_xf}{f_x}f_{xx})
\end{array}   \]
Setting $f_2= id_2 +t{\xi}_2 + ...$ and passing to the limit when $ t \rightarrow 0$, we finally obtain:  \\
\[  \begin{array}{lcl}
\delta {\chi}_{,x}  &  = &  ({\partial}_x{\xi} - {\xi}_x) + (\xi {\partial}_x{\chi}_{,x} + {\chi}_{,x}{\partial}_x\xi - {\chi}_{,x}{\xi}_x)   \\
\delta {\chi}_{x,x} & = &  ({\partial}_x{\xi}_x - {\xi}_{xx}) + (\xi {\partial}_x{\chi}_{x,x} + {\chi}_{x,x}{\partial}_x\xi - {\chi}_{,x}{\xi}_{xx})
\end{array}  \]
If we use the standard euclidean metric $\omega=1 \Rightarrow \gamma =0$, we may thus introduce the pure $1$-form $\alpha= {\chi}_{x,x} + \gamma{\chi}_{,x}$. We should consider the defining formula ${\chi}'_1=f^{-1}_2\circ {\chi}_1\circ j_1(f_1) + \bar{D}f_2$ and have to introduce the addidtional term $f^{-1}_2(\gamma){\chi}_{,x}$ which is only leading to the additional infinitesimal term $(L({\xi}_2)\gamma) {\chi}_{,x}={\xi}_{xx}{\chi}_{,x}$ because $\gamma=0$. We finally obtain:   
\[   \delta \alpha= \delta {\chi}_{x,x} + {\xi}_{xx}{\chi}_{,x} + \gamma \delta {\chi}_{,x}= ({\partial}_x{\xi}_x - {\xi}_{xx}) + (\xi {\partial}_x\alpha + \alpha {\partial}_x{\xi})  \]
and this result can be easily extended to an arbitrary dimension with the formula:  
\[  {\alpha}_i={\chi}^r_{r,i} + {\gamma}^s_{sr}{\chi}^r_{,i} \,\, \Rightarrow \,\,   (\delta \alpha)_i= ({\partial}_i{\xi}^r_r - {\xi}^r_{ri}) + 
({\xi}^r{\partial}_r{\alpha}_i + {\alpha}_r {\partial}_i{\xi}^r)  \]

Comparing this procedure with the one we have adopted in the previous exampes, we have:  \\
\[  {\chi}_{,x}=\frac{{\partial}_xf}{f_x} - 1=A - 1 \Rightarrow \delta {\chi}_{,x}=\frac{{\partial}_x\delta f}{f_x} -\frac{{\partial}_xf}{(f_x)^2} \delta f_x=
\frac{1}{f_x}(\frac{\partial \eta}{\partial y} - {\eta}_y){\partial}_xf  \]
However, taking into account the formulas $\eta=\xi {\partial}_xf$ and ${\eta}_y={\xi}_x + \frac{{\partial}_xf_x}{f_x} \xi$, we also get:  \\
\[  \begin{array}{rcl}
\delta {\chi}_{,x}  &= &  \frac{1}{f_x} ({\partial}_x\xi{\partial}_xf + \xi{\partial}_{xx}f) - \frac{{\partial}_xf}{(f_x)^2}({\xi}_xf_x+\xi {\partial}_xf_x)  \\
   & =  & A({\partial}_x\xi - {\xi}_x)  +  \xi {\partial}_x {\chi}_{,x}   \\
   &  =  &({\partial}_x\xi - {\xi}_x)  + ( \xi {\partial}_x {\chi}_{,x}+ {\chi}_{,x}{\partial}_x\xi - {\chi}_{,x} {\xi}_x)
   \end{array}   \]
Working over the target is more difficult. Indeed, we have successively ({\it care to the first step}):  \\
\[{\sigma}_{,y}=\frac{f_x}{{\partial}_xf}- 1 \Rightarrow  \delta {\sigma}_{,y} + \eta \frac{\partial {\sigma}_{,y}}{\partial y}=\frac{\delta f_x}{{\partial}_xf} - \frac{f_x}{({\partial}_xf)^2} {\partial}_x\delta f = - \frac{f_x}{({\partial}_xf) ^2}(\frac{\partial \eta}{\partial y} - {\eta}_y)  \]
\[ \begin{array}{rcl}
 \delta {\sigma}_{,y} &  =  &  - [ \frac{f_x}{{\partial}_xf}(\frac{\partial \eta }{\partial y} - {\eta}_y) + \eta \frac{\partial {\sigma}_{,y}}{\partial y}]\\
                                &  =  &  -[(\frac{\partial \eta}{\partial y} - {\eta}_y) + (\eta \frac{\partial {\sigma}_{,y}}{\partial y}+ {\sigma}_{,y}\frac{\partial \eta}{\partial y} - {\sigma}_{,y} {\eta}_y ) ]
\end{array}   \]
More generaly, we let the reader prove that the variation of ${\sigma}_q$ over the target (respectively the source) is described by "{\it minus}" the same formula as the variation of ${\chi}_q$ over the source (respectively the target). In any case, the reader must not forget that the word "variation" just means that the section $f_{q+1}$ is changed, {\it not} that the source is moved. Accordingly, getting in mind this example and for simplicity, we  shall always prefer to work with vertical bundles over the source, closely following the purely mathematical definitions, contrary to Weyl ([54],\S 28, formulas (17) to (27), p 233-236). The reader must be now ready for comparing the variations of ${\chi}_{x,x}$ and ${\sigma}_{y,y}$. \\

In order to conclude this section, we provide without any proof two results and refer the reader to ([28]) for details.  \\

\noindent
{\bf PROPOSITION 4.18}: Changing slightly the notation while setting ${\sigma}_{q-1}={\bar{D}}'{\chi}_q$, we have:  \\
\[   {\chi}'_q= f^{-1}_{q+1}\circ {\chi}_q \circ j_1(f) + \bar{D}f_{q+1}  \,\,\, \Rightarrow \,\,\, {\sigma}'_{q-1}= f^{-1}_q \circ {\sigma}_{q-1}\circ j_1(f) \]
where $f^{-1}_q$ acts on $J_{q-1}(T)$ and $j_1(f)$ acts on ${\wedge}^2T^*$. It follows that gauge transformations exchange the solutions of 
${\bar{D}}'$ among themselves.\\  
 
 \noindent
 {\bf COROLLARY 4.19}: Denoting by ${\cal{C}}( )$ the cyclic sum, we have the so-called {\it Bianchi identity}:   \\
 \[   D{\sigma}_{q-1}(\xi,\eta,\zeta) + {\cal{C}}(\xi,\eta,\zeta)\{{\sigma}_{q-1}(\xi,\eta) , {\chi}_{q-1}(\zeta)=0   \]  \\   \\

 \noindent
{\bf 5) APPLICATIONS}   \\ 

Before studying in a specific way electromagnetism and gravitation, we shall come back to Example $4.10$ and provide a technical result which, though looking like evident at first sight, is at the origin of a deep misunderstanding done by the brothers Cosserat and Weyl on the variational procedure used in the study of physical problems (Compare to [20] and [50]).  \\

Setting $dx=dx^1 \wedge ... \wedge dx^n$ for simplicity while using Lemma $4.11$ and the fact that the standard Lie derivative is commuting with any diffeomorphism, we obtain at once:   \\
\[   y=f(x) \,\, \Rightarrow \,\, dy= det({\partial}_if^k(x)) dx = {\Delta}(x) dx  \]
\[  \eta =T(f)\xi \,\, \Rightarrow \,\, {\cal{L}}(\eta)dy={\cal{L}}(\xi)(\Delta (x)dx)\,\, \Rightarrow \,\,   
\delta \Delta  =  \Delta div_y(\eta) = \Delta div_x(\xi) + {\xi}^r{\partial}_r\Delta   \]
The interest of such a presentation is to provide the right correspondence between the source/target and the Euler/Lagrange choices. Indeed, if we use the way followed by most authors up to now in continuum mechanics, we should have source=Lagrange, target=Euler, a result leading to the conservation of mass $dm= \rho dy={\rho}_0dx=dx$ when ${\rho}_0$ is the original initial mass per unit volume. We may set ${\rho}_0=1$ and obtain therefore $\rho(f(x))=1/\Delta (x)$, a choice leading to:  \\
\[ \delta  \rho + {\eta}^k\frac{\partial \rho}{\partial y^k}= - \frac{1}{{\Delta}^2}\delta \Delta \,\, \Rightarrow \,\, \delta \rho= - \rho \frac{\partial {\eta}^k}{\partial y^k} - {\eta}^k\frac{\partial \rho}{\partial y^k}= - \rho \frac{\partial {\xi}^r}{\partial x^r} \,\,  \Rightarrow  \,\,  \delta \rho= - \frac{\partial (\rho {\eta}^k)}{\partial y^k}       \]
but the concept of "{\it variation} " is not mathematically well defined, though this result is coherent with the classical formulas that can be found for example in ([7],[27]) or ([53], (17) and (18) p 233, (20) to (21) p 234, (76) p 289, (78) p 290) where " {\it points are moved} ". \\
On the contrary, if we adopt the {\it unusual} choice source=Euler, target=Lagrange, we should get $\rho (x)= \Delta (x)$, a choice leading to $\delta \rho= \delta \Delta$ and thus:  \\
\[\delta \rho= \rho \frac{\partial {\eta}^k}{\partial y^k}=\rho \frac{\partial {\xi}^r}{ \partial x^r} + {\xi}^r{\partial}_r\rho = {\partial}_r(\rho {\xi}^r) \]
which is the right choice agreeing, up to the sign, with classical formulas but with the important improvement that this result becomes a purely mathematical one, obtained from a well defined variational procedure involving only the so-called " {\it vertical} " machinery. This result fully explains why we had doubts about the sign involved in the variational formulas of ([27], p 383) but without being able to correct them at that time.  We may finally revisit Lemma $4.11$ in order to obtain the {\it fundamental identity over the source}:   \\ 
 \[    \frac{\partial}{\partial x^i}(\Delta (x) \frac{\partial g^i}{\partial y^k}(f(x))) \equiv 0 , \hspace{4mm}\forall x\in X \]    
which becomes the conservation of mass when $n=4$ and $k=4$.  \\  

In addition, as many chases will be used through many diagrams in the sequel, we invite the reader not familiar with these technical tools to consult the books ([3],[21],[48]) that we consider as the best references for learning about homological algebra. A more elementary approach can be found in ([32]) that has been used during many intensive courses on the applications of homological algebra to control theory. As for {\it differential homological algebra}, one of the most difficult tools existing in mathematics today, and its link with applications, we refer the reader to the various references provided in ([37],[45]).\\

Finally, for the reader interested by a survey on more explicit applications, we particularly refer to ([2],[24],[34],[42]) for analytical mechanics and hydrodynamics, ([31],[33],[46]) for coupling phenomenas, ([36],[38],[42],[55-56]) for the foundations of Gauge Theory, ([35],[39],[41],[42-43]) for the foundations of General Relativity, ([40],[44]) for unusual explicit computations of compatibility conditions (CC) for linear differential operators.  \\   \\

\noindent
{\bf A) POINCARE, WEYL AND CONFORMAL GROUPS}  \\
 
When constructing inductively the Janet and Spencer sequences for an involutive system $R_q \subset J_q(E)$, we have to use the following commutative and exact diagrams where we have set $F_0=J_q(E)/R_q$ and used a diagonal chase :  \\
\[  \begin{array}{rcccccl} 
  &  0  &  &  &  &  &  \\
   & \downarrow &  &  &  & &  \\
    &  0  &  &  0 &  &  0  &  \\
   & \downarrow &  &  \downarrow  & & \downarrow  \\
   0 \rightarrow & \delta ({\wedge}^{r-1}T^*\otimes g_{q+1} ) & \rightarrow & {\wedge}^rT^*\otimes R_q & \rightarrow & C_r & \rightarrow 0   \\
              & \downarrow & & \downarrow & & \downarrow  &  \\
  0 \rightarrow & \delta({\wedge}^{r-1}T^*\otimes S_{q+1}T^*\otimes T)  &                                                                                                         \rightarrow & {\wedge}^rT^*\otimes J_q(E) & \rightarrow & C_r(E) & \rightarrow 0  \\
    &  \downarrow &  &  \downarrow  & \searrow &  \downarrow  &    \\
   0 \rightarrow & {\wedge}^rT^*\otimes R_q + \delta ({\wedge}^{r-1}T^*\otimes S_{q+1}T^*\otimes E) & \rightarrow & {\wedge}^rT^*\otimes F_0 & \rightarrow & F_r & \rightarrow 0  \\
  &\downarrow &  & \downarrow &  & \downarrow &  \\  
 & 0   &  &  0   &  & 0 
\end{array}  \] 
 It follows that the short exact sequences $0 \rightarrow C_r \rightarrow C_r(E) \stackrel{{\Phi}_r}{\longrightarrow} F_r \rightarrow 0$ are allowing to define the Janet and Spencer bundles inductively. If we consider {\it two} involutive systems $0 \subset R_q \subset  {\hat{R}}_q \subset J_q(E)$, it follows that the {\it kernels} of the induced canonical epimorphisms $F_r \rightarrow {\hat{F}}_r \rightarrow 0$ are isomorphic to the {\it cokernels} of the canonical monomorphisms $0 \rightarrow C_r \rightarrow {\hat{C}}_r \subset C_r(E)$ and we may say that {\it Janet and Spencer play at see-saw} because we have the formula $dim(C_r) + dim(F_r)=dim(C_r(E))$.\\
 
When dealing with applications, we have set $E=T$ and considered systems of finite type Lie equations determined by Lie groups of transformations. Accordingly, we have obtained in particular $C_r={\wedge}^rT^*\otimes R_2 \subset {\wedge}^rT^*\otimes {\hat{R}}_2 ={\hat{C}}_r \subset C_r(T)$ when comparing the classical and conformal Killing systems, but {\it these bundles have never been used in physics}. However, instead of the classical Killing system $R_1\subset J_1(T)$ defined by the infinitesimal first order PD Lie equations $\Omega \equiv {\cal{L}}(\xi)\omega=0$ {\it and} its first prolongations  $R_2 \subset J_2(T)$ defined by the infinitesimal additional second order PD Lie equations $\Gamma\equiv {\cal{L}}(\xi)\gamma=0$ or the conformal Killing system ${\hat{R}}_2\subset J_2(T)$ defined by $\Omega\equiv {\cal{L}}(\xi)\omega=2 \, A(x)\omega$ and ${\Gamma} \equiv {\cal{L}}(\xi)\gamma= ({\delta}^k_iA_j(x) +{\delta} ^k_j A_i(x) -{\omega}_{ij}{\omega}^{ks}A_s(x)) \in S_2T^*\otimes T$ but we may also consider the formal Lie derivatives for geometric objects:  \\
\[   {\Omega}_{ij}\equiv (L({\xi}_1)\omega)_{ij} \equiv {\omega}_{rj}{\xi}^r_i + {\omega}_{ir}{\xi}^r_j + {\xi}^r{\partial}_r{\omega}_{ij}=0  \]
\[  {\Gamma}^k_{ij} \equiv (L({\xi}_2)\gamma)^k_{ij} \equiv {\xi}^k_{ij} + {\gamma}^k_{rj}{\xi}^r_j  + {\gamma}^k_{ir}{\xi}^r_j - 
{\gamma}^r_{ij}{\xi}^r_k +{\xi}^r{\partial}_r {\gamma}^k_{ij}=0  \]                                                                                                                                            

We may now introduce the {\it intermediate differential system} ${\tilde{R}}_2 \subset J_2(T)$ defined by ${\cal{L}}(\xi)\omega=2A(x) \omega$ and 
$\Gamma \equiv {\cal{L}}(\xi)\gamma=0 $, for the {\it Weyl group} obtained by adding the only dilatation with infinitesimal generator $x^i{\partial}_i$ to the Poincar\'e group. We have the relations $R_1\subset {\tilde{R}}_1={\hat{R}}_1$ and the strict inclusions $ R_2 \subset {\tilde{R}}_2 \subset {\hat{R}}_2$ when $R_2 ={\rho}_1(R_1) , {\tilde{R}}_2={\rho}_1({\tilde{R}}_1),{\hat{R}}_2={\rho}_1({\hat{R}}_1) $ but we have to notice that we must have ${\partial}_iA - A_i=0 $ for the conformal system and thus $A_i=0 \Rightarrow A=cst$ if we do want to deal again with an involutive second order system ${\tilde{R}}_2 \subset J_2(T)$. However, we must not forget that the comparison between the Spencer and the Janet sequences can only be done for involutive operators, that is we can indeed use the involutive systems $R_2 \subset {\tilde{R}}_2$ but we have to use ${\hat{R}}_3$ even if it is isomorphic to ${\hat{R}}_2$. Finally, as ${\hat{g}}_2\simeq T^*$ and ${\hat{g}}_3=0, \forall n\geq 3$, the first Spencer operator ${\hat{R}}_2\stackrel{D_1}{\longrightarrow} T^*\otimes {\hat{R}}_2$ is induced by the usual Spencer operator ${\hat{R}}_3 \stackrel{D}{\longrightarrow} T^*\otimes {\hat{R}}_2:(0,0,{\xi}^r_{rj},{\xi}^r_{rij}=0) \rightarrow (0,{\partial}_i0-{\xi}^r_{ri}, {\partial}_i{\xi}^r_{rj}- 0)$ and thus projects by cokernel onto the induced operator $T^* \rightarrow T^*\otimes T^*$. Composing with $\delta$, it projects therefore onto $T^*\stackrel{d}{\rightarrow} {\wedge}^2T^*:A \rightarrow dA=F$ as in EM and so on by using he fact that $D_1$ {\it and} $d$ {\it are both involutive} or the composite epimorphisms ${\hat{C}}_r \rightarrow {\hat{C}}_r/{\tilde{C}}_r\simeq {\wedge}^rT^*\otimes ({\hat{R}}_2/{\tilde{R}}_2) \simeq {\wedge}^rT^*\otimes {\hat{g}}_2\simeq {\wedge}^rT^*\otimes T^*\stackrel{\delta}{\longrightarrow}{\wedge}^{r+1}T^*$. The main result we have obtained is thus to be able to increase the order and dimension of the underlying jet bundles and groups as we have ([47],[53-55]): \\
\[   POINCARE \hspace{2mm} GROUP \subset WEYL \hspace{2mm} GROUP \subset CONFORMAL\hspace{2mm} GROUP  \]
that is  $10 < 11 < 15$ when $n=4$ and our aim is now to prove that {\it the mathematical structures  of electromagnetism and gravitation only depend on the second order jets}.  \\

With more details, the {\it Killing system} $R_2 \subset J_2(T)$ is defined by the infinitesimal Lie equations in {\it Medolaghi form} with the well known {\it Levi-Civita isomorphism} $(\omega,\gamma )\simeq j_1(\omega)$ for geometric objects:  \\
\[ \left\{  \begin{array}{rcl}
{\Omega}_{ij} & \equiv & {\omega}_{rj}{\xi}^r_i + {\omega}_{ir}{\xi}^r_j + {\xi}^r{\partial}_r{\omega}_{ij}=0  \\
{\Gamma}^k_{ij} & \equiv &  {\gamma}^k_{rj}{\xi}^r_i + {\gamma}^k_{ir}{\xi}^r_j - {\gamma}^r_{ij}{\xi}^k_r + 
     {\xi}^r {\partial}_r{\gamma}^k_{ij} = 0
     \end{array} \right.  \]    
      
We notice that $R_2(\bar{\omega})=R_2(\omega) \Leftrightarrow \bar{\omega}= a \, \omega, a=cst, \bar{\gamma}=\gamma$ and refer the reader to ([LAP]) for more details about the link between this result and the deformation theory of algebraic structures. We also notice that $R_1$ is formally integrable and thus $R_2$ is involutive if and only if $\omega$ has constant Riemannian curvature along the results of L. P. Eisenhart ([11]). The 
only structure constant $c$ appearing in the corresponding Vessiot structure equations is such that $\bar{c}= c/a$ and the normalizer of $R_1$ is $R_1$ if and only if $c\neq 0$. Otherwise $R_1$ is of codimension $1$ in its normalizer ${\tilde{R}}_1$ as we shall see below by adding the only dilatation. In all what follows, $\omega$ is assumed to be flat with $c=0$ and vanishing Weyl tensor. \\

 The {\it Weyl system} ${\tilde{R}}_2 \subset J_2(T)$ is defined by the infinitesimal Lie equations:  \\
 \[   \left \{  \begin{array}{rcl}
   {\omega}_{rj}{\xi}^r_i + {\omega}_{ir}{\xi}^r_j + {\xi}^r{\partial}_r{\omega}_{ij}& =  & 2A(x){\omega}_{ij}   \\
   {\xi}^k_{ij}  + {\gamma}^k_{rj}{\xi}^r_i + {\gamma}^k_{ri}{\xi}^r_j - {\gamma}^r_{ij}{\xi}^k_r  + {\xi}^r{\partial}_r{\gamma}^k_{ij} &=&0 \end{array} \right.    \]
and is involutive if and only if ${\partial}_iA=0\Rightarrow A=cst$. Introducing for convenience the {\it metric density}  ${\hat{\omega}}_{ij}= {\omega}_{ij} / (\mid det(\omega)\mid )^{\frac{1}{n}}$ , we obtain the {\it Medolaghi form} for $(\hat{\omega}, \gamma)$ with $\mid det(\hat{\omega})\mid = 1$ :   \\
\[   \left \{  \begin{array}{rcl}
 {\hat{\Omega}}_{ij} & \equiv & {\hat{\omega}}_{rj}{\xi}^r_i + {\hat{\omega}}_{ir}{\xi}^r_j - \frac{2}{n}{\hat{\omega}}_{ij}{\xi}^r_r 
 + {\xi}^r{\partial}_r{\hat{\omega}}_{ij} = 0   \\         
   {\Gamma}^k_{ij} &  \equiv  &  {\xi}^k_{ij}  + {\gamma}^k_{rj}{\xi}^r_i + {\gamma}^k_{ri}{\xi}^r_j - {\gamma}^r_{ij}{\xi}^k_r  + {\xi}^r{\partial}_r{\gamma}^k_{ij} =0 \end{array} \right.    \]                  
           
 Finally, the {\it conformal system} ${\hat{R}}_2 \subset J_2(T)$ is defined by the following infinitesimal Lie equations:   \\
 \[   \left \{  \begin{array}{rcl}
   {\omega}_{rj}{\xi}^r_i + {\omega}_{ir}{\xi}^r_j + {\xi}^r{\partial}_r{\omega}_{ij}& =  & 2A(x){\omega}_{ij}   \\
  {\xi}^k_{ij}  + {\gamma}^k_{rj}{\xi}^r_i + {\gamma}^k_{ri}{\xi}^r_j - {\gamma}^r_{ij}{\xi}^k_r  + {\xi}^r{\partial}_r{\gamma}^k_{ij} &  =  & 
  {\delta}^k_i A_j(x) + {\delta}^k_jA_i(x) - {\omega}_{ij}{\omega}^{kr}A_r(x)
 \end{array} \right.    \]
 and is involutive if and only if ${\partial}_iA-A_i= 0$ or, equivalently, if $\omega$ has vanishing Weyl tensor.
 
 However, introducing again the {\it metric density}  $\hat{\omega}$ while substituting, we obtain after prolongation and division by $(\mid det(\omega\mid)^{\frac{1}{n}}$ the second order system ${\hat{R}}_2 \subset J_2(T)$ in {\it Medolaghi form} ad the Levi-Civita isomorphim $(\omega,\gamma \simeq j_1(\omega)$ {\it restricts} to an isomorphism $(\hat{\omega}, \hat{\gamma})\simeq j_1(\hat{\omega})$ if we set:  \\
 \[{\hat{\gamma}}^k_{ij}= {\gamma}^k_{ij} - \frac{1}{n}( {\delta}^k_i{\gamma}^r_{rj} + {\delta}^k_j{\gamma}^r_{ri} - {\omega}_{ij}{\omega}^{ks}{\gamma}^r_{rs}) \Rightarrow {\hat{\gamma}}^r_{ri}=0 \,\,\, ( tr(\hat{\gamma})=0) \]
 \[   \left \{  \begin{array}{rcl}
 {\hat{\Omega}}_{ij} & \equiv & {\hat{\omega}}_{rj}{\xi}^r_i + {\hat{\omega}}_{ir}{\xi}^r_j - \frac{2}{n}{\hat{\omega}}_{ij}{\xi}^r_r 
 + {\xi}^r{\partial}_r{\hat{\omega}}_{ij} = 0 \,\, \Rightarrow \,\,   {\omega}^{ij} {\bar{\Omega}}^{ij}=0  \\
 {\hat{\Gamma}}^k_{ij} &  \equiv & {\xi}^k_{ij} - \frac{1}{n}({\delta}^k_i{\xi}^r_{rj} + {\delta}^k_j{\xi}^r_{ri} - {\hat{\omega}}_{ij}{\hat{\omega}}^{kr}{\xi}^s_{rs}) + {\hat{\gamma}}^k_{rj}{\xi}^r_i +  {\hat{\gamma}}^k_{ri}{\xi}^r_j - {\hat{\gamma}}^r_{ij}{\xi}^k_r  + {\xi}^r{\partial}_r{\hat{\gamma}}^k_{ij}=0 \,\, \Rightarrow {\hat{\Gamma}}^r_{ri}=0
 \end{array} \right.    \]
 Contracting the first equations by ${\hat{\omega}}^{ij}$ we notice that ${\xi}^r_r$ is {\it no longer vanishing} while, contractig in $k$ and $j$ the second equations, we now notice that ${\xi}^r_{ri}$ is {\it no longer vanishing}. It is also essential to notice that the symbols ${\hat{g}}_1$ and ${\hat{g}}_2$ only depend on $\omega$ and not on any conformal factor.  \\
 
The following Proposition does not seem to be known: \\

\noindent
{\bf PROPOSITION 5.A.1}: $(id, - \hat{\gamma})$ is the only symmetric ${\hat{R}}_1$-connection wih vanishing trace.\\

\noindent
{\it Proof}: Using a direct substitution, we have to study:  \\
\[  - {\hat{\omega}}_{ir}{\hat{\gamma}}^r_{jt}  - {\hat{\omega}}_{rj} {\hat{\gamma}}^r_{it} +
 \frac{2}{n} {\hat{\omega}}_{ij}{\hat{\gamma}}^r_{rt} + {\partial}_t{\hat{\omega}}_{ij}   \]
 Multiplying by $ (\mid det(\omega)\mid)^{\frac{1}{n}}$, we have to study:   \\
 \[  - {\omega}_{ir}{\hat{\gamma}}^r_{jt}  - {\omega}_{rj} {\hat{\gamma}}^r_{it} +
 \frac{2}{n} {\omega}_{ij}{\hat{\gamma}}^r_{rt} + (\mid det(\omega) \mid)^{\frac{1}{n}}{\partial}_t{\hat{\omega}}_{ij}  \]
 or equivalently:   \\
 \[    - {\omega}_{ir}{\hat{\gamma}}^r_{jt}  - {\omega}_{rj} {\hat{\gamma}}^r_{it} +
 \frac{2}{n} {\omega}_{ij}{\hat{\gamma}}^r_{rt} + {\partial}_t{\omega}_{ij} -\frac{1}{n}{\omega}_{ij} (\mid det(\omega) \mid)^{-1}{\partial}_t(\mid det(\omega)\mid)  \]
that is to say:   \\
\[    - {\omega}_{ir}{\hat{\gamma}}^r_{jt}  - {\omega}_{rj} {\hat{\gamma}}^r_{it} +
 {\partial}_t{\omega}_{ij} -\frac{2}{n}{\omega}_{ij} {\gamma}^s_{st}  \]
 Now, we have:  \\
 \[  - {\omega}_{ir} ({\gamma}^r_{jt} - \frac{1}{n} ({\delta}^r_j{\gamma}^s_{st} + {\delta}^r_t{\gamma}^s_{sj} - {\omega}_{jt}{\omega}^{ru}{\gamma}^s_{su}))= -{\omega}_{ir}{\gamma}^r_{jt} +
  \frac{1}{n} {\omega}_{ij}{\gamma}^s_{st} + \frac{1}{n}{\omega}_{it}{\gamma}^s_{sj} - \frac{1}{n}{\omega}_{jt}{\gamma}^s_{si}       \]
 Finally, taking into account that $(id, - \gamma)$ is a $R_1$-connection, we have:   \\
 \[    - {\omega}_{ir}{\gamma}^r_{jt} - {\omega}_{rj}{\gamma}^r_{it} + {\partial}_t {\omega}_{ij}=0  \]
 Hence, collecting all the remaining terms, we are left with $\frac{2}{n} {\omega}_{ij} {\gamma}^s_{st} - \frac{2}{n} {\omega}_{ij} {\gamma}^s_{st}=0$.  \\
 As for the unicity, it comes from a chase in the commutative and exact diagram:   \\
 \[  \begin{array}{rcccccl}
   & 0  &  &  0 &  &  0  &   \\
   &  \downarrow &   &  \downarrow &   &  \downarrow  &      \\
0 \longrightarrow  & {\hat{g}}_2 & \stackrel{\delta}{\longrightarrow} & T^*\otimes {\hat{g}}_1 & \stackrel{\delta}{\longrightarrow} & {\wedge}^2T^*\otimes T & \longrightarrow 0 \\
  &  \downarrow &   &  \downarrow &   &  \parallel & \\
 0 \longrightarrow & S_2T^* \otimes T& \stackrel{\delta}{\longrightarrow} & T^*\otimes T^* \otimes T & \stackrel{\delta}{\longrightarrow} & {\wedge}^2T^*\otimes T & \longrightarrow 0  \\
        &   &   &   &   &  \downarrow   &   \\
       &   &   &    &   &  0  &
 \end{array}  \]
obtained by counting the respective dimensions with $dim({\hat{g}}_1)= (n(n-1)/2)+1 = (n^2 - n +2)/2$ and $dim({\hat{g}}_2)=n$ while checking that $ - n + n(n^2 - n +2) - n^2(n-1)/2 = 0 $.  The lower sequence splits because the short exact $\delta$-sequence $0 \rightarrow S_2T^* \stackrel{\delta}{\longrightarrow} T^*\otimes T¬* \stackrel{\delta}{\longrightarrow} {\wedge}^2T^* \rightarrow 0$ splits and the upper sequence also splits because we have a composite monomorphism $ {\wedge}^2T^*\otimes T \simeq T^* \otimes g_1 \rightarrow T^*\otimes {\hat{g}}_1 $.   \\
 \hspace*{12cm}  Q.E.D.   \\     
 
\noindent
{\bf COROLLARY 5.A.2}: The $R_1$-connection $(id, - \gamma)$ is alo a ${\hat{R}}_1$-connection.   \\

\noindent
{\it Proof}: This result first follows from the fact that $(id, - \gamma) \in T^*\otimes R_1$ is over $id\in T^*\otimes T$ and $R_1 \subset {\hat{R}}_1$. However, we may also check such a property directly. Indeed, mutiplying  $- {\hat{\omega}}_{rj} {\gamma}^r_{it} - {\hat{\omega}}_{ir} {\gamma}^r_{rt}+ \frac{2}{n} {\hat{\omega}}_{ij} {\gamma}^r_{rt} + {\partial}_t{\hat{\omega}}_{ij}$ by $(\mid det(\omega)\mid)^{\frac{1}{n}}$ as in the last Proposition, we obtain:  \\
\[  - {\omega}_{rj} {\gamma}^r_{it} - {\omega}_{ir}{\gamma}^r_{jt} + \frac{2}{n}{\omega}_{ij}{\gamma}^r_{rt} + {\partial}_t {\hat{\omega}}_{ij} = - {\omega}_{rj} {\gamma}^r_{it} - {\omega}_{rj}{\gamma}^r_{jt} +{\partial}_t{\omega}_{ij} = 0  \]
because $(id, - \gamma)$ is a $R_1$-connection.     \\

\hspace*{12cm}      Q.E.D.    \\

\noindent
{\bf REMARK 5.A.3}: If one is using $(id, - \gamma)$, then $(L({\xi}_2)\gamma)^k_{ij}{\xi}^k_{ij}$ when $\gamma = 0$ locally and we have $(\delta \alpha)_i = ({\partial}_i{\xi}^r_r - {\xi}^r_{ri}) + ({\alpha}_r{\partial}_i{\xi}^r + {\xi}^r{\partial}_r{\alpha}-i)$ as the simplest variation . However, we have $f^{-1}_2(\gamma)= \bar{\gamma} \neq \gamma$ and we cannot thus split the Spencer operator over the target by means of a pull-back. Nevertheless, if one is using $(id, - \hat{\gamma})$, then $L({\xi}_2)\hat{\gamma}=0$ when ${\xi}_2\in {\hat{R}}_2$ and the variation $(\delta \alpha)_i$ contains an additional term ${\xi}^s_{sr}{\chi}^r_{,i}$ but $f^{-1}_2(\hat{\gamma})=\hat{\gamma}$ and one can split the Spencer operator over the source and over the target with the help of $\hat{\gamma}$ but we have to point out that $\gamma = 0 \Rightarrow \hat{\gamma}=0$ {\it locally}.   \\
\hspace*{12cm}     Q.E.D.    \\

 We let the reader exhibit similarly the finite {\it Lie forms} of the previous equations that will be presented when needed. We have to distinguish the strict inclusions $\Gamma \subset \tilde{\Gamma} \subset \hat{\Gamma} \subset aut(X)$ with:  \\ \\
 \noindent
 $\bullet$ The Lie pseudogroup $\Gamma \subset aut(X)$ of isometries which is preserving the metric $\omega\in S_2T^*$ with $ det(\omega)\neq 0$ {\it and thus also} $\gamma$.  \\
 \noindent
 $\bullet$ The Lie pseudogroup $\tilde{\Gamma}$  which is preserving $\hat{\omega}$ {\it and} $\gamma$.  \\
\noindent
$\bullet$ The Lie pseudogroup $\hat{\Gamma}$ of conformal isometries which is preserving $\hat{\omega}$ {\it and thus also} $\hat{\gamma}$ with:   \\
\[  g^k_l(x)(f^l_{ij}(x)+ {\gamma}^l_{rs}(f())f^r_i(x)f^s_j(x))={\bar{\gamma}}^k_{ij}(x)={\gamma}^k_{ij}(x) +{\delta}^k_ia_j(x) +{\delta}^k_ia_j(x) - {\omega}_{ij}(x){\omega}^{kr}(x)a_r(x) \]
 where $a_i(x)dx^i\in T^*$ and thus $ \bar{\gamma} - \gamma \in {\hat{g}}_2 \subset S_2T^*\otimes T^* \otimes T $.  \\

\noindent 
{\bf B) ELECTROMAGNETISM}   \\

The key idea, still never ackowledged, is that, even if we shall prove that {\it electromagnetism only depends on the elations of the conformal group} which are clearly non-linear transformations, we shall see that {\it electromagnetism has " by chance "  a purely linear behaviour}.

Indeed, setting as we already did ${\chi}_0=A - id$ and defining ${\chi}^k_{lr,j}=A^s_j{\tau}^k_{lr,s}$, we may rewrite the defining equation of the second non-linear Spencer operator ${\bar{D}}'$ in the form:   \\
\[  \left \{ \begin{array}{lcl}
{\partial}_iA^k_j -{\partial}_jA^k_i & = &  A^r_i{\chi}^k_{r,j} - A^r_j{\chi}^k_{r,i}   \\
                                                           & = &  A^r_iA^s_j( {\tau}^k_{r,s} - {\tau}^k_{s,r})  \\
{\partial}_i{\chi}^k_{l,j} - {\partial}_j{\chi}^k_{l,i} - {\chi}^r_{l, i}{\chi}^k_{r,j} + {\chi}^r_{l,j}{\chi}^k_{r,i} 
&  =  & A^r_i{\chi}^k_{lr,j} - A^r_j{\chi}^k_{lr,i}  \\
&  =  &  A^r_iA^s_j({\tau}^k_{lr,s} -{\tau}^k_{ls,r})  
\end{array} \right.   \]
Hence, contracting in $k$ and $l$, {\it the quadratic terms in} $\chi$ {\it disappear} and we get:  \\
\[     {\partial}_i{\chi}^r_{r,j} -{\partial}_j{\chi}^r_{r,i} =A^r_iA^s_j({\tau}^k_{kr,s} - {\tau}^k_{ks,r})  \] 
By analogy with EM it should be tempting to introduce ${\alpha}_i={\chi}^r_{r,i}$ and denote by ${\varphi}_{ij}$ the right member of the last formula but the relation ${\partial}_i{\alpha}_j - {\partial}_j{\alpha}_i={\varphi}_{ij}$ thus obtained has no intrinsic meaning because $\alpha$ is far from being a $1$-form while $\varphi$ is far from being a $2$-form.  \\

\noindent
{\bf REMARK 5.B.1}: The {\it target}  "$y$" could be called " {\it hidden variable} " as it is just used in order to construct  objects over the {\it source} "$x$". As a byproduct, the changes of local coordinates are of the form $\bar{x}=\varphi (x), \bar{y}=\psi(y)$ but the second one does not appear through the implicit summations over the target because the first order transition rules are:  \\
\[   {\bar{y}}^l_j\frac{\partial {\varphi}^j}{\partial x^i}(x)=\frac{\partial {\psi}^l}{\partial y^k}(y)y^k_i  \Rightarrow {\bar{f}}^l_j(\varphi (x))\frac{\partial {\varphi}^j}{\partial x^i}(x)=\frac{\partial {\psi}^l}{\partial y^k}(f(x))f^k_i (x)   \]
It follows therefore that $A\in T^*\otimes T$ indeed and is thus a well defined object over the source.  \\

\noindent
{\bf LEMMA 5.B.2}: The short exact $\delta$-sequence $ 0 \longrightarrow S_2T^* \stackrel{\delta}{\longrightarrow} T^*\otimes T^* \stackrel{\delta}{\longrightarrow} {\wedge}^2T^* \longrightarrow 0 $ admits a canonical splitting, that is a splitting coherent with the tensor nature of the vector bundles involved.  \\

\noindent
{\it Proof}: The splitting of the above sequence is obtained by setting $({\tau}_{i,j})\in T^*\otimes T^* \rightarrow (\frac{1}{2}({\tau}_{i,j} + {\tau}_{j,i}))\in S_2T^*$ in such a way that $({\tau}_{i,j}={\tau}_{j,i}={\tau}_{ij})\in S_2T^* \Rightarrow \frac{1}{2}({\tau}_{ij}+{\tau}_{ji})={\tau}_{ij}$. \\
Similarly, $({\varphi}_{ij}= - {\varphi}_{ji})\in {\wedge}^2T^*  \rightarrow (\frac{1}{2}{\varphi}_{ij})\in T^*\otimes T^*$ and 
$ (\frac{1}{2}{\varphi}_{ij} - \frac{1}{2}{\varphi}_{ji})=({\varphi}_{ij}) \in {\wedge}^2T^*$.  \\ 
\hspace*{12cm}   Q.E.D.  \\

 We shall revisit the previous results by showing that, {\it in fact}, all the maps and splittings existing for the Killing operator are coming from maps and splittings existing for the conformal Killing operator, {\it though surprising it may look like}. As these results are based on a systematic use of the Spencer $\delta$-map, they are neither known nor acknowledged. \\
 
We now recall the commutative diagrams allowing to define the (analogue) of the first Janet bundle and their dimensions when $n=4$:  \\
 
 \noindent
           \[ \small { \begin{array}{rccccccccl}
   & 0 &  & 0  & &  0  &  &   &       \\
   & \downarrow &  &  \downarrow & & \downarrow & & & &   \\
0 \rightarrow & g_3 & \rightarrow &S_3T^*\otimes T& \rightarrow &S_2T^ *\otimes F_0 &\rightarrow & F_1 & \rightarrow 0 \\
 & \downarrow  &  &  \downarrow & & \downarrow & &  &   \\
0 \rightarrow & T^*\otimes g_2 & \rightarrow &T^*\otimes S_2T^*\otimes T& \rightarrow &T^*\otimes T^ *\otimes F_0 &\rightarrow & 0 & &  \\
 & \downarrow  &  &  \downarrow & & \downarrow & & & &  \\
0 \rightarrow &{\wedge}^2T^*\otimes g_1 & \rightarrow &\underline{{\wedge}^2T^*\otimes T^*\otimes T}& \rightarrow &{\wedge}^2T^*\otimes F_0 &\rightarrow &0&& \\
& \downarrow  &  &  \downarrow & & \downarrow & &  & &  \\
0 \rightarrow &{\wedge}^3 T^*\otimes T & = &{\wedge}^3T^*\otimes T& \rightarrow & 0 & &  &  &    \\
  & \downarrow  &  &  \downarrow & &  & & & &  \\
 & 0  &  & 0  & &    &  &   &   &   
\end{array} } \]   \\

\noindent
           \[ \small { \begin{array}{rccccccccl}
   &  &  & 0  & &  0  &  &   &  &     \\
   &  &  &  \downarrow & & \downarrow & & & &   \\
  & 0 &\rightarrow &80& \rightarrow & 160 &\rightarrow & 20 & \rightarrow 0 \\
 &   &  &  \downarrow & & \downarrow & & & &  \\
   & 0 & \rightarrow & 160 & \rightarrow & 160 &\rightarrow & 0 &  &  \\
 &  &  &  \downarrow & & \downarrow & & & &   \\
 0 \rightarrow &36 & \rightarrow &96 & \rightarrow &60&\rightarrow &0&& \\
& \downarrow  &  &  \downarrow & &\downarrow & &  & &  \\ 
0 \rightarrow &16& = &16& \rightarrow &0 & & & &  \\
 & \downarrow  &  &  \downarrow & &  & & & &  \\
 & 0  &  & 0  & &    &  &   &   &   
\end{array} } \]   \\

\noindent
{\bf PROPOSITION 5.B.3}: Recalling that we have $F_1=H^2(g_1)=Z^2(g_1)$ in the Killing case and ${\hat{F}}_1=H^2({\hat{g}}_1)\neq Z^2({\hat{g}}_1)$ in the conformal Killing case, we have the unusual commutative diagram:  \\ 

 \[   \begin{array}{cccccccc}
   & 0 & & 0  &   &  0  &  &  0   \\
   &  \downarrow & & \downarrow  &   &  \downarrow  &  &  \downarrow  \\
 0\rightarrow & Z^2(g_1) & \rightarrow & Z^2({\hat{g}}_1) & \subset &  Z^2(T^*\otimes T) &  \rightarrow  & S_2T^*  \\
  & \downarrow &  &  \downarrow  &   &  \downarrow  &  & \hspace{3mm} \downarrow  \delta  \\
 0 \rightarrow & {\wedge}^2T^* \otimes g_1 & \rightarrow  & {\wedge}^2T^*\otimes {\hat{g}}_1  &  \subset  &  {\wedge}^2T^*\otimes T^*\otimes T  &  
 \rightarrow &  T^*\otimes T^*   \\
 & \hspace{3mm} \downarrow \delta &  & \hspace{3mm} \downarrow  \delta  &   & \hspace{3mm}  \downarrow  \delta  &  &  \hspace{3mm}  
\downarrow  \delta   \\
 0 \rightarrow &  {\wedge}^3T^*\otimes T &  =  &  {\wedge}^3T^*\otimes T  & =  &  {\wedge}^3T^*\otimes T    & \rightarrow  &  {\wedge}^2T^*  \\
  &   \downarrow &  &   \downarrow  &   &  \downarrow  &  &  \downarrow  \\
    &  0  &  &  0  &   &  0  &  &  0  
  \end{array}  \]
 
 \noindent
{\it Proof}: First of all, we must point out that the surjectivity of the bottom $\delta$ in the central column is well known from the exactness of the $\delta$-sequence for $S_3T^*$ and thus also after tensoring by $T$. However, the surjectivity of the bottom $\delta$ in the left column is {\it not evident at all} as it comes from a delicate circular chase in the preceding diagram, using the fact that the Riemann and Weyl operators are second order operators. Then, setting ${\varphi}_{ij}={\rho}^r_{r,ij}= - {\varphi}_{ji}$ and ${\rho}_{ij}={\rho}^r_{i,rj}\neq {\rho}_{ji}$, we may define the right central horizontal map by ${\rho}^k_{l,ij} \rightarrow {\rho}_{ij} - \frac{1}{2}{\varphi}_{ij}$ and the right bottom horizontal map by $\omega\otimes \xi \rightarrow - i(\xi)\omega$ by introducing the interior product $i( )$. We obtain at once:   \\
\[-({\rho}^r_{r,ij}+{\rho}^r_{i,jr}+{\rho}^r_{j,ri})= - {\varphi}_{ij}+ {\rho}_{ij} - {\rho}_{ji} = ({\rho}_{ij} - \frac{1}{2}{\varphi}_{ij}) - 
({\rho}_{ji} - \frac{1}{2}{\varphi}_{ji})\] 
and the right bottom diagram is commutative, clearly inducing the upper map. If we restrict to the Killing symbol, then ${\varphi}_{ij}=0$ and we obtain ${\rho}_{ij} - {\rho}_{ji}=0 \Rightarrow ({\rho}_{ij}={\rho}_{ji})\in S_2T^*$, that is the classical contraction allowing to obtain the Ricci tensor from the Riemann tensor but {\it there is no way to go backwards with a canonical lift}. A similar comment may be done for the conformal Killing symbol and the $\frac{1}{2}$ coefficient. \\
\hspace*{12cm}    Q.E.D.   \\

Using the previous diagram allowing to define both $F_1=H^2(g_1)=Z^2(g_1))$ if we use $\omega$ or ${\hat{F}}_1=H^2({\hat{g}}_1)= Z^2({\hat{g}}_1) / \delta (T^*\otimes {\hat{g}}_2)$ if we use $\hat{\omega}$ while taking into account that $dim({\hat{g}}_1/g_1)=1$ and ${\hat{g}}_2\simeq T^*$, we obtain the crucial theorem {\it which is in fact only depending on} $\omega$:  \\

\noindent
{\bf THEOREM 5.B.4}: We have the commutative and exact "{\it fundamental diagram II} ":  \\
 \[ \begin{array}{rcccccccl}
 & & & & & & & 0 & \\
 & & & & & & & \downarrow & \\
  & & & & & 0& & S_2T^* &  \\
  & & & & & \downarrow & & \downarrow  &  \\
   & & & 0 &\longrightarrow & Z^2(g_1) & \longrightarrow & H^2(g_1)  & \longrightarrow 0  \\
   & & & \downarrow & & \downarrow & &  \downarrow  &  \\
   & 0 &\longrightarrow & T^*\otimes {\hat{g}}_2 & \stackrel{\delta}{\longrightarrow} & Z^2({\hat{g}}_1) & \longrightarrow & H^2({\hat{g}}_1) & \longrightarrow 0  \\
    & & & \downarrow & & \downarrow & & \downarrow     &   \\
 0 \longrightarrow & S_2T^* & \stackrel{\delta}{\longrightarrow}& T^*\otimes T^* &\stackrel{\delta}{\longrightarrow} & {\wedge}^2T^* & \longrightarrow & 0 &   \\
   & & & \downarrow &  & \downarrow & & &  \\
   & & & 0 & & 0 & & &  \\
  & & & & &  & & &     
   \end{array}  \]
\[   \vspace{2mm}\]
The following theorem will provide {\it all} the classical formulas of both Riemannian and conformal geometry in a totally unusual framework {\it not depending on any conformal factor}:    \\
 
\noindent
{\bf THEOREM 5.B.5}: All the short exact sequences of the preceding diagram split in a canonical way, that is in a way compatible with the underlying tensorial properties of the vector bundles involved. With more details:   \\
\[  \begin{array}{ccccccl}
T^*\otimes T^* \simeq S_2T^* \oplus {\wedge}^2 T^* &  \Rightarrow & Z^2({\hat{g}}_1) & \simeq & Z^2(g_1) & + &  \delta (T^*\otimes  {\hat{g}}_2) \simeq Z^2(g_1)  \oplus   {\wedge}^2T^*  \\
 & \Rightarrow  & H^2(g_1)  & \simeq & H^2({\hat{g}}_1) & \oplus & S_2T^* \\
  & \Rightarrow &   F_1 & \simeq & {\hat{F}}_1 &  \oplus &  S_2T^*
 \end{array}  \]
 
\noindent
{\it Proof}: First of all, we recall that:   \\
\[  g_1 =\{{\xi}^k_i \in T^*\otimes T\mid {\omega}_{rj}{\xi}^r_i+{\omega}_{ir}{\xi}^r_j=0 \}\subset {\hat{g}}_1=\{{\xi}^k_i\in T^*\otimes T \mid {\omega}_{rj}{\xi}^r_i+{\omega}_{ir}{\xi}^r_j - \frac{2}{n}{\omega}_{ij}{\xi}^r_r=0\}  \]
\[  \Rightarrow   \hspace{1cm} 0=g_2 \subset {\hat{g}}_2= \{ {\xi}^k_{ij}\in S_2T^*\otimes T \mid  n{\xi}^k_{ij}={\delta}^k_i{\xi}^r_{rj} +{\delta}^k_j{\xi}^r_{ri} - {\omega}_{ij}{\omega}^{ks}{\xi}^r_{rs} \}     \]
Now, if $({\tau}^k_{li,j})\in T^*\otimes {\hat{g}}_2$, then we have:  \\
\[  n {\tau}^k_{li,j}={\delta}^k_l{\tau}^r_{ri,j} + {\delta}^k_i {\tau}^r_{rl,j}-{\omega}_{li}{\omega}^{ks}{\tau}^r_{rs,j}  \] 
 and we may set ${\tau}^r_{ri,j}={\tau}_{i,j}\neq {\tau}_{j,i}$ with $({\tau}_{i,j})\in T^*\otimes T$ and such a formula does not depend on any conformal factor. Taking into account Proposition 4.B.5, we have:  \\
\[  \delta ({\tau}^k_{li,j})=({\tau}^k_{li,j} - {\tau}^k_{lj,i})=({\rho}^k_{l,ij}) \in B^2( {\hat{g}}_1)\subset Z^2({\hat{g}}_1)  \]
 with:   \\
 \[  Z^2({\hat{g}}_1)= \{ ({\rho}^k_{l,ij})\in {\wedge}^2T^*\otimes {\hat{g}}_1)\mid \delta ({\rho}^k_{l,ij})=0 \}\Rightarrow {\varphi}_{ij}={\rho}^r_{r,ij}\neq 0 \]
 \[   \delta ({\rho}^k_{l,ji}) = ({\cal{C}}_{(l,i,j)}{\rho}^k_{l,ij}={\rho}^k_{l,ij} + {\rho}^k_{i,jl} + {\rho}^k_{j,li}) \in {\wedge}^3T^*\otimes T  \]
 \noindent

\noindent
$\bullet$ The splitting of the central vertical column is obtained from a lift of the epimorphism $Z^2({\hat{g}}_1) \rightarrow {\wedge}^2T^* \rightarrow 0$ which is obtained by lifting $({\varphi}_{ij})\in {\wedge}^2T^*$ to $(\frac{1}{2}{\varphi}_{ij})\in T^*\otimes T^*$, setting ${\tau}^r_{ri,j}=\frac{1}{2} {\varphi}_{ij}$ and applying $\delta$ to obtain $({\tau}^r_{ri,j}-{\tau}^r_{rj,i}=\frac{1}{2} {\varphi}_{ij} - \frac{1}{2}{\varphi}_{ji}={\varphi}_{ij})\in B^2({\hat{g}}_1)\subset Z^2({\hat{g}}_1)$.\\

\noindent
$\bullet$ Now, let us define $({\rho}_{i,j}={\rho}^r_{i,rj}\neq {\rho}_{j,i})\in T^*\otimes T^*$. Hence, elements of $Z^2(g_1)$ are such that:  \\
\[  {\varphi}_{ij}={\rho}^r_{r,ij}=0, \,\,\, {\varphi}_{ij}-{\rho}_{i,j}+ {\rho}_{j,i}=0 \Rightarrow ({\rho}_{ij}={\rho}_{i,j}={\rho}_{j,i}={\rho}_{ji}) \in S_2T^*\]
while elements of $Z^2({\hat{g}}_1)$ are such that:  \\
\[ ({\rho}^r_{r,ij}={\varphi}_{ij}={\rho}_{i,j}- {\rho}_{j,i}={\tau}_{i,j}-{\tau}_{j,i}\neq 0 ) \in {\wedge}^2T^*  \]
Accordingly, $({\rho}_{i,j}- \frac{1}{2}{\varphi}_{ij}={\rho}_{j,i}- \frac{1}{2}{\varphi}_{ji}) \in S_2T^*$. More generally, we may consider ${\rho}^k_{l,ij}- ({\tau}^k_{li,j}-{\tau}^k_{lj,i})$ with ${\tau}^r_{ri,j}=\frac{1}{2}{\varphi}_{ij}$. Such an element is killed by $\delta$ and thus belongs to $Z^2({\hat{g}}_1)$ because each member of the difference is killed by $\delta$. However, we have 
${\rho}^r_{r,ij}-({\tau}^r_{ri,j}-{\tau}^r_{rj,i})={\varphi}_{ij} - {\varphi}_{ij}=0$ and the element does belong indeed to $Z^2(g_1)$, providing a lift $Z^2({\hat{g}}_1) \rightarrow Z^2(g_1) \rightarrow 0$.  \\

\noindent
$\bullet$ Of course, {\it the most important result is to split the right column}. As this will be the hard step, we first need to describe the monomorphism $0 \rightarrow S_2T^* \rightarrow H^2(g_1)$ which is in fact produced by a north-east diagonal snake type chase. Let us choose $({\tau}_{ij}={\tau}_{i,j}={\tau}_{j,i}={\tau}_{ji})\in S_2T^* \subset T^* \otimes T^*$. Then, we may find $({\tau}^k_{li,j})\in T^* \otimes {\hat{g}}_2$ by deciding that ${\tau}^r_{ri,j}={\tau}_{i,j}={\tau}_{j,i}={\tau}^r_{rj,i}$ in $Z^2({\hat{g}}_1)$ and apply 
$\delta$ in order to get ${\rho}^k_{l,ij}={\tau}^k_{li,j} - {\tau}^k_{k,lj,i}$ such that ${\rho}^r_{r,ij}={\varphi}_{ij}=0$ and thus $({\rho}^k_{l,ij})  \in Z^2(g_1)=H^2(g_1)$. We obtain:  \\
\[ \begin{array}{rcl}
n{\rho}^k_{l,ij} & = &  {\delta}^k_l{\tau}^r_{ri,j}-{\delta}^k_l{\tau}^r_{rj,i}+{\delta}^k_i {\tau}^r_{rl,j}-{\delta}^k_j{\tau}^r_{rli}-{\omega }^{ks}({\omega}_{li}{\tau}^r_{rs,j} - {\omega}_{lj}{\tau}^r_{rs,i })   \\
  & =  & ({\delta}^k_i{\tau}_{lj} - {\delta}^k_j{\tau}_{li}) -{\omega}^{ks}({\omega}_{li}{\tau}_{sj} - {\omega}_{lj}{\tau}_{si})                          \\
\end{array}  \]
Contracting in $k$ and $i$ while setting simply $tr(\tau) ={\omega}^{ij}{\tau}_{ij}, tr(\rho)={\omega}^{ij}{\rho}_{ij}$, we get:  \\
\[  n {\rho}_{ij}=n{\tau}_{ij}-{\tau}_{ij}-{\tau}_{ij}+{\omega}_{ij} tr(\tau)=(n-2){\tau}_{ij}+{\omega}_{ij}tr(\tau)=n{\rho}_{ji} \Rightarrow ntr(\rho)=2(n-1)tr(\tau) \]              
Substituting, we finally obtain ${\tau}_{ij}=\frac{n}{(n-2)}{\rho}_{ij} - \frac{n}{2(n-1)(n-2)}{\omega}_{ij}tr(\rho)$ and thus the tricky formula:  \\
\[  {\rho}^k_{l,ij}=\frac{1}{(n-2)}(({\delta}^k_i{\rho}_{lj} - {\delta}^k_j{\rho}_{li}) - {\omega}^{ks}({\omega}_{li}{\rho}_{sj}-{\omega}_{lj}{\rho}_{si})) - \frac{1}{(n-1)(n-2)}({\delta}^k_i{\omega}_{lj}-{\delta}^k_j{\omega}_{li})tr(\rho)  \]
Contracting in $k$ and $i$, we check that ${\rho}_{ij}={\rho}_{ij}$ indeed, obtaining therefore the desired canonical lift $H^2(g_1) \rightarrow S_2T^* \rightarrow 0: {\rho}^k_{i,lj} \rightarrow  {\rho}^r_{i,rj}={\rho}_{ij}$. Finally, using again Proposition 3.4, the epimorphism $H^2(g_1) \rightarrow H^2({\hat{g}}_1) \rightarrow 0$ is just described by the formula:  \\
\[ \small \begin{array}{c}{\sigma}^k_{l,ij}={\rho}^k_{l,ij}-\frac{1}{(n-2)}(({\delta}^k_i{\rho}_{lj} - {\delta}^k_j{\rho}_{li}) - {\omega}^{ks}({\omega}_{li}{\rho}_{sj}-{\omega}_{lj}{\rho}_{si})) + \frac{1}{(n-1)(n-2)}({\delta}^k_i{\omega}_{lj}-{\delta}^k_j{\omega}_{li})tr(\rho) \end{array}   \]
which is just the way to define the Weyl tensor. We notice that ${\sigma}^r_{r,ij}={\rho}^r_{r,ij}=0$ and ${\sigma}^r_{i,rj}=0$ by using indices or a circular chase showing that $Z^2({\hat{g}}_1)=Z^2(g_1) + \delta (T^*\otimes {\hat{g}}_2)$. This purely algebraic result only depends on the metric $\omega$ and does not depend on any conformal factor. In actual practice, the lift $H^2(g_1) \rightarrow S_2T^*$ is described by ${\rho}^k_{l,ij}\rightarrow {\rho}^r_{i,rj}={\rho}_{ij}={\rho}_{ji}$ but it is not evident at all that the lift $H^2({\hat{g}}_1) \rightarrow H^2(g_1)$ is described by the strict inclusion ${\sigma}^k_{l,ij} \rightarrow {\rho}^k_{l,ij}={\sigma}^k_{l,ij}$ providing a short exact sequence as in Proposition $3.4$ because ${\rho}_{ij}={\rho}^r_{i,rj}={\sigma}^r_{i,rj}=0$ by composition.\\
 \hspace*{12cm}   Q.E.D.   \\

 \noindent
 {\bf PROPOSITION 5.B.6}: We have the following comutative and exact diagram:  \\
 \[ \begin{array}{rcccccccl}
  &  &  &  &  0  &  &  0  &  \\
  &  &  &  &  \downarrow &  &  \downarrow &  \\
  &  &  0 & \longrightarrow & {\hat{g}}_2 & \longrightarrow & T^* & \longrightarrow & 0   \\
   &  &   \downarrow &  &  \downarrow &  & \parallel  &     \\
 0  &  \longrightarrow &{\tilde{R}}_2 & \longrightarrow & {\hat{R}}_2 & \longrightarrow &  T^*  & \longrightarrow & 0  \\
 &  &   \downarrow &  &  \downarrow &  & \downarrow  &     \\ 
 0  & \longrightarrow & {\tilde{R}}_1 & =  &  {\hat{R}}_1 & \longrightarrow & 0 &  &     \\
   &  & \downarrow &  & \downarrow &  &  &    \\
   &  &  0  &  &  0  &  &    & 
\end{array}   \] 
leading thus to a short exact sequence:  \\
\[ 0    \longrightarrow  T^*\otimes {\tilde{R}}_2  \longrightarrow  T^*\otimes {\hat{R}}_2  \longrightarrow  T^*\otimes T^*  \longrightarrow  0  \]
with a canonical splitting $T^* \otimes T^*\simeq S_2T^* \oplus {\wedge}^2T^*$.  \\

\noindent
{\it Proof}:  According to the definition of the Christoffel symbols $\gamma$ for te metric $\omega$, we have:   \\
 \[    2 {\omega}_{rk} {\gamma}^k_{ij}= {\partial}_i{\omega}_{rj} + {\partial}_j{\omega}_{ri} - {\partial}_r{\omega}_{ij} \,\,\, \Leftrightarrow \,\,\,  {\omega}_{kj}{\gamma}^k_{ir} +  {\omega}_{ik}{\gamma}^k_{jr} -  {\partial}_r{\omega}_{ij}=0  \]
 It follows that $ - \,  \gamma$ ({\it care}) is the unique symmetric $R_1$-connection, that is a map $T\rightarrow R_1$ considered as an element of $T^*\otimes R_1$ projecting onto ${id}_T\in T^*\otimes T$. Accordingly, any ${\chi}_1\in T^*\otimes J_1(T)$ provides $({\chi}^k_{j,i} + {\gamma}^k_{jr}{\chi}^r_{,i})\in T^*\otimes T^*\otimes T$ and thus a true $1$-form $({\alpha}_i= {\chi}^r_{r,i} +  {\gamma}^r_{r,s}{\chi}^s_{,i})\in T^*$. However, such an approach cannot be extended to higher orders and we prefer to consider half of the morphism defining the Killing operator, namely the morphism $J_1(T) \rightarrow S_2T^*: {\xi}_1 \rightarrow \frac{1}{2} L({\xi}_1)\omega$, tensor it by $T^*$ and contract it by ${\omega}^{-1}$ in order to get:  \\
 \[  \frac{1}{2} {\omega}^{st}( {\omega}_{rt}{\chi}^r_{s,i} + {\omega}_{sr}{\chi}^r_{t,i} + {\chi}^r_{,i}{\partial}_r{\omega}_{st}) = {\chi}^r_{r,i} + \frac{1}{2}{\chi}^r_{,i} {\omega}^{st}{\partial}_r{\omega}_{st}= {\alpha}_i  \]     
where we notice that:  \\
\[  2{\gamma}^r_{ri}= {\omega}^{st}{\partial}_r{\omega}_{st}= (1/det(\omega)){\partial}_idet(\omega)  \Rightarrow {\partial}_i{\gamma}^r_{rj}-{\partial}_j{\gamma}^r_{r,i} = 0  \]
Similarly, there is a well defined map $J_2(T) \rightarrow S_2T^*\otimes T:{\xi}_2 \rightarrow L({\xi}_2)\gamma$ that can be tensored by $T^*$ and restricted to $T^*\otimes {\hat{R}}_2$ in order to obtain a well defined map $ T^*\otimes {\hat{R}}_2 \rightarrow  T^*\otimes S_2T^*\otimes T$ that can be contracted to $T^*\otimes T^*$ according to the following local formulas: \\
\[    \begin{array}{rcl}
 {\beta}^k_{lr,s}  & = & {\tau}^k_{lr,s} + {\gamma}^k_{ur}{\tau}^u_{l,s} +{\gamma}^k_{lu}{\tau}^u_{r,s} - {\gamma}^u_{lr}{\tau}^k_{u,s} + {\tau}^u_{,s}{\partial}_u{\gamma}^k_{lr}  \\
 {\beta}^k_{kr,s} & = & {\tau}^k_{kr,s} +{\gamma}^k_{ku}{\tau}^u_{r,s} +{\tau}^u_{,s}{\partial}_u{\gamma}^k_{kr}
 \end{array}   \]
We can "{\it twist}" by $A$ and apply $\delta:T^*\otimes S_2T^*\otimes T \rightarrow {\wedge}^2T^*\otimes T^*\otimes T$ that can be contracted to 
${\wedge}^2T^*$ according to the following local formulas:  \\
\[   {\varphi}^k_{l,ij}= A^r_iA^s_j ({\beta}^k_{lr,s} - {\beta}^k_{ls,r}) \,\,  \Rightarrow \,\, {\varphi}_{ij}={\varphi}^r_{r,ij}=
A^r_iA^s_j ({\beta}^k_{kr,s} -{\beta}^k_{ks,r})        \]
\hspace*{12cm}  Q.E.D.   \\    
      
As ${\varphi} \in {\wedge}^2T^*$ though it comes from the $1$-form ${\chi}_2 \in T^*\otimes {\hat{R}}_2$, we obtain the following crucial theorem ([27-28]):  \\    
     
\noindent
{\bf THEOREM  5.B.7}: The non-linear Spencer sequence for the conformal group of transformations projects onto a part of the Poincar\'{e} sequence for the exterior derivative according to the following commutative and locally exact diagram:  
\[  \begin{array}{rcccccccl}
0 \longrightarrow & \hat{\Gamma} & \stackrel{j_2}{\longrightarrow} & {\hat{{\cal{R}}}}_2 & \stackrel{{\bar{D}}_1}{\longrightarrow} & T^* \otimes  {\hat{R}}_2 & \stackrel{{\bar{D}}_2}{\longrightarrow} &  {\wedge}^2T^*\otimes {\hat{R}}_2   \\
&  &  &  \downarrow & \swarrow &  \downarrow  &  &  \downarrow   \\
&  &  &  T^* & \stackrel{d}{\longrightarrow}  & {\wedge}^2T^* & \stackrel{d}{\longrightarrow} &  {\wedge}^3T^* \\
&  &  &  \alpha &  &  d \alpha=\varphi &  &d\varphi=0   
\end{array}   \]
Accordingly, {\it this purely mathematical result contradicts classical gauge theory}.  \\

\noindent
{\it Proof}: Substituting the previous results in the last formula, we obtain successively:  \\
\[   \begin{array}{rcl}
{\varphi}_{ij}  &  =  &  A^r_iA^s_j({\tau}^k_{kr,s} - {\tau}^k_{ks,r}) +{\gamma}^k_{ku}A^r_iA^s_j({\tau}^u_{r,s} - {\tau}^u_{s,r}) + A^r_iA^s_j({\tau}^u_{,s}{\partial}_u{\gamma}^k_{kr} -  {\tau}^u_{,r}{\partial}_u
{\gamma}^k_{ks} ) \\
    & =  & ({\partial}_i{\chi}^r_{r,j} - {\partial}_j{\chi}^r_{r,i}) + {\gamma}^r_{ru}({\partial}_iA^u_j - {\partial}_jA^u_i) + ({\delta}^r_i + {\chi}^r_{,i}){\chi}^s_{,j} {\partial}_r{\gamma}^k_{ks} - ({\delta}^s_j + {\chi}^s_{,j}){\chi}^r_{,i}{\partial}_s{\gamma}^k_{kr} \\
   & = & ({\partial}_i{\chi}^r_{r,j} - {\partial}_j{\chi}^r_{r,i}) + {\gamma}^r_{rs}({\partial}_i{\chi}^s_{,j} - {\partial}_j{\chi}^s_{,i})  + ({\chi}^s_{,j}{\partial}_i{\gamma}^r_{rs} - {\chi}^s_{,i}{\partial}_j{\gamma}^r_{rs})\\
   & = & {\partial}_i{\alpha}_j - {\partial}_j{\alpha}_i
\end{array}         \]
because ${\partial}_i{\gamma}^r_{rj} - {\partial}_j{\gamma}^r_{ri}=0$. It follows that $d\alpha=\varphi \in {\wedge}^2T^*$ and thus $d\varphi=0$, that is $ {\partial}_i{\varphi}_{jk} + {\partial}_j{\varphi}_{ki} + {\partial}_k{\varphi}_{ij}=0 $, has an intrinsic meaning in ${\wedge}^3T^*$. It 
is important to notice that the corresponding EM Lagrangian is defined on sections of ${\hat{C}}_1$ killed by ${\bar{D}}_2$ but {\it not} on ${\hat{C}}_2$, {\it contrary to gauge theory}. Finally, the south west arrow in the left square is the composition:   \\
\[  f_2 \in {\hat{\cal{R}}}_2 \stackrel{{\bar{D}}_1}{\longrightarrow} {\chi}_2 \in T^* \otimes {\hat{R}}_2 \stackrel{{\pi}^2_1}{\longrightarrow } {\chi}_1 \in T^*\otimes {\hat{R}}_1 \stackrel{(\gamma)}{\longrightarrow} \alpha \in  T^*    \]
Accordingly, though $\alpha$ is a potential for $\varphi$, it can also be considered as a part of the {\it field} but the important fact is that the first set of ({\it linear}) Maxwell equations $d\varphi =0$ is induced by the ({\it nonlinear}) operator ${\bar{D}}_2$. The linearized framework will explain this point.   \\
One of the most important but difficult result of this paper will be the following direct proof of the existence of the right square in the previous diagram.

Supposing for simplicity that $\omega$ is a (locally) constant metric (in fact the Minkowski metric !) and thus $\gamma=0$. When we are considering the conformal group of space-time, it first follows that the jets of order three vanish and the formula $(3^*)$ can be now written:  \\
\[  {\partial}_i{\chi}^k_{lr,j} - {\partial}_j{\chi}^k_{lr,i} 
 -  ( {\chi}^s_{r,i} {\chi}^k_{ls,j} + {\chi}^s_{l,i}{\chi}^k_{rs,j} + {\chi}^s_{lr,i} {\chi}^k_{s,j} - {\chi}^s_{r,j}{\chi}^k_{ls,i} - {\chi}^s_{l,j}{\chi}^k_{rs,i} - {\chi}^s_{lr,j}{\chi}^k_{s,i}) = 0  \]

Contracting in $k=l=u$ and replacing $r$ by $t$, we obtain the simple formula:  \\
\[  {\partial}_i{\chi}^u_{ut,j} - {\partial}_j{\chi}^u_{ut,i} 
 -   {\chi}^s_{t,i} {\chi}^u_{us,j} +  {\chi}^s_{t,j}{\chi}^u_{us,i} = 0  \]   
Multiplyig by $A^t_k$ the {\it two last terms} and replacing $\chi$ by $\tau$, we get {\it for these terms only}: \\
\[   A^r_i A^s_jA^t_k ( {\tau}^v_{t,s}{\tau}^u_{uv,r} - {\tau}^v_{t,r}{\tau}^u_{uv,s})   \]
Now, denoting by ${\cal{C}}(i,j,k)$ the cyclic sum on the permutation $(i,j,k)\rightarrow (j,k,i) \rightarrow (k,i,j)$ and proceeding in this way on the last formula, we obtain easily:  \\
\[   {\cal{C}}(i,j,k)  A^r_iA^s_jA^t_k ( {\tau}^v_{t,s} - {\tau}^v_{s,t}){\tau}^u_{uv,r}   \]
or, equivalently:  \\
\[       A^r_iA^s_jA^t_k {\cal{C}}(r,s,t) ( {\tau}^v_{t,s} - {\tau}^v_{s,t}){\tau}^u_{uv,r} =   A^r_iA^s_jA^t_k {\cal{C}}(r,s,t) ( {\tau}^v_{s,r} - {\tau}^v_{r,s}){\tau}^u_{uv,t}   \]

Let us now similarly consider only the {\it two first terms}. After multiplication by $A^t_k$ and integration by part,  we get for the first:  \\
\[  A^t_k({\partial}_i(A^s_j{\tau}ž_{ut,s})={\partial}_i(A^s_jA^t_k{\tau}^u_{ut,s}) - A^s_j{\tau}^u_{ut,s}{\partial}_iA^t_k \]
Applying the same procedure to the second term and considering the sum ${\cal{C}}(i,j,k)$ while rearranging the six terms of the summation two by two, we obtain:   \\
\[ {\cal{C}}(i,j,k) ({\partial}_k ( A^t_jA^s_i {\tau}^u_{ut,s} - A^t_iA^r_j {\tau}^u_{ut,r})+ A^t_k{\tau}^u_{ur,t}{\partial}_jA^r_i- A^t_k{\tau}^u_{ut,r}{\partial}_iA^r_j) \]
Exchanging the dumb indices between themselves, we finally obtain:  \\
\[  {\cal{C}}(i,j,k) ({\partial}_k(A^r_iA^s_j({\tau}^u_{us,r}- {\tau}^u_{ur,s}) + A^t_k{\tau}^u_{ur,t}({\partial}_iA^r_j - {\partial}_jA^r_i))  \]
that is to say, taking into account the equations $(1^*)$:  \\
\[ - {\cal{C}}(i,j,k) ({\partial}_k{\varphi}_{ij}) + {\cal{C}}(i,j,k) (A^r_iA^s_jA^t_k( {\tau}^v_{r,s} -{\tau}^v_{s,r}) {\tau}^u_{uv,t})  \]
or, equivalently:   \\
\[ - {\cal{C}}(i,j,k) ({\partial}_k{\varphi}_{ij}) + A^r_iA^s_jA^t_k {\cal{C}}(r,s,t)( {\tau}^v_{r,s} -{\tau}^v_{s,r}) {\tau}^u_{uv,t})  \]
Collecting all the results, we are only left, up to sign, with $ {\cal{C}}(i,j,k) ({\partial}_k{\varphi}_{ij}) = 0$ as we wished.  \\
\hspace*{12cm}     Q.E.D.     \\

\noindent
{\bf COROLLARY  5.B.8}: The linear Spencer sequence for the conformal group of transformations projects onto a part of the Poincar\'{e} sequence for the exterior derivative according to the following commutative and locally exact diagram:  
\[  \begin{array}{rcccccccl}
0 \longrightarrow & \hat{\Theta} & \stackrel{j_2}{\longrightarrow} & {\hat{R}}_2 & \stackrel{D_1}{\longrightarrow} & T^* \otimes  {\hat{R}}_2 & \stackrel{D_2}{\longrightarrow} &  {\wedge}^2T^*\otimes {\hat{R}}_2   \\
&  &  &  \downarrow & \swarrow &  \downarrow  &  &  \downarrow   \\
&  &  &  T^* & \stackrel{d}{\longrightarrow}  & {\wedge}^2T^* & \stackrel{d}{\longrightarrow} &  {\wedge}^3T^* \\
&  &  &  A &  &  d A=F &  &d F=0   
\end{array}   \]
Accordingly, {\it this purely mathematical result also contradicts classical gauge theory} because it proves that EM only depends on the structure of the conformal group of space-time but not on $U(1)$.  \\

\noindent
{\it Proof}: Considering $\omega$ and $\gamma$ as geometric objects, we obtain at once the formulas:\\
\[{\bar{\omega}}_{ij}=e^{2a(x)}{\omega}_{ij} \hspace{5mm}  \Rightarrow  \hspace{5mm}  {\bar{\gamma}}^r_{ri}={\gamma}^r_{ri} + 
{\partial}_ia  \]
Though looking like the key formula ($69$)in ([54], p 286), this transformation is quite different because the sign is not coherent and the second object has nothing to do with a $1$-form. Moreover, if we use $n=2$ and set ${\cal{L}}(\xi)\omega=2A\omega$ for the standard euclidean metric, we should have $({\partial}_{11}+{\partial}_{22})A=0$, contrary to the assumption that $A$ is arbitrary which is {\it only} agreeing with the following jet formulas improving the ones already provided in ([38],[42],[47]) in order to point out the systematic use of the Spencer operator:  \\
\[ L({\xi}_1)\omega=2A\omega\hspace{3mm} \Rightarrow \hspace{3mm} ({\xi}^r_r+{\gamma}^r_{ri}{\xi}^i)=nA, \hspace{3mm}(L({\xi}_2)\gamma)^r_{ri}=nA_i , \hspace{3mm} \forall {\xi}_2\in {\hat{R}}_2\]
Now, if we make a change of coordinates $\bar{x}=\varphi (x)$ on a function $a\in {\wedge}^0T^*$, we get:\\
\[  \bar{a}(\varphi(x))=a(x) \hspace{4mm} \Rightarrow \hspace{4mm} \frac{\partial\bar{a}}{\partial {\bar{x}}^j}\frac{\partial {\varphi}^j}{\partial x^i}=\frac{\partial a}{\partial x^i}  \]
We obtain therefore an isomorphism $J_1({\wedge}^0T^*)\simeq {\wedge}^0T^*{\times}_XT^*$, a result leading to the following commutative diagram: \\
\[  \begin{array}{rcccccl}
0 \longrightarrow & R_2 & \longrightarrow & {\hat{R}}_2 & \longrightarrow  & J_1({\wedge}^0T^*) & \longrightarrow 0  \\
  & \hspace{3mm}\downarrow D  &  & \hspace{3mm}  \downarrow D  & &\hspace{3mm} \downarrow D  &   \\
0 \longrightarrow & T^* \otimes R_1 & \longrightarrow & T^* \otimes {\hat{R}}_1 & \longrightarrow  & T^* & \longrightarrow 0     
\end{array}    \]
where the rows are exact by counting the dimensions. The operator $D:(A,A_i) \longrightarrow ({\partial}_iA-A_i)$ on the right is induced by the central Spencer operator, a result that could not have been even imagined by Weyl and followers. This result provides a good transition towards the conformal origin of electromagnetism. \\
As the nonlinear aspect has been already presented, we restrict our study to the linear framework. A first problem to solve is to construct vector bundles from the components of the image of $D_1$. Using the corresponding capital letter for denoting the linearization, let us introduce:   \\
\[ (B^k_{l,i}=X^k_{l,i}+{\gamma}^k_{ls}X^s_{,i}) \in T^*\otimes T^*\otimes T \Rightarrow  (B^r_{r,i}=B_i)\in T^*\]
\[  (B^k_{lj,i}=X^k_{lj,i}+{\gamma}^k_{sj}X^s_{l,i}+{\gamma}^k_{ls}X^s_{j,i}-{\gamma}^s_{lj}X^k_{s,i}+X^r_{,i}{\partial}_r{\gamma}^k_{lj}) \in T^*\otimes S_2T^*\otimes T \Rightarrow (B^r_{ri,j}-B^r_{rj,i}=F_{ij})\in {\wedge}^2T^*  \]
We obtain from the relations ${\partial}_i{\gamma}^r_{rj}={\partial}_j{\gamma}^r_{ri}$ and the previous results:  \\
\[ \begin{array}{rcl}
F_{ij}=B^r_{ri,j}-B^r_{rj,i} & = & X^r_{ri,j}-X^r_{rj,i}+{\gamma}^r_{rs}X^s_{i,j}-{\gamma}^r_{rs}X^s_{j,i}+X^r_{,j}{\partial}_r{\gamma}^s_{si}-X^r_{,i}{\partial}_r{\gamma}^s_{sj}  \\
  &  =  & {\partial}_iX^r_{r,j}-{\partial}_jX^r_{r,i}+{\gamma}^r_{rs}(X^s_{i,j}-X^s_{j,i})+X^r_{,j}{\partial}_i{\gamma}^s_{sr}-X^r_{,i}{\partial}_j{\gamma}^s_{sr} \\
    &  =  & {\partial}_i(X^r_{r,j}+{\gamma}^r_{rs}X^s_{,j})-{\partial}_j(X^r_{r,i}+{\gamma}^r_{rs}X^s_{s,i})  \\
      &  =  &  {\partial}_iB_j-{\partial}_jB_i
      \end{array}   \]
Now, using the contracted formula ${\xi}^r_{ri}+ {\gamma}^r_{rs}{\xi}^s_i + {\xi}^s{\partial}_s{\gamma}^r_{ri}=nA_i$ from section $A$, we obtain:  \\
\[ \begin{array}{rcl}
 B_i & =  & ({\partial}_i{\xi}^r_r - {\xi}^r_{ri})+{\gamma}^r_{rs}({\partial}_i{\xi}^s - {\xi}^s_i)\\
    &  =  &{\partial}_i{\xi}^r_r + {\gamma}^r_{rs}{\partial}_i{\xi}^s+
{\xi}^s {\partial}_s{\gamma}^r_{ri} - nA_i \\
  &  =  &{\partial}_i({\xi}^r_r + {\gamma}^r_{rs}{\xi}^s) - nA_i \\
    &  =  &n ({\partial}_iA - A_i) 
\end{array}   \]  
and we finally get $F_{ij}=n({\partial}_jA_i-{\partial}_iA_j)$ {\it which is no longer depending on} $A$, a result fully solving the dream of Weyl. Of course, when $n=4$ and $\omega$ is the Minkowski metric, then we have $\gamma=0$ in actual practice and the previous formulas become particularly simple. \\     

It follows that $d B=F \Leftrightarrow - ndA=F$ in ${\wedge}^2T^*$ and thus $d F=0$, that is $ {\partial}_i{F}_{jk} + {\partial}_j{F}_{ki} + {\partial}_k{F}_{ij}=0 $, has an intrinsic meaning in ${\wedge}^3T^*$. It is finally important to notice that the usual EM Lagrangian is defined on sections of ${\hat{C}}_1$ killed by $D_2$ but {\it not} on ${\hat{C}}_2$. Finally, the south west arrow in the left square is the composition:   \\
\[  {\xi}_2 \in {\hat{R}}_2 \stackrel{D_1}{\longrightarrow} X_2 \in T^* \otimes {\hat{R}}_2 \stackrel{{\pi}^2_1}{\longrightarrow } X_1 \in T^*\otimes {\hat{R}}_1 \stackrel{(\gamma)}{\longrightarrow} (B_i) \in  T^*  \Leftrightarrow {\xi}_2 \in {\hat{R}}_2 \rightarrow (nA_i)\in T^*  \]
Accordingly, though $A$ and $B$ are potentials for $F$, then $B$ can also be considered as a part of the {\it field} but the important fact is that the first set of ({\it linear}) Maxwell equations $d F=0$ is induced by the ({\it linear}) operator $D_2$ because we are only dealing with involutive and thus formally integrable operators, a {\it fact} justifying the commutativity of the square on the left of the diagram.   \\
\hspace*{12cm}     Q.E.D.     \\

\noindent
{\bf REMARK 5.B.9}: Taking the determinant of each term of the non-linear second order PD equations defining 
$\hat{\Gamma}$, we obtain successively:  \\
\[    det(\omega)(det(f^k_i(x)))^2  = e^{2na(x)} det(\omega) \Rightarrow det(f^k_i(x))= e^{na(x)} \]  
in such a way that we may define $b(f(x))= a(x) \Leftrightarrow b(y)=a(g(y))$ and set $\Theta(y)= e^{- b(y)}>0$ over the target when caring only about the connected component $0[ \rightarrow 1 \rightarrow \infty$ of the dilatation group. The problem is thus to change at the same time the numerical value of the section and /or the nature of the geometric object cosifered, passing therefore from a (metric) tensor to a (metric) tensor density, exactly what also happens with the contact structure when it was necessary to pass from a $1$-form to a $1$-form density ([27-28],[39],[45]). In a more specific way, the idea has been to consider successively the two non-linear systems of finite defining Lie equations:   \\
\[  {\omega}_{kl}(y)y^k_i y^l_j ={\omega}_{ij}(x) \,\,\, \rightarrow \,\,\, {\hat{\omega}}_{kl}y^k_i y^l_j (det(y^k_i))^{\frac{-2}{n}}={\hat{\omega}}_{ij}(x) \]
 Now, with $\gamma=0$ we have ${\chi}^r_{r,i} = g^s_k ( {\partial}_i f^k_s - A^r_if^k_{rs})$ and:   \\
 \[    g^s_k{\partial}_i f^k_s= (1/det(f^k_i)){\partial}_i det(f^k_i) =n{\partial}_ia , \,\,\,   g^s_k f^k_{rs}= na_r(x)  \]
 Finally, we have the jet compositions and contractions:  \\
\[   g^r_kf^k_i={\delta}^r_i \Rightarrow g^r_kf^k_{ij}= - g^r_{kl}f^k_if^l_j \,\, \Rightarrow \,\,  n\, a_i(x)=g^s_kf^k_{is}= - f^k_if^l_r g^r_{kl}= 
- n\, f^k_i(x)b_k(f(x))  \] 
It follows that ${\alpha}_i= n({\partial}_i a(x) - A^r_ia_r(x))$ but we may also set $a_i(x)=f^k_i(x)b_k(f(x))$ in order to obtain ${\alpha}_i=  n ( \frac{\partial b}{\partial y^k} - b_k) {\partial}_if^k$ as a way to pass from source to target (Compare to [28]). We have:   \\

\noindent
{\bf PROPOSITION  5.B.10}: EM does not depend on the choice between source and target.  \\

\noindent
{\it Proof}: Replacing the groupoid by its inverse in each formula, we may introduce:  \\
\[ \alpha = {\alpha}_i(x)dx^i, \,\,\, {\alpha}_i=n({\partial}_ia - A^r_ia_r) \,\,\,  \Leftrightarrow \,\,\, 
\beta ={\beta}_k(y)dy^k, \,\,\,{\beta}_k= n(\frac{\partial b}{\partial y^k} - b_k)  \]
and compare:  
\[x\stackrel{a}{\longrightarrow} (\alpha, \varphi)  \,\,\,  \Leftrightarrow  \,\,\,  y \stackrel{b}{\longrightarrow} (\beta, \psi)  \]
while setting ${\psi}_{kl}=\frac{\partial{\beta}_l}{\partial y^k} - \frac{\partial {\beta}_k}{\partial y^l}$. We have successively:  \\
\[  \begin{array}{rcl}
{\varphi}_{ij}= {\partial}_i{\alpha}_j - {Ê\partial}_j{\alpha}_i & = & - n({\partial}_i(A^s_ja_s)- {\partial}_j(A^r_ia_r))  \\
  &  =  & - n({\partial}_i(b_l{\partial}_jf^l) - {\partial}_j(b_k{\partial}_if^k))  \\
     &  = & - n( \frac{\partial b_l}{\partial y^k} - \frac{\partial b_k}{\partial y^l}) {\partial}_if^k{\partial}_jf^l   \\
     &    =  & (\frac{\partial {\beta}_l}{\partial y^k} - \frac{\partial {\beta}_k}{\partial y^l}) {\partial}_if^k{\partial}_jf^l \\
        &  =  &  {\psi}_{kl} {\partial}_if^k{\partial}_jf^l 
\end{array}  \]
and we notice that $\varphi$ {\it does not depend any longer on} $a$ while $\psi$ {\it does not depend any longer on} $b$. Accordingly, we have the equivalences:  \\
\[  NO \,\,  EM \Leftrightarrow \varphi=0 \,\,\Leftrightarrow \,\, \psi=0 \,\, \Leftrightarrow \,\, \frac{\partial b_l}{\partial y^k} - \frac{\partial b_k}{\partial y^l} =0 \,\, \Leftrightarrow \,\, {\partial}_i(A^r_ja_r) - {\partial}_j(A^r_ia_r)=0  \]
\hspace*{12cm}   Q.E.D.   \\

\noindent
{\bf REMARK 5.B.11}: If we use {\it only} the conformal group, we must use the metric density $\hat{\omega}$ instead of the metric $\omega$. However, if we can define $\hat{\omega}$ from $\omega$ by setting ${\hat{\omega}}_{ij}={\omega}_{ij}/(\mid det(\omega)\mid)^{\frac{1}{n}}$, we cannot recover $\omega$ from $\hat{\omega}$. The way to escape from such a situation is to notice that:  \\
\[   \omega \rightarrow e^{2a(x)}\omega \,\,\,  \Rightarrow \,\,\, {\gamma}^k_{ij} \rightarrow {\gamma}^k_{ij} + {\delta}^k_i{\partial}_ja(x) + {\delta}^k_j {\partial}_ia(x) - {\omega}_{ij}{\omega}^{kr}{\partial}_ra(x) \,\,\, \Rightarrow \,\,\, {\gamma}^r_{ri} \rightarrow {\gamma}^r_{ri}  
+ n{\partial}_i a(x)   \]
a result showing that the conformal symbols ${\hat{g}}_1$ and ${\hat{g}}_2$ do not depend on any conformal factor.  \\

\noindent
{\bf REMARK 5.B.12}: In fact, our purpose is quite different now though it is also based on the combined use of group theory and the Spencer operator. The idea is to notice that the brothers are {\it always} dealing with the same group of rigid motions because the lines, surfaces or media they consider are all supposed to be in the same $3$-dimensional background/surrounding space which is acted on by the group of rigid motions, namely a group with $6$ parameters ($3$ {\it translations} + $3$ {\it rotations}). In 1909 it should have been {\it strictly impossible} for the two brothers to extend their approach to bigger groups, in particular to include the only additional {\it dilatation} of the Weyl group that will provide the virial theorem and, {\it a fortiori}, the {\it elations} of the conformal group considered later on by H.Weyl ([47],[53]). In order to explain the reason for using Lie equations, we provide the explicit form of the $n$ finite elations and their infinitesimal counterpart with $ 1\leq r,s,t \leq n$:\\
\[  y=\frac{x-x^2b}{1-2(bx)+b^2x^2} \, \Rightarrow  \, {\theta}_ s = - \frac{1}{2} x^2 {\delta}^r_s{\partial}_r+{\omega}_{st}x^tx^r{\partial}_r   \, \Rightarrow  \, {\partial}_r{\theta}^r_s=n{\omega}_{st}x^t, \,\, [{\theta}_s,{\theta}_t]=0  \]
where the underlying metric is used for the scalar products $x^2,bx,b^2$ involved. It is easy to check that ${\xi}_2 \in S_2T^*\otimes T$ defined by 
$  {\xi}^k_{ij}(x)= {\lambda}^s(x){\partial}_{ij} {\theta}^k_s(x)$ belongs to ${\hat{g}}_2$ with $A_i={\omega}_{si}{\lambda}^s$. In view of these local formulas, we understand how important it is to use "{\it equations} " rather than "{\it solutions} ".  \\

\noindent
{\bf REMARK 5.B.13}: Setting ${\sigma}_{q-1}={\bar{D}}' {\chi}_q \in {\wedge}^2T^* \otimes J_{q-1}(T)$, we let the reader prove, as an exercise, that the following so-called {\it Bianchi identities} hold ([28], p 221):  \\
\[  D{\sigma}_{q-1}(\xi, \eta, \zeta) + {\cal{C}}(\xi,\eta,\zeta) \{ {\sigma}_{q-1}(\xi,\eta),{\chi}_{q-1}(\zeta)\}=0, \,\,\,\,\, \forall \xi,\eta,\zeta \in T \]
In the nonlinear conformal framework, it follows that the first set of Maxwell equations has only to do with ${\bar{D}}'$ in the nonlinear Spencer sequence and thus nothing to do with the Bianchi identities, contrary to what happens with $U(1)$ in classical gauge theory. Similarly, in the linear conformal framework, the first set of Maxwell equations has only to do with $D_2$ and thus nothing to do with $D_3$ in the linear Spencer sequence. Indeed, the EM potential $A$ is a section of ${\hat{C}}_0$ while the EM field $F$ is a section of ${\hat{C}}_1$ killed by $D_2$. This " {\it shift by one step to the left} " is the most important result of this section and could not be even imagined with any other approach. \\  \\

\noindent
{\bf C) GRAVITATION}  \\

In the subsection B, we proved that the EM field ${\wedge}^2T^*$ could be described by $n(n-1)/2$ components of the bundle $T^*\otimes 
{\hat{g}}_2 $ of $1$-forms with value in the conformal symbol ${\hat{g}}_2$ which is a sub-bundle of the first Spencer bundle for the conformal group  described by the bundle $T^* \otimes {\hat{R}}_2$ of $1$-forms with value in the Lie algebroid ${\hat{R}}_2$, with no relation at all with the second Spencer bundle ${\wedge}^2T^* \otimes {\hat{R}}_2$ that can be identified with the Cartan curvature. Similarly, in this subsection C, which is by far the most difficult of the whole paper beause third order jets are involved, we shall prove that the substitute for the Riemann curvature is only described by $n(n+1)/2$ other linearly independent components of $T^*\otimes {\hat{g}}_2 \subset T^* \otimes {\hat{R}}_2$ in such a way that $n(n-1)/2 + n(n+1)/2 = n^2 = dim (T^*\otimes {\hat{g}}_2)$.  \\

Let us start with a preliminary mathematical comment, independently of what has already been said in the previous subsection B, and explain the main differences existing between the initial part of the Janet sequence for a formally integrable  system $C_0 = R_q\subset J_q(E) = C_0(E)$ with a $2$-acyclic symbol $g_q \subset S_qT^*\otimes E$ such that $g_{q+1}=0$ and the initial part of the corresponding Spencer sequence for the first order involutive system $R_{q+1} \subset J_1(R_q)$ (See [47] for examples). First of all, we recall the following commutative diagram with short exact vertical sequences, only depending on the left lower commutative square:  \\
\[  \begin{array}{rcccccccccl}
 & &  &  &  &  & 0 & & 0  &  & 0  \\
  & &  &  &  & &  \downarrow &  &  \downarrow &  &  \downarrow   \\
 &  & 0 & \longrightarrow  & \Theta & \underset 2{\stackrel{j_q}{\longrightarrow}} & C_0 & \underset 1{\stackrel{D_1}{\longrightarrow}} & C_1 & \underset 1{\stackrel{D_2}{\longrightarrow}} & C_2  \\ 
    &  &  &  & &  &  \downarrow &  & \downarrow & & \downarrow    \\
 &  & 0 &\longrightarrow & E & \stackrel{j_q}{\longrightarrow} & C_0(E) & \underset 1{\stackrel{D_1}{\longrightarrow}} & C_1(E) & \underset 1{\stackrel{D_2}{\longrightarrow}} & C_2(E)  \\ 
  &  &  &  & \parallel &  & \hspace{5mm} \downarrow {\Phi}_0 &  & \hspace{5mm} \downarrow {\Phi}_1 & &   \\
0 & \longrightarrow  & \Theta & \longrightarrow & E & \underset q {\stackrel{{\cal{D}}}{\longrightarrow}} & F_0 & 
\underset 1{\stackrel{{\cal{D}}_1}{\longrightarrow}} & F_1 &  &  \\
 &  &  &  &  &  &  \downarrow &  &  \downarrow &  &  \\
 & &  &  &  &  & 0 & & 0  &  &  
\end{array}   \]
In this diagram, $\Phi = {\Phi}_0$ is defined by the canonical projection $ \Phi:J_q(E) \rightarrow J_q(E)/R_q = F_0$ with kernel $R_q$ and 
$F_1= T^*\otimes J_q(E)/(T^*\otimes R_q + \delta (S_{q+1}T^*\otimes E))$ is induced by ${\Phi}_0$ after {\it only one} (care) prolongation. As $D_1$ is a first order operator because it is induced by the Spencer operator, it is essential to notice that such a result is coming from the fact that ${\cal{D}}_1$ is of order $1$ because $R_q$ is formally integrable and $g_q$ is $2$-acyclic (See [28], p 116,120, 165). This very delicate result cannot be extended to the right with ${\cal{D}}_2: F_1 \rightarrow F_2$ unless $g_q$ is involutive, a situation fulfilled by $j_q$ which is an involutive injective operator. Also the first order operator $ D_1:R_q \rightarrow J_1(R_q)/R_{q+1}= C_1 \simeq T^*\otimes R_q /\delta (g_{q+1})= T^*\otimes R_q$ is trivially involutive because $g_{q+1}=0$ and $C_1\subset C_1(E)$ while $C_2={\wedge}^2T^* \otimes R_q \subset C_2(E)$. Hence, the upper sequence is formally exact, a result that can be extended to the right side ( See [40] for a nice counterexample). From a snake chase in this diagram, it follows that the (local) cohomology at $C_1$ in the upper sequence is the same as the (local) cohomology at $F_0$ in the lower sequence though there is no link at all between $C_1$ and $F_0$ from a purely group theoretical point of view. In the present situation, we have an isomorphism $R_{q+1} \simeq R_q$ and obtain therefore $D_1 {\xi}_q =D {\xi}_{q+1}, \forall {\xi}_q \in R_q$.  \\

For helping the reader, we provide the two long exact sequences allowing to define $C_1$ and $C_2$ in the Spencer sequence while proving the formal exactness of the upper sequence on the jet level if we set $J_r(E)=0, \forall r< 0$ and $J_0(E)=E$ for any vector bundle $E$:  \\
\[ \begin{array}{rccccccc}
   &   &  &  0  &  &  0  &  & 0    \\
   &  &  &  \downarrow &  &  \downarrow &  &  \downarrow   \\
                      &  0 & \rightarrow & S_{r+1}T^*\otimes C_0 &  \rightarrow & S_rT^* \otimes C_1 & \rightarrow & S_{r-1}T^* \otimes C_2  \\ 
                      &  \downarrow  &  &  \downarrow &  &  \downarrow & & \downarrow   \\
0  \rightarrow & R_{q+r+1} & \rightarrow & J_{r+1}(C_0) & \rightarrow & J_r(C_1) & \rightarrow & J_{r-1}(C_2)  \\
                      &  \downarrow &   &  \downarrow  &  & \downarrow  &  &  \downarrow   \\
0  \rightarrow & R_{q+r}& \rightarrow & J_r(C_0) & \rightarrow & J_{r-1}(C_1) &  \rightarrow   & J_{r-2}(C_2)   \\
   &   \downarrow  &  &  \downarrow &  &  \downarrow  &  &     \\
   &  0  &  &  0  &  &  0  &   &  
\end{array}  \]
It just remains to apply inductively the Spencer $\delta$-operator to the various upper symbol sequences obtained by successive prolongations, starting 
from the case $r=0$ already considered. \\
Similarly, if we define $F_2$ in the Janet sequence by the following commutative and exact diagram:   \\
\[  \begin{array}{rcccccccccl}
   &   &  &  0  &  &  0  &  & 0  & & 0 & \\
   &  &  &  \downarrow &  &  \downarrow &  &  \downarrow & &\downarrow & \\
                      &  0 & \rightarrow & S_{q+2}T^*\otimes C_0 &  \rightarrow & S_2T^* \otimes F_0 & \rightarrow & T^* \otimes F_1 & \rightarrow & F_2& \rightarrow 0 \\ 
                      &  \downarrow  &  &  \downarrow &  &  \downarrow & & \downarrow & & \parallel &  \\
0  \rightarrow & R_{q+2} & \rightarrow & J_{q+2}(E) & \rightarrow & J_2(F_0) & \rightarrow & J_1(F_1) & \rightarrow & F_2 & \rightarrow 0  \\
                      &  \downarrow &   &  \downarrow  &  & \downarrow  &  &  \downarrow & &   \downarrow  & \\
0  \rightarrow & R_{q+1}& \rightarrow & J_{q+1}(E) & \rightarrow & J_1(F_0) &  \rightarrow   & F_1 & \rightarrow & 0 & \\
   &   \downarrow  &  &  \downarrow &  &  \downarrow  &  &  \downarrow & &  & \\
   &  0  &  &  0  &  &  0  &   & 0 & & &
\end{array}  \]
we have $F_2 \simeq C_2(E)/C_2$. If we apply the Spencer $\delta$-operator to the long symbol sequence:  \\
\[  0 \rightarrow S_{q+3}T^*\otimes E \rightarrow S_3T^*\otimes F_0 \rightarrow  S_2T^*\otimes F_1 \rightarrow T^*\otimes F_2  \]
we discover, through a standard snake diagonal chase, that such a sequence may not be exact at $S_2T^*\otimes F_1$ with a cohomology equal to $H^3(g_q)$ {\it that may not vanish}.  \\

With $n=4, q=2$ and the conformal system ${\hat{R}}_2\subset J_2(T)$, we provide below the fiber dimensions:   \\
\[  \begin{array}{rcccccccl}
   & & &  & 0 & & 0  &  & 0  \\
  & &  & &  \downarrow &  &  \downarrow &  &  \downarrow   \\
 &  & &  & 15 & \underset 1{\stackrel{D_1}{\longrightarrow}} & 60 & \underset 1{\stackrel{D_2}{\longrightarrow}} & 90  \\ 
   &  & &  &  \downarrow &  & \downarrow & & \downarrow    \\
  0 &\longrightarrow & 4 & \stackrel{j_2}{\longrightarrow} & 60 & \underset 1{\stackrel{D_1}{\longrightarrow}} & 160 & \underset 1{\stackrel{D_2}{\longrightarrow}} & 180  \\ 
   &  & \parallel &  & \hspace{5mm} \downarrow {\Phi}_0 &  & \hspace{5mm} \downarrow {\Phi}_1 & &   \\
  &  & 4 & \underset 2 {\stackrel{{\cal{D}}}{\longrightarrow}} & 45 & 
\underset 1{\stackrel{{\cal{D}}_1}{\longrightarrow}} & 100 &  &  \\
  &  &  &  &  \downarrow &  &  \downarrow &  &  \\
   &  &  &  & 0 & & 0  &  &  
\end{array}   \]  \\   \\
Now $H^3({\hat{g}}_2)\neq 0$ when $n=4$ as ${\hat{g}}_2$ is $3$-acyclic only when $n\geq 5$ but no classical approach could even allow to imagine such a specific cohomological importance of $n=4$ ([38], p 26-28).  \\

The {\it large infinitesimal equivalence principle} initiated by the Cosserat brothers becomes natural in this framework, namely an observer cannot measure sections of $R_q$ but can only measure their images by $D_1$ or, equivalently, can only measure sections of $C_1$ killed by $D_2$. Accordingly, for a free falling particle in a constant gravitational field, we have successively:    \\
 \[{\partial}_4 {\xi}^k - {\xi}^k_4=0, \,\,\, {\partial}_4 {\xi}^k_4 - {\xi}^k_{44}=0, \,\,\,{\partial}_i{\xi}^k_{44} - 0=0, \,\,\,1 \leq i,k\leq 3  \]
 
Our purpose is now to extend these comments to the nonlinear sequences and we start with a few useful but technical local computations ([28]). First of all, we have:  \\
\[  {\chi}^k_{l,i}=g^k_u({\partial}_if^u_l-A^r_if^u_{rl}) \,\, \Rightarrow \,\, {\tau}^k_{s,r}=B^i_r{\chi}^k_{s,i}=g^k_u(B^i_r{\partial}_if^u_s -f^u_{rs}) = g^k_uB^i_r{\partial}_if^u_s- {\Gamma}^k_{rs}  \]
\[   A^r_iA^s_j({\tau}^k_{r,s} - {\tau}^k_{s,r})= {\partial}_iA^k_j - {\partial}_jA^k_i \,\, \Rightarrow \,\,  
{\tau}^k_{r,s} - {\tau}^k_{s,r}= B^i_rB^j_s( {\partial}_iA^k_j - {\partial}_jA^k_i ) \]
and let the reader ckeck these formulas directly as an exercise. \\

\noindent
{\bf LEMMA 5.C.1}: Summing on $k$ and $r$ when $\gamma =0$, we get successively:   \\
\[  \begin{array}{rcl}
({\tau}^r_{i,r} - {\tau}^r_{r,i})\, det(A) & =  & B^r_iB^j_k({\partial}_rA^k_j - {\partial}_jA^k_r) \, det(A) \\
                                            & = &  B^r_i(B^j_k{\partial}_rA^k_j  + A^j_r{\partial}_sB^s_j)\, det(A)   \\
                                            & = & B^r_i(B^j_k{\partial}_rA^k_j) \, det(A) + \, det(A) {\partial}_rB^r_i \\
                                            & = & B^r_i {\partial}_r \, det(A)+ \, det(A) {\partial}_rB^r_i   \\
                                            & = & {\partial}_r(B^r_i \, det(A))
\end{array}   \]  
\[  \begin{array}{rcl}
- {\omega}^{ij}{\partial}_r(B^r_idet(A)){\xi}^s_{sj} - {\omega}^{ij}{\tau}^r_{j,i}det(A) {\xi}^s_{sr} 
& = & [{\omega}^{ij}({\tau}^r_{r,i} - {\tau}^r_{i,r}){\xi}^s_{sj} -{\omega}^{ij}{\tau}^r_{j,i}{\xi}^s_{sr}] det(A)  \\
& = & [{\omega}^{ij}{\tau}^r_{r,i} - ({\omega}^{ij}{\tau}^r_{i,r}+{\omega}^{ir}{\tau}^j_{r,i})] det(A){\xi}^s_{sj} \\
& = & [{\omega}^{ij}{\tau}^r_{r,i} - ({\omega}^{ij}{\tau}^r_{i,r}+{\omega}^{ir}{\tau}^j_{i,r})] det(A){\xi}^s_{sj} \\
& = & [{\omega}^{ij} {\tau}^r_{r,i} - \frac{2}{n}{\omega}^{jr} {\tau}^t_{t,r}]det(A){\xi}^s_{sj} \\
 & = & \frac{(n-2)}{n}{\omega}^{ij} {\tau}^r_{r,i}det(A){\xi}^s_{sj}
\end{array}  \]   \\

Using the " {\it vertical machinery} ", namely the isomorphism $V(J_q({\cal{E}})) \simeq J_q(V({\cal{E}}))$, like in the preceding sections, we shall vary the sections $\delta f_q=(\delta f^k_{\mu}(x))$ while setting $\delta ({\partial}_if^k_{\mu}(x))={\partial}_i \delta (f^k_{\mu}(x))$ as it is done in analytical mechanics with the notations $\delta \dot{q}=\dot{\delta q}$ when studying the variation of a Lagrangian $L(t,q,\dot{q})$ 
([26],[27],[28],[30]).  \\

\noindent
{\bf LEMMA 5.C.2}: Let us compute directly the variation of the $1$-form $\alpha$ over the target and over the source, recalling that $\alpha={\alpha}_idx^i$ with ${\chi}^r_{r,i}= g^r_k{\partial}_i f^k_r - A^r_i g^s_k f^k_{rs} = n({\partial}_ia - A^r_i a_r)$, $na_i=g^r_kf^k_{ri}$ and ${\alpha}_i={\chi}^r_{r,i} + {\gamma}^r_{rs} {\chi}^s_{,i}$, even if $\gamma = 0$ locally. We have successively, exchanging source with target:  \\   
\[ \delta f^k={\eta}^k={\xi}^r{\partial}_r f^k, \hspace{2mm}  \delta f^k_i={\eta}^k_uf^u_i={\xi}^r{\partial}_rf^k_i + f^k_r {\xi}^r_i  \]
\[\delta f^k_{ij}= {\eta}^k_{uv}f^u_if^v_j+{\eta}^k_uf^u_{ij}={\xi}^r{\partial}_rf^k_{ij}+f^k_{rj}{\xi}^r_i + f^k_{ri}{\xi}^r_j + f^k_r{\xi}^r_{ij}\]
\[ n \delta a_i=g^r_k\delta f^k_{ri}+ f^k_{ri}\delta g^r_k = g^r_k({\eta}^k_{uv}f^u_if^v_r+{\eta}^k_uf^u_{ir}) - f^u_{ri}g^r_k{\eta}^k_u=
  f^r_i{\eta}^s_{sr}  \]
\[ \begin{array}{rcl}
n \delta a_i &= &g^s_k ( {\xi}^r{\partial}_rf^k_{is}+f^k_{rs}{\xi}^r_i + f^k_{ri}{\xi}^r_s + f^k_r{\xi}^r_{si}) - f^u_{si}g^s_k(g^t_u({\xi}^r{\partial}_rf^k_t+f^k_r{\xi}^r_t))  \\
   & =  &  n ({\xi}^r{\partial}_ra_i + a_r{\xi}^r_i ) +{\xi}^r_{ri}
\end{array}   \]
\[ \begin{array}{rcl}
a_i=f^k_ib_k  & \Rightarrow  & n \delta a_i= n ({\xi}^r{\partial}_ra_i + a_r{\xi}^r_i ) +{\xi}^r_{ri}= nf^k_i (\delta b_k +{\eta}^l{\partial b_k}/{\partial y^l})+ n b_k ({\xi}^r{\partial}_rf^k_i + f^k_r {\xi}^r_i )         \\
                     & \Rightarrow &  n {\xi}^r{\partial}_ra_i  +{\xi}^r_{ri}= nf^k_i (\delta b_k +{\xi}^r{\partial b_k}/{\partial x^r})+ n b_k {\xi}^r{\partial}_rf^k_i    \\
                     & \Rightarrow & n \delta b_k= g^i_k{\xi}^r_{ri}   
\end{array}   \]

Then, using the definition of $a$, namely $det(f^k_i(x))= e^{na(x)}$, we have {\it over the source}:   \\
\[  n \delta a= (1/det(f^k_i))\delta det(f^k_i)= g^i_k\delta f^k_i= {\eta}^s_s = g^i_k({\xi}^r{\partial}_rf^k_i + 
f^k_r{\xi}^r_i)=  n \,{\xi}^r{\partial}_ra + {\xi}^r_r      \]
Using the variation $\delta A^k_i={\xi}^r{\partial}_rA^k_i + A^k_r{\partial}_i{\xi}^r - A^r_i{\xi}^k_r$, we finally get when $\gamma=0$: \\
\[  \begin{array}{rcl}
\delta {\chi}^r_{r,i} & = & n\, \delta {Ê\partial}_ia - nA^r_i\delta a_r -n \, a_r \delta A^r_i   \\
    &  =  &  ({\partial}_i{\xi}^r_r - {\xi}^r_{ri}) + ({\xi}^r{\partial}_r {\chi}^r_{r,i} + 
    {\chi}^s_{s,r}{\partial}_i{\xi}^r ) - {\chi}^s_{,i}{\xi}^r_{rs}   
\end{array}   \]
\[  \Rightarrow \delta {\alpha}_i  =  \delta {\chi}^r_{r,i} + {\xi}^r_{rs}{\chi}^s_{,i} =  ({\partial}_i{\xi}^r_r - {\xi}^r_{ri}) + 
( {\alpha}_r{\partial}_i{\xi}^r + {\xi}^r{\partial}_r {\alpha}_i )   \] 
The terms ${\partial}_i{\xi}^r_r + ( {\alpha}_r{\partial}_i{\xi}^r + {\xi}^r{\partial}_r {\alpha}_i )$ of the variation, including the variation of 
$\alpha={\alpha}_idx^i$ as a $1$-form, are {\it exactly} the ones introduced by Weyl in ([54] formula (76), p 289). We also recognize the variation $\delta A_i$ of the $4$-potential used by engineers now expressed by means of second order jets but the use of the Spencer operator sheds a new light on EM.  \\
Similarly,when $\gamma = 0$, we have over the target:  \\
\[ f^k_rA^r_i = {\partial}_i f^k \,\, \Rightarrow  \,\, f^k_r \delta A^r_i + A^r_i {\eta}^k_uf ^u_r  = \frac{\partial {\eta}^k}{ \partial y^u} 
{\partial}_if^u \,\,  \Rightarrow \,\,  \delta A^r_i = g^r_l ( \frac{\partial {\eta}^l}{\partial y^k} - {\eta}^l_k){\partial}_i f^k  \]  
\[  \begin{array}{lcl}
  \delta {\chi}^r_{r,i} & = & [ \frac{\partial {\eta}^s_s}{\partial y^k} - ng^r_l (\frac{\partial {\eta}^l}{\partial y^k} - {\eta}^l_k ) a_r ] {\partial}_if^k - 
A^r_i f^k_r {\eta}^s_{sk}   \\  \\
        & = & [(\frac{\partial {\eta}^s_s}{\partial y^k} - {\eta}^s_{sk}) - n\, b_l(\frac{\partial {\eta}^l}{\partial y^k}- {\eta}^l_k)]\frak{\partial}_i f^k  
\end{array}  \]
a result only depending on the components of the Spencer operator, in a coherent way with the general variational formulas that could have been used otherwise. We notice that these formulas, which have been obtained with difficulty for second order jets, could not even be obtained by hand for third order jets. They  show the importance and usefulness of the general formulas providing the Spencer non-linear operators for an arbitrary order, in particular for the study of the conformal group which is defined by second order lie equations with a $2$-acyclic symbol. When $\gamma = 0$ locally, it is also important to notice that:\\
 \[{\alpha}_i(x)=n(\frac{\partial b}{\partial y^k} - b_k){\partial}_if^k={\beta}_k(f(x)){\partial}_if^k(x) \Rightarrow \delta {\alpha}_i = (\delta {\beta}_k + {\eta}^l \frac{\partial {\beta}_k}{\partial y^l}+ {\beta}_r \frac{\partial {\eta}^r}{\partial y^k} ) {\partial}_if^k \]
and thus $\delta \beta$ does not only depend linearly on the Spencer operator, contrary to $\delta \alpha$.  \\

\noindent
{\bf LEMMA 5.C.3}: We have {\it over the source}:   \\
\[  \begin{array}{lcl}
\delta det(A) & = & det(A)B^i_k\delta A^k_i \\
                   & = & det(A)B^i_k({\xi}^r{\partial}_rA^k_i +A^k_r{\partial}_i{\xi}^r - A^r_i{\xi}^k_r)   \\
                   & = &  {\xi}^r{\partial}_r(det(A)) + det(A) ({\partial}_r{\xi}^r - {\xi}^r_r)
\end{array}   \]

Now, we recall the identities:  \\
\[ {\partial}_i{\chi}^k_{l,j}-{\partial}_j{\chi}^k_{l,i}-{\chi}^r_{l,i}{\chi}^k_{r,j}+{\chi}^r_{l,j}{\chi}^k_{r,i}-A^r_i{\chi}^k_{lr,j}+A^r_j{\chi}^k_{lr,i}=0   \]
that we may rewrite in the equivalent form:  \\
\[ \begin{array}{rcl}
{\tau}^k_{lr,s} - {\tau}^k_{ls,r} & =  & B^i_rB^j_s({\partial}_i{\chi}^k_{l,j}-{\partial}_j{\chi}^k_{l,i}-{\chi}^r_{l,i}{\chi}^k_{r,j}+
{\chi}^r_{l,j}{\chi}^k_{r,i})  \\
                                                 & = & B^i_rB^j_s({\partial}_i{\chi}^k_{l,j}-{\partial}_j{\chi}^k_{l,i}) - ({\tau}^t_{l,r}{\tau}^k_{t,s} - 
 {\tau}^t_{l,s}{\tau}^k_{t,r} )                                               
\end{array}   \] 
Looking only at the terms not containing the jets of order $2$ in the right member, we have {\it separately}:  \\
\[  B^i_rB^j_s(({\partial}_i(g^k_u {\partial}_jf^u_l) - {\partial}_j(g^k_u{\partial}_if^u_l))=B^i_rB^j_s(({\partial}_ig^k_u)({\partial}_jf^u_l ) - 
({\partial}_jg^k_u)({\partial}_if^u_l) )  \]
\[  \begin{array}{rcl}
(g^t_uB^i_r{\partial}_if^u_l)(g^k_vB^i_s{\partial}_if^v_t) - (r\leftrightarrow s) & = &B^i_rB^j_s((g^t_u{\partial}_if^u_l)(g^k_v{\partial}_jf^v_t)) - 
(r\leftrightarrow s)  \\
           & = & - (B^i_rB^j_s (g^t_u{\partial}_if^u_l)(f^v_t{\partial}_jg^k_v) -  (r\leftrightarrow s ) ) \\
           & = & - (B^i_rB^j_s ({\partial}_jg^k_u)({\partial}_if^u_l) - (r \leftrightarrow s))
\end{array}   \]
and the total sum vanishes.  \\

Looking at the terms linear in the second order jets $g^k_uf^u_{ij}$, we have {\it separately} (care to the sign):   \\
\[ B^i_rB^j_s({\partial}_jA^t_i - {\partial}_iA^t_j) g^k_uf^u_{tl} = ({\tau}^t_{r,s}- {\tau}^t_{s,r})g^k_uf^u_{tl}= g^t_v(B^i_s{\partial}_if^v_r - 
B^i_r{\partial}_if^v_s)g^k_uf^u_{tl}   \]
\[    (g^t_uB^i_r({\partial}_if^u_l )g^k_vf^v_{st} + g^k_vB^j_s({\partial}_jf^v_t)g^t_uf^u_{rl} ) - (r \leftrightarrow s)   \]

The simplest and final checking concerns the derivatives of the second order jets. We get:  \\
\[  \begin{array}{rcl} 
B^i_rB^j_s ({\partial}_i{\chi}^k_{l,j} - {\partial}_j{\chi}^k_{l,i}) & = & B^i_rB^j_s(A^t_i{\partial}_j(g^k_uf^u_{tl}) - 
A^t_j{\partial}_i(g^k_uf^u_{tl})) + ...  \\
                                              &  =  & B^j_s {\partial}_j(g^k_uf^u_{rl}) - B^i_r {\partial}_i(g^k_uf^u_{sl}) + ...                  
\end{array}    \]
With $y=f(x)\leftrightarrow x=g(y)$,  it remains to substitute the formulas $B^i_r=f^k_r \partial g^i/\partial y^k$ while taking into account that we have ${\Gamma}^k_{ij}=g^k_uf^u_{ij}={\delta}^k_ia_j + {\delta}^k_ja_i - {\omega}_{ij}{\omega}^{kr}a_r$ because $\gamma=0$ in the conformal case which only depends on the Minkowski metric $\omega$ and not on a conformal factor.

The novelty and most tricky point is to notice that we have now only $n^2$ components for $({\tau}^k_{li,j}) \in T^*\otimes {\hat{g}}_2$ and no longer the $n^2(n^2-1)/12$ components of the classical Riemannian curvature. As we have already used the $n(n-1)/2$ components ${\varphi}_{ij}={\tau}^r_{ri,j} - {\tau}^r_{rj,i}= - {\varphi}_{ji}$, we may choose the $n(n+1)/2$ symmetric components ${\tau}_{ij}=\frac{1}{2}({\tau}^r_{ri,j} + {\tau}^r_{rj,i})={\tau}_{ji}$ that should involve the third order jets which are only vanishing in the linear case but do not vanish at all in the non-linear case. To avoid such a situation, we shall use the following key proposition that must be compared to the procedure used in classical GR:  \\

\noindent
{\bf PROPOSITION  5.C.4}: Defining ${\rho}^k_{l,ij}={\tau}^k_{li,j} - {\tau}^k_{lj,i}$ it is just sufficient to study ${\rho}_{i,j}={\rho}^r_{i,rj}\neq {\rho}_{j,i}$ and $tr(\rho)={\omega}^{ij}{\rho}_{i,j}$ or ${\tau}_{i,j}={\tau}^r_{ri,j}$ and $tr(\tau )= {\omega}^{ij}{\tau}_{i,j}$. Setting:  \\
\[ {\rho}_{ij} = \frac{(n-2)}{n}{\tau}_{ij} + \frac{1}{n} {\omega}_{ij}tr(\tau)  \]
in a way not depending on any conformal factor, we have the equivalences:  \\
\[  {\tau}^k_{li,j}=0  \Leftrightarrow {\rho}^k_{l,ij} \Leftrightarrow {\varphi}_{ij}=0 \oplus {\tau}_{ij}=0 \Leftrightarrow 
{\varphi}_{ij}=0 \oplus {\rho}_{ij}=0   \]

\noindent
{\it Proof} : As ${\hat{g}}_2\simeq T^*$, we have successively (Compare to Proposition 4.B.3): \\
\[ n{\rho}^k_{l,ij}= {\delta}^k_l{\tau}^r_{ri,j} + {\delta}^k_i{\tau}^r_{rl,j}- {\omega}_{li}{\omega}^{ks}{\tau}^r_{rs,j} - {\delta}^k_l{\tau}^r_{rj,i} - {\delta}^k_j {\tau}^r_{rl,i} +{\omega}_{lj}{\omega}^{ks}{\tau}^r_{rs,i} \,\, \Rightarrow \,\, {\rho}^r_{r,ij}= {\tau}_{i,j} - {\tau}_{j,i}  \]
\[  n {\rho}_{i,j}= (n-1) {\tau}^r_{ri,j} - {\tau}^r_{rj,i} + {\omega}_{ij} {\omega}^{st} {\tau}^r_{rs,t}= (n-1){\tau}_{i,j} - {\tau}_{j,i} + {\omega}_{ij}tr(\tau) \,\, \Rightarrow \,\, {\rho}_{i,j} - {\rho}_{j,i}= {\tau}_{i,j} - {\tau}_{j,i}\]
\[   ntr(\rho)= 2(n-1){\omega}^{ij }{\tau}^r_{ri,j}= 2(n-1) tr(\tau )  \]
When we suppose that there is no EM, that is:  \\
\[  {\varphi}_{ij}=0 \,\, \Leftrightarrow \,\,  {\tau}_{i,j}={\tau}_{j,i}= {\tau}_{ij}={\tau}_{ji} \,\, \Leftrightarrow \,\,  {\rho}_{i,j}={\rho}_{j,i}={\rho}_{ij}={\rho}_{ji} \] 
the above formulas become simpler with:  \\
\[  n{\rho}_{ij}=(n-2){\tau}_{ij} + {\omega}_{ij} tr(\tau ) \,\, \Leftrightarrow \,\,  {\tau}_{ij} = \frac{n}{n-2} {\rho}_{ij} - 
\frac{n}{2(n-1)(n-2)}{\omega}_{ij}tr(\rho) \]
Surprisingly, in the general situation, we have:  \\
\[  n {\rho}_{i,j}=(n-1) {\tau}_{i,j}  - {\tau}_{j,i} + {\omega}_{ij} tr(\tau), \,\,\,\,\,\,  
n {\rho}_{j,i}= (n-1) {\tau}_{j,i} - {\tau}_{i,j}  + {\omega}_{ij} tr(\tau)       \]
Summing, we discover that the same formula is still valid.

We may thus express ${\rho}^k_{l,ij}$ by means of ${\rho}_{i,j}$ or by means of ${\tau}_{i,j}$ while using the relations 
${\varphi}_{ij}={\rho}^r_{r,ij}={\tau}_{i,j} - {\tau}_{j,i}={\rho}_{i,j} - {\rho}_{j,i}$. As ${\hat{g}}_2$ is $2$-acyclic when $n\geq 4$ in the conformal case ([28],[38]), we have the short exact sequence:  \\
 \[  0  \longrightarrow  {\hat{g}}_3  \stackrel{\delta}{\longrightarrow} T^* \otimes {\hat{g}}_2 \stackrel{\delta}{\longrightarrow} \delta(T^*\otimes {\hat{g}}_2 ) \longrightarrow 0  \]
Moreover, as ${\hat{g}}_3=0$ when $n\geq 3$, we have an isomorphism $T^*\otimes {\hat{g}}_2 \simeq \delta (T^*\otimes {\hat{g}}_2)$, both vector bundles having the same fiber dimension $n^2= \frac{n(n-1)}{2} + \frac{n(n+1)}{2}$ when $n\geq 4$ and thus ${\tau}^k_{li,j}=0 
\Leftrightarrow  {\rho}^k_{l,ij}=0 $.  \\
When there is no EM, that is when $\varphi=0$, then one can express ${\rho}^k_{l,ij}$ by means of ${\rho}_{ij}={\rho}_{i,j}={\rho}_{j,i}={\rho}_{ji}$ but there is no longer the Levi-Civita isomorphism $(\omega, \gamma)\simeq j_1(\omega)$ in the Spencer sequence and the above proposition is quite different from the concept of curvature in GR as it just amounts to the vanishing of the Weyl tensor according to Theorem 5.B.5.  \\
\hspace*{12cm}     Q.E.D.   \\

We notice that no one of the preceding results could be obtained by classical methods because they crucially depend on the Spencer $\delta$-cohomology. As a byproduct, the same formulas provide:  \\

\noindent
{\bf COROLLARY 5.C.5}: The corresponding Weyl tensor vanishes.  \\

Supposing again that there is no EM and looking for the derivatives of the second order jets, contracting in $k$ and $r$ while replacing $l$ by $i$ and 
$s$ by $j$, we get with $a_i= f^k_ib_k$:   \\
\[  \begin{array}{rcl}
{\rho}_{ij}= {\rho}_{ji}= {\tau}^r_{ri,j} - {\tau}^r_{ij,r} & = & B^t_j{\partial}_t (g^r_uf^u_{ri}) - B^t_r{\partial}_t(g^r_uf^u_{ij}) + ... \\
                  & = & n B^t_j{\partial}_ta_i - B^t_r{\partial}_t({\delta}^r_ia_j + {\delta}^r_ja_i - {\omega}_{ij}{\omega}^{rs}a_s ) + ...  \\
    & = &  nf^l_j \frac{\partial a_i}{\partial y^l} - f^k_i\frac{\partial a_j}{\partial y^k} - f^l_j\frac{\partial a_i}{\partial y^l} +
    {\omega}_{ij}{\omega}^{rs}f^k_r \frac{\partial a_s}{\partial y^k} + ...   \\
    & = & f^k_if^l_j[ (n-2)\frac{\partial b_k}{\partial y^l} + {\omega}_{kl}(y){\omega}^{rs}(y)\frac{\partial b_s}{\partial y^r}] + ... 
\end{array}   \]
with bracket symmetric under the exchange of $k$ and $l$. We have to take into account the following terms linear in the $b_k$, left aside in the derivations:   \\
\[     [(n-1) f^l_j\frac{\partial f^k_i}{\partial y^l} + f^l_i\frac{\partial f^k_j}{\partial y^l} +
{\omega}_{ij}{\omega}^{rs}f^l_r \frac{\partial f^k_s}{\partial y^l}] b_k =[(n-1)B^t_j{\partial}_tf^k_i +  B^t_i{\partial}_tf^k_j + {\omega}_{ij}{\omega}^{rs}f^l_r{\partial}_tf^k_s]b_k  \]

Under the same assumption, let us work out the quadratic terms in $b_k$ as follows:   \\
\[   ( {\tau}^t_{l,r}{\tau}^k_{t,s} - {\tau}^t_{l,s}{\tau}^k_{t,r} )   =  (g^t_uf^u_{rl}) (g^k_vf^v_{st}) - (g^t_uf^u_{sl})(g^k_vf^v_{rt})   \]                                
Contracting in $k$ and $r$ as above while replacing $l$ by $i$ and $s$ by $j$, we get:  \\
\[({\tau}^t_{i,r}{\tau}^r_{t,j}- {\tau}^t_{i,j}{\tau}^r_{t,r}) = (g^t_uf^u_{ri})(g^r_vf^v_{jt}) -(g^t_uf^u_{ij})(g^r_vf^v_{rt}) \]
that is:  \\
\[ ({\delta}^t_ra_i + {\delta}^t_ia_r - {\omega}_{ri}{\omega}^{st}a_s)({\delta}^r_ja_t + {\delta}^r_ta_j - {\omega}_{jt}{\omega}^{rs}a_s) - 
n({\delta}^t_ia_j + {\delta}^t_ja_i - {\omega}_{ij}{\omega}^{st}a_s)a_t   \]
Effecting all the contractions, we get:  \\
\[  (n a_ia_j ) + ( 2a_ia_j  - {\omega}_{ij}{\omega}^{rs}a_ra_s) - ({\omega}_{ij}{\omega}^{rs}a_ra_s ) - n(2 a_ia_j -{\omega}_{ij}{\omega}^{rs}a_ra_s)  \]
and obtain the unexpected very simple formula:  \\
\[ na_ia_j + 2a_ia_j -2{\omega}_{ij} {\omega}^{rs}a_ra_s - 2n a_ia_j+n{\omega}_{ij}{\omega}^{rs}a_ra_s = (2-n)a_ia_j+(n-2){\omega}_{ij}{\omega}^{rs}a_ra_s  \]
or, equivalently $f^k_if^l_j[(2-n)b_kb_l +(n-2) {\omega}_{kl}(y){\omega}^{rs}b_rb_s] $. Collecting these results, we finally get:  \\

\noindent
{\bf THEOREM 5.C.6}: When there is no EM, we have over the target the formulae:  \\   \\
\hspace*{1cm} \fbox{  $      {\rho}_{ij} =  f^k_if^l_j[ (n-2)\frac{\partial b_k}{\partial y^l} +  {\omega}_{kl}(y){\omega}^{rs}(y)\frac{\partial b_s}{\partial y^r} + (n-2)b_kb_l - (n-2) {\omega}_{kl}(y){\omega}^{rs}b_rb_s ]   $} \\  \\  
\hspace*{35mm} \fbox{$ {\tau}_{ij}= nf^k_if^l_j[\frac{\partial  b_k}{\partial y^l} + b_kb_l - 
\frac{1}{2}{\omega}_{kl}(y){\omega}^{rs}(y)b_rb_s] $}  \\    \\
that {\it do not depend on any conformal factor for} $\omega$ and thus simply:  \\   \\
\hspace*{15mm} \fbox{ $ \tau= \frac{n}{{\Theta}^2}[ {\omega}^{kl}(y)\frac{\partial b_k}{\partial y^l} - 
\frac{(n-2)}{2}{\omega}^{kl}(y) b_kb_l] = n[{\bar{\omega}}^{kl}(y)\frac{\partial b_k}{\partial y^l} - 
\frac{(n-2)}{2}{\bar{\omega}}^{kl}(y) b_kb_l]  $}  \\     \\
that only depends on the new metric $\bar{\omega}={\Theta}^2\omega $ defined over the target.  \\
 
\noindent
{\it Proof} : We have to prove the following technical result which is indeed the hardest step, namely that ${\rho}_{ij}$ does not contain terms 
linear in $b_k$ {\it over the target}. The main problem is that, if we have any derivative of the second order jets {\it over the source}, like ${\partial}_ra_i$, we obtain therefore a term like ${\partial}_r(f^k_i b_k)=f^k_i{\partial}_rb_k + ({\partial}_rf^k_i)b_k$ which is bringing a term linear in the $b_k$ and we have to prove that {\it such terms may not exist if we work only over the target}. \\
For this, let us set over the source:  \\
\[  {\tau}^k_{s,r}=T^k_{s,r} -{\Gamma}^k_{rs}, \,\, \,T^k_{s,r}= g^k_uB^i_r{\partial}_if^u_s\neq T^k_{r,s},\,\,\,
{\Gamma}^k_{rs}={\delta}^k_ra_s + {\delta}^k_sa_r - {\omega}_{rs}{\omega}^{kt}a_t = {\Gamma}^k_{sr}  \]
Looking for the derivatives of the second order jets, we already saw that they can only appear through the terms:   \\
\[ B^i_s{\partial}_i{\Gamma}^k_{rl}- B^i_r{\partial}_i{\Gamma}^k_{sl}= f^v_s\frac{\partial {\Gamma}^k_{rl}}{\partial y^v}- f^u_r\frac{\partial {\Gamma}^k_{sl}}{\partial y^u} \]
Contracting in $k$ and $r$, we get when there is no EM:  \\
\[ \begin{array}{rcl}
   f^v_s\frac{\partial {\Gamma}^r_{rl}}{\partial y^v} -f^u_r\frac{\partial {\Gamma}^r_{sl}}{\partial y^u}& =&f^v_s\frac{\partial}{\partial y^v}(na_l)
   - f^u_r\frac{\partial}{\partial y^u}({\delta}^r_sa_l + {\delta}^r_la_s - {\omega}_{sl}{\omega}^{rt}a_t)   \\
     &  =  &  (n-1)f^v_s\frac{\partial (f^u_lb_u)}{\partial y^v} - f^u_l\frac{\partial (f^v_sb_v)}{\partial y^u} 
     + {\omega}_{sl}{\omega}^{rt}f^u_r\frac{\partial (f^v_tb_v)}{\partial y^u}   \\
& = & (n-2)f^u_lf^v_s \frac{\partial b_v}{\partial y^u}+ {\omega}_{sl}{\omega}^{rt}f^u_rf^v_t\frac{\partial b_v}{\partial y^u} + ...
\end{array}   \]
but we have to take into account the linear terms produced by an integration by parts:   \\
\[    (n-1)f^v_s\frac{\partial f^u_l}{\partial y^v}b_u - f^u_l\frac{\partial f^v_s}{\partial y^u} b_v
     + {\omega}_{ls}{\omega}^{rt}f^u_r\frac{\partial f^v_t}{\partial y^u} b_v  \]
that is to say, we have to {\it substract}:  \\
\[  (n-1)g^t_uB^i_s{\partial}_if^u_la_t - g^t_vB^i_l{\partial}_if^v_s a_t + {\omega}_{ls}{\omega}^{rt}g^u_vB^i_r{\partial}_if^v_ta_u
   = (n-1) T^t_{l,s}a_t  -  T^t_{s,l}a_t  -  {\omega}_{ls}{\omega}^{rt} T^u_{t,r}a_u                 \]

\noindent
Meanwhile, as we already saw, we have to compute ({\it care to the signs involved} ):   \\
\[   (T^t_{r,s} - T^t_{s,r}) {\Gamma}^k_{lt} - (T^t_{l,r}{\Gamma}^k_{st} + T^k_{t,s}{\Gamma}^t_{lr}) 
  + (T^t_{l,s}{\Gamma}^k_{rt} + T^k_{t,r}{\Gamma}^t_{ls}) \]
and to contract in $k$ and $r$ in order to get:   \\ 
\[   (T^t_{r,s} - T^t_{s,r}) {\Gamma}^r_{lt} - (T^t_{l,r}{\Gamma}^r_{st} + T^r_{t,s}{\Gamma}^t_{lr}) 
  + (T^t_{l,s}{\Gamma}^r_{rt} + T^r_{t,r}{\Gamma}^t_{ls}) \]
However, {\it two terms are disappearing} and we are left with:  \\
\[    - T^t_{s,r} {\Gamma}^r_{lt} - T^t_{l,r}{\Gamma}^r_{st} 
  + (T^t_{l,s}{\Gamma}^r_{rt} + T^r_{t,r}{\Gamma}^t_{ls}) \]
that is to say:  \\
\[   \begin{array}{l} 
- T^t_{s,r}({\delta}^r_la_t +{\delta}^r_ta_l -{\omega}_{lt}{\omega}^{ru}a_u) -T^t_{l,r}({\delta}^r_s a_t +{\delta}^r_t a_s - {\omega}_{st}{\omega}^{ru}a_u)) \\
+ n T^t_{l,s}a_t + T^r_{t,r}({\delta}^t_la_s +{\delta}^t_sa_l - {\omega}_{ls}{\omega}^{tu}a_u) 
\end{array}    \]
and thus:  \\
\[ \begin{array}{l}
 -T^t_{s,l} - T^r_{s,r}a_l + {\omega}_{lt}{\omega}^{ru}T^t_{r,s}a_u - T^t_{l,s}a_t - T^t_{l,t}a_s + {\omega}_{st}{\omega}^{ru}T^t_{l,r}a_u  \\
 + n T^t_{l,s} a_t + T^r_{l,r}a_s + T^r_{s,r}a_l - {\omega}_{ls}{\omega}^{tu}T^r_{t,r}a_u 
 \end{array}   \]
The four terms containing $a_l$ and $a_s$ are disappearing and we are left with:   \\
\[   (n - 1)T^t_{l,s}a_t - T^t_{s,l}a_t     {\omega}_{lt}{\omega}^{ru}T^t_{r,s}a_u + {\omega}_{st}{\omega}^{ru}T^t_{l,r}a_u 
- {\omega}_{ls}{\omega}^{tu}T^r_{t,r}a_u  \]
Taking into account {\it twice successively} the conformal Killing equations, we obtain:   \\
\[  \begin{array}{l}
  (n - 1)T^t_{l,s}a_t - T^t_{s,l}a_t + \frac{2}{n}{\omega}_{ls}{\omega}^{ru}T^t_{t,r}a_u  - {\omega}_{ls}{\omega}^{tu}T^r_{t,r}a_u  \\
    = (n - 1)T^t_{l,s}a_t - T^t_{s,l}a_t   - {\omega}_{ls} {\omega}^{rt}T^u_{t,r}a_u     
     \end{array}  \]
that is {\it exactly the terms we had to substract} and there is thus no term linear in $a_i$ in the Ricci tensor over the target, a quite difficult result indeed because {\it no concept of classical Riemannian geometry could be used}.  \\

\noindent
We finally obtain from the definition of $\Theta$ while taking inverse matrices:  \\
\[  {\Theta}^2 {\omega}_{kl}(y)f^k_if^l_j = {\omega}_{ij}(x) \,\, \Rightarrow \,\, {\Theta}^{-2}{\omega}^{kl}(y)g^i_kg^j_l={\omega}^{ij}(x) \,\, \Rightarrow \,\, {\Theta}^{-2}{\omega}^{kl}(y)= {\omega}^{ij}(x)f^k_if^l_j  \]
and just need to set $\tau={\omega}^{ij}{\tau}_{ij}$ in order to get the last formula.  \\
\hspace*{12cm}   Q.E.D.   \\

\noindent
{\bf REMARK 5.C.7}: When $A^r_i={\delta}^r_i$, we get ${\rho}^k_{l, ij}= {\partial}_i{\chi}^k_{l,j}-{\partial}_j{\chi}^k_{l,i}-{\chi}^r_{l,i}{\chi}^k_{r,j}+{\chi}^r_{l,j}{\chi}^k_{r,i} $ with ${\chi}^k_{j,i}= g^k_u{\partial}_if^u_j - g^k_uf^u_{ij}$. However, in such a situation, we have:  \\
\[   {\omega}_{kl}(f(x))f^k_if^l_j= e^{2a(x)}{\omega}_{ij}(x) \,\, \Rightarrow \,\, {\omega}_{kl}(f(x)){\partial}_if^k_(x){\partial}_jf^l(x)= e^{2a(x)}{\omega}_{ij}(x) ={\bar{\omega}}_{ij}(x)  \]
Using the Minkowski metric $\omega$ which is locally constant and thus flat, it follows from the Vessiot structure equations that $\bar{\omega}$ {\it must } also be flat but we may have $f_2\neq j_2(f)$ even though $f_1=j_1(f)$. As $\bar{\omega}$ is conformally equivalent to $\omega$, then both metric have vanishing Weyl tensor and the integrability condition for $\bar{\omega}$ is thus to have a vanishing Ricci tensor, that is to say, prolonging once the system $j_1(f)^{-1}(\omega)=\bar{\omega}$, we get $j_2(f)^{-1}(\gamma)=\bar{\gamma}$ and obtain:  \\
\[  \gamma=0 \,\,\,  \Rightarrow \,\,\,  {\bar{\gamma}}^k_{ij}= {\delta}^k_i{\partial}_ja + {\delta}^k_j{\partial}_ia - {\omega}_{ij}{\omega}^{kr}{\partial}_ra \]
\[  (n-2){\partial}_{ij}a  +  {\omega}_{ij}{\omega}^{rs}{\partial}_{rs}a +
        (n-2){\partial}_ia{\partial}_ja - (n-2) {\omega}_{ij}{\omega}^{rs}{\partial}_ra{\partial}_sa = 0   \]   
This is a very striking result showing out for the first time that there may be links between the non-linear Spencer sequence and classical conformal geometry as the above result is just the variation of the classical Ricci tensor under a conformal change of the metric and the reason for which we introduced exponentials for describing conformal factors. \\

With $\phi=\frac{GM}{r}$ and thus $\frac{\phi}{c^2}\ll 1$, we have thus been able to replace $1-\frac{\phi}{c^2}$ by $1 + \frac{\phi}{c^2}$, suppressing therefore the {\it horizon} $r=GM/c^2$ when $G$ is the gravitational constant and $M$ the central attractive mass, along with the following scheme: \\ 
\[   source \stackrel{inversion}{\longleftrightarrow} target   \]
\[    ATTRACTION  \stackrel{inversion}{\longleftrightarrow}  REPULSION     \]    \\
As it is based on the inversion rule for the second order jets of the conformal Lie groupoid, we get:  \\
{\it such a procedure could not be even imagined in any classical framework dealing with Lie groups of transformations}.  \\

\noindent
{\bf THEOREM 5.C.8}: We have the variation over the source:  \\
\[  \delta {\tau}_{j,i}=B^r_i{\partial}_r{\xi}^s_{sj} + {\xi}^r{\partial}_r{\tau}_{j,i} + 
{\tau}_{j,r}{\xi}^r_i + {\tau}_{r,i}{\xi}^r_j  - {\tau}^r_{j,i}{\xi}^s_{sr}   \]   \\

\noindent
{\it Proof} : Using the general variational formulas one obtains:
\[  \begin{array}{rcl}
\delta {\chi}^k_{lj,i}& = &  ({\partial}_i {\xi}^k_{lj}- {\xi}^k_{lij}) +{\xi}^r{\partial}_r{\chi}^k_{lj,i}+{\chi}^k_{lj,r}{\partial}_i{\xi}^r \\
                              &    & + {\chi}^k_{lr,i}{\xi}^r_j + ({\chi}^k_{rj,i}{\xi}^r_l -  {\chi}^r_{lj,i}{\xi}^k_r )  \\
                              &    & + {\chi}^k_{r ,i}{\xi}^r_{lj} - {\chi }^r_{l,i}{\xi}^k_{rj}- {\chi}^r_{j.i}{\xi}^k_{lr} - {\chi}^r_{,i}{\xi}^k_{lrj}
\end{array}  \]
where one must take into account that the third order jets of conformal vector fields vanish, that is to say ${\xi}^k_{lrj}=0$. Contracting in $k$ and $l$, we get:  \\
\[  \delta {\chi}^s_{sj,i} = {\partial}_i {\xi}^s_{sj} +{\xi}^r{\partial}_r{\chi}^s_{sj,i}+{\chi}^s_{sj,r}{\partial}_i{\xi}^r +{\chi}^s_{sr,i}{\xi}^r_j   \\
                             - {\xi}^r_{j.i}{\xi}^s_{sr} - {\chi}^r_{,i}{\xi}^s_{srj}     \]
\[  {\chi}^k_{lj,i}= A^r_i{\tau}^k_{lj,r}  \,\,\, \Rightarrow \,\,\, \delta {\chi}^k_{lj,i}= A^r_i\delta {\tau}^k_{lj,r} + {\tau}^k_{lj,r}\delta A^r_i 
\,\,\, A^r_i\delta {\tau}^s_{sj,r}= \delta {\chi}^s_{sj,i} - {\tau}_{j,r} \delta A^r_i   \]
\[  A^r_i \delta {\tau}_{j,r}= {\partial}_i {\xi}^s_{sj} +{\xi}^r{\partial}_r{\chi}^s_{sj,i} + {\chi}^s_{sj,r}{\partial}_i{\xi}^r - 
{\chi}^r_{j.i}{\xi}^s_{sr} - {\tau}_{j,r}({\xi}^s{\partial}_sA^r_i + A^r_i{\partial}_i{\xi}^s - A^s_i{\xi}^r_s ) \]
\[   \delta {\tau}_{j,i}= B^r_i {\partial}_r{\xi}^s_{sj} + {\xi}^r{\partial}_r{\tau}_{j,i}+ ({\tau}_{j,r} {\xi}^r_i+ {\tau}_{r,i} {\xi}^r_j)
+ {\tau}^r_{j,i}{\xi}^s_{sr} \]
\hspace*{12cm}    Q.E.D.   \\

Using the fact that $\omega$ is locally constant and not varied ({\it care}), we have at once:   \\
\[ \begin{array}{rcl}
 \delta {\tau} & = & {\omega}^{ij}(B^r_i{\partial}_r{\xi}^s_{sj}) + {\xi}^r{\partial}_r{\tau} + 
{\omega}^{ij}({\tau}_{j,r}{\xi}^r_i + {\tau}_{r,i}{\xi}^r_j ) - {\omega}^{ij}{\tau}^r_{j,i}{\xi}^s_{sr}  \\
                    & = & {\omega}^{ij}(B^r_i{\partial}_r{\xi}^s_{sj}) + {\xi}^r{\partial}_r{\tau} + 
{\tau}_{r,s}({\omega}^{is}{\xi}^r_i + {\omega}^{js}{\xi}^r_j ) - {\omega}^{ij}{\tau}^r_{j,i}{\xi}^s_{sr}  \\
                    & = & {\omega}^{ij}(B^r_i{\partial}_r{\xi}^s_{sj}) + {\xi}^r{\partial}_r{\tau} + 
\frac{2}{n}{\omega}^{rs}{\tau}_{r,s} {\xi}^t_t - {\omega}^{ij}{\tau}^r_{j,i}{\xi}^s_{sr}                       
\end{array}  \] 
and thus: \\

\noindent
{\bf COROLLARY 5.C.9}:  \hspace{1cm} \fbox{ $ \delta \tau = {\omega}^{ij}B^r_i{\partial}_r{\xi}^s_{sj} + {\xi}^r{\partial}_r{\tau} + 
\frac{2}{n}{\tau}{\xi}^r_r - {\omega}^{ij}{\tau}^r_{j,i}{\xi}^s_{sr} $ }  \\

Combining this result with the three preceding Lemmas, we finally obtain:  \\

\noindent
{\bf COROLLARY 5.C.10}: The action variation over the source is:  \\

\fbox{ $     \delta (\tau det(A)) = {\partial}_r({\xi}^r\tau det(A) + {\omega}^{ij}(x)B^r_idet(A){\xi}^s_{sj}) - \frac{(n-2)}{n}\tau det(A){\xi}^r_r + 
\frac{(n-2)}{n}{\omega}^{ij}(x){\tau}^r_{r,i}det(A){\xi}^s_{sj}  $  }  \\

\noindent
{\it Proof} : According to Lemma 5.C.3, we have:  \\
\[  \begin{array}{rcl}
\delta (\tau det(A)) & = & (\delta \tau)det(A) + \tau \delta det(A)  \\
        & = & {\omega}^{ij}B^r_i det(A){\partial}_r {\xi}^s_{sj}+{\partial}_r({\xi}^r\tau det(A)) - \frac{(n-2)}{n}\tau det(A) {\xi}^r_r -
        {\omega}^{ij}{\tau}^r_{j,i} det(A){\xi}^s_{sj} \\
        & = & {\partial}_r({\xi}^r\tau det(A) + {\omega}^{ij}(x)B^r_idet(A){\xi}^s_{sj}) -\frac{(n-2)}{n}\tau det(A) {\xi}^r_r  \\
         &    &  - {\omega}^{ij} ({\partial}_r(B^r_i det(A)){\xi}^s_{sj} - {\omega}^{ij}{\tau}^r_{j,i}det(A){\xi}^s_{sr}
\end{array}   \]
and we just need to use Lemma 5.C.1.  \\
\hspace*{12cm}      Q.E.D.      \\

\noindent
{\bf THEOREM 5.C.11}: We have the following Euler-Lagrange equations when $n=4$ {\it only}:  \\
\[   \left\{\begin{array}{lcl}
{\xi}^r_{ri}  & \rightarrow & \exists  \,\, gravitational \,\, potential  \\
{\xi}^r_r    & \rightarrow & \exists \,\, Poisson \,\, equation  \\
{\xi}^r  & \rightarrow & \exists \,\, Newton \,\, law
\end{array}   \right.\]
In particular \fbox{${\tau}^r_{r,i}=0 \Leftrightarrow {\chi}^r_{r,i}=0 \Leftrightarrow {\alpha}_i=0 \Leftrightarrow b_k= 
- \frac{1}{\Theta}\frac{\partial \Theta}{\partial y^k}$} .  \\

\noindent
{\it Proof} : For $n$ arbitrary, we have:   \\
\[ \begin{array}{rcl}
\tau det(A) & = & n{\Theta}^{(n-2)}\Delta ({\omega}^{kl} \frac{\partial b_k}{\partial y^l}- \frac{(n-2)}{2}{\omega}^{kl} b_kb_l)  \\
       &  =  & - n {\Theta}^{(n-2)} \Delta (  {\Theta}^{-1}{\omega}^{kl}\frac{{\partial}^2 \Theta}{\partial  y^k \partial y^l}+ 
       \frac{(n-4)}{2}{\Theta}^{-2}{\omega}^{kl}\frac{\partial \Theta}{\partial y^k}\frac{\partial \Theta}{\partial y^l})
\end{array}   \]
Hence, {\it for} $n=4$ {\it only}, we have $\tau det(A) = - 4 \Delta \Theta {\omega}^{kl}\frac{{\partial}^2 \Theta}{\partial  y^k \partial y^l}$. In the static case the gravity vector {\it must} be in first approximation $g^k\simeq - {\gamma}^k_{44}= {\omega}_{44}{\omega}^{kl}b_l= - b_k <0 \Leftrightarrow  b_k>0, \forall k=1,2,3$ (care to the minus sign coming from the inversion of the elations). If we introduce the gravitational potential $\phi= \frac{GM}{r}$ where $r$ is the distance at the central attractive mass $M$ and $G$ is the gravitational constant, then we have $\frac{\phi}{c^2}\ll 1$ as a dimensionless number and $\Theta=1$ when there is no gravity. When there is static gravity, the conformal factor $\Theta$ must be therefore close to $1$ with vanishing Laplacian and $\frac{\partial \Theta}{\partial y}<0$. The only coherent possibility is to set $\Theta = 1 + \frac{\phi}{c^2}$ in order to correct the value $\Theta= 1- \frac{\phi}{c^2}$ we found in ([28], p 450) and we have already explained the confusion we made on the physical meaning of {\it source} and {\it target}. Hence, gravity in vacuum only depends on the conformal isotropy groupoid through the conformal factor but this new approach is quite different from the ideas of G. Nordstr\"{o}m ([15],[53]), H. Weyl ([54]) or even Einstein-Fokker ([10],[23]). Indeed, it has only to do with the nonlinear Spencer sequence and {\it not at all} with the nonlinear Janet sequence, contrary to all these theories, as we just said, and the conformal factor 
$\Theta$ is now well defined everywhere apart from the origin of coordinates where is the central attractive mass. We have thus no longer any need to introduce the so-called {\it horizon} $r=GM/c^2$ and gravitation only depends on the structure of the conformal group theory like electromagnetism, with the only experimental need to fix the gravitational constant. Such a " {\it philosophy} " has been first proposed by the Cosserat brothers in ([1],[8],[18],[27]) for elasticity with the only experimental need to measure the elastic constants and extended to electromagnetism in the last section with the same comments (See [27],[30] and [46],[50] for details). An additional dynamical term must be added for the Newton law but this rather physical question will be studied in another paper as we already said in the Introduction.  \\
 \hspace*{12cm}   Q.E.D.   \\
 
\noindent
{\bf REMARK 5.C.12}: We shall find back the same Euler-Lagrange variational equations by using the variation over the target. With $dy=\Delta dx$ by definition, we have indeed for $n$ arbitrary:  \\
\[  \int \tau det(A)dx= \int n{\Theta}^{(n-2)} [{\omega}^{kl}(y) \frac{\partial b_k}{\partial y^l} -\frac{(n-2)}{2}{\omega}^{kl}(y)b_kb_l ]dy  \]
If we are only interested by the variation of the second order jets, we may equivalently vary the $b_k $ alone and get after integration by parts:  \\
 \[  \delta b_l \,\,\, \longrightarrow \,\,\,(n-2){\Theta}^{(n-3)}{\omega}^{kl} \frac{\partial \Theta}{\partial y^k} + (n-2){\Theta}^{(n-2)} {\omega}^{kl} b_k=0 \,\,\, \Rightarrow \,\,\, \fbox{$ b_k=  - \frac{1}{\Theta}\frac{\partial \Theta}{\partial y^k}$}  \]
Now, with $dx= dx^1 \wedge ... \wedge dx^n$ and $dy=dy^1 \wedge ... \wedge dy^n$, we have:  \\
\[ \begin{array}{rcl}
\int \tau det(A)dx  & =  &  - \int [n {\Theta}^{(n-3)} {\omega}^{kl}(y)\frac{{\partial}^2 \Theta}{ \partial y^k\partial y^l} +\frac{n(n-4)}{2}{\Theta}^{(n-4)}{\omega}^{kl}(y)\frac{\partial \Theta}{\partial y^k}\frac{\partial \Theta}{\partial y^l}]dy  \\
    &    &     \\
    &  = &   - \int \frac{\partial}{\partial y^l}( n{\Theta}^{(n-3)} {\omega}^{kl}(y)\frac{\partial \Theta}{ \partial y^k})dy 
   -  \int \frac{n(n-2)}{2} {\Theta}^{(n-4)}{\omega}^{kl}(y)\frac{\partial \Theta}{\partial y^k}\frac{\partial \Theta}{\partial y^l}dy     
 \end{array}  \]  \\
 \noindent   
If we only vary the section $y=f(x)$ of $X\times Y $ over $X$, we have $ dy= \Delta dx, \delta \Delta= \Delta \frac{\partial {\eta}^u}{\partial y^u}$ and :   \\
\[ {\Theta}^ndet(f^k_i(x))=1 \,\, \Rightarrow \,\,  0=\delta ({\partial}_i\Theta) = \delta (\frac{\partial \Theta}{\partial y^k}) {\partial}_if^k + \frac{\partial \Theta}{\partial y^u}\frac{\partial {\eta}^u}{\partial y^k} {\partial}_if^k \Rightarrow \delta (\frac{\partial \Theta}{\partial y^k})= -  \frac{\partial \Theta}{\partial y^u}\frac{\partial {\eta}^u}{\partial y^k}  \]
It follows that the variation of the last integral is:   \\
\[ - \int n(n-2) {\Theta}^{(n-4)}{\omega}^{kl}(y)(\frac{\partial \Theta}{\partial y^l}\frac{\partial \Theta}{\partial y^u}\frac{\partial {\eta}^u}{\partial y^k} - \frac{1}{2}\frac{\partial \Theta}{\partial y^k}\frac{\partial \Theta}{\partial y^l}\frac{\partial {\eta}^u}{\partial y^u}) dy  \]
After integration by parts, we get, up to a divergence:  \\
\[   - n(n-2) \int  \frac{\partial}{\partial y^k}[{\Theta}^{(n-4)}({\omega}^{rk}(y)\frac{\partial \Theta}{\partial y^r}\frac{\partial \Theta}{\partial y^u} - \frac{1}{2}{\delta}^k_u {\omega}^{rs}(y)\frac{\partial \Theta}{\partial y^r}\frac{\partial \Theta}{\partial y^s})]{\eta}^u) dy    \]
When $n=4$, the direct computation becomes simpler because a part of the integral disappears. We are left with $\tau det(A)= - 4 \Theta \Box \Theta$ and we recognize the well known {\it Abraham tensor} in the bracket ([28]), without any other assumption. Accordingly, we may finally say as in the previous section:  \\

{\it The whole gravitational scheme only depends on the structure of the conformal group}. \\

\noindent
{\bf REMARK 5.C.13}: Proceeding as in GR, we may consider the variation:   \\
\[   \delta \int {\mathfrak{g}}^{kl}(y) [ \frac{\partial b_l}{\partial y^k} + b_kb_l -\frac{1}{2}{\omega}_{kl}(y){\omega}^{rs}(y)b_rb_s]dy=0  \]
Varying only the second order jets $b_k$, we get equivalently through an integration by parts:   \\
\[  (2 {\mathfrak{g}}^{kl} - {\omega}^{kl}{\omega}_{rs}{\mathfrak{g}}^{rs}) b_l    =\frac{\partial {\mathfrak{g}}^{kl}}{\partial y^l}   \]
If  we set $b(f(x))=a(x)$ and $ {\Theta}(y)=e^{- b(y)}$, then ${\Theta}^2 (y)=e^{-2b(y)}$ and we have successively:  \\
\[        {\omega}_{kl}(y)f^k_if^l_j=e^{2a(x)}{\omega}_{ij}(x)  \Leftrightarrow e^{-2b(y)}{\omega}_{kl}(y)f^k_if^l_j={\omega}_{ij}(x)  \Leftrightarrow {\Theta}^2(y){\omega}_{kl}(y)f^k_if^l_j={\omega}_{ij}(x)   \]
Inverting the matrices, we obtain equivalently:   \\
\[ {\Theta}^{- 2}(y){\omega}^{kl}(y)g^i_kg^j_l={\omega}^{ij}(x) \Leftrightarrow {\Theta}^{- 2}(y){\omega}^{kl}(y)={\omega}^{ij}(x)f^k_if^l_j \]
and thus:   \\
\[{\Theta}^{2n}det(f^k_i)^2=1\,\,\Rightarrow \,\,{\Theta}^{n}det(f^k_i)=1\,\,\Rightarrow \,\,det(A)= det({\partial}_if^k)/det(f^k_i)={\Theta}^n\Delta  \]
Hence, if we set ${\mathfrak{g}}^{kl}(y)={\Theta}^{(n-2)}(y){\omega}^{kl}(y)$, we finally obtain:  \\
\[  (n-2){\Theta}^{(n-2)} b_k= - (n-2){\Theta}^{(n-3)}\frac{\partial \Theta}{\partial y^k} \Rightarrow  \fbox{ $b_k= - \frac{1}{\Theta}\frac{\partial \Theta}{\partial y^k}$} \] 
in a coherent way with the logarithmic derivatives:   \\
\[ {\beta}_k=0 \,\, \Leftrightarrow \,\,  \frac{\partial b}{\partial y^k}= - \frac{1}{\Theta}\frac{\partial \Theta}{\partial y^k}=b_k \,\, \Leftrightarrow \,\, 
{\partial}_ia= - \frac{1}{\Theta} \frac{\partial \Theta}{\partial y^k}{\partial}_if^k=b_k{\partial}_if^k=a_rg^r_k{\partial}_if^k=A^r_ia_r 
\,\, \Leftrightarrow \,\, {\alpha}_i=0   \]    \\    \\

\newpage

\noindent
{\bf 6) CONCLUSION}  \\

This paper is the achievement of a lifetime research work on the common conformal origin of electromagnetism and gravitation. Roughly speaking, the Cosserat brothers have only been dealing with the $3$ translations and $3$ rotations of the group of rigid motions of space with $6$ parameters while 
Weyl has only been dealing with the dilatation and the $4$ elations of the conformal group of space-time with now $4 + 6 + 1 +4=15$ parameters ([47]). 
Among the most striking results obtained from this conformal extension, we successively notice:  \\

\noindent
$\bullet$ The generating nonlinear {\it first order} (care) compatibility conditions (CC) for the Cosserat fields are {\it exactly} described by the {\it first order} nonlinear second Spencer operator ${\bar{D}}_2$. Accordingly, there is no conceptual difference between these nonlinear CC and the first set $d:{\wedge}^2T^* \rightarrow {\wedge}^3T^*$ of Maxwell equations where $d$ is the exterior derivative. However, the classical CC of elasticity are described by the nonlinear {\it second order} (care) {\it Riemann} operator existing in the nonlinear Janet sequence but this different canonical nonlinear differential sequence could not explain the existence of field-matter couplings like piezzoelectricity or photoelasticity ([31],[46]). On the contrary, in the conformal approach, it is essential to notice that {\it the elastic and electromagnetic fields are both specific sections of} ${\hat{C}}_1=T^* \otimes {\hat{R}}_2$ {\it killed by} ${\bar{D}}_2$. They can thus be coupled in a natural way but {\it cannot be associated to the concept of curvature} described by ${\hat{C}}_2$. This {\it shift by one step to the left}, even in the nonlinear framework, can be considered as the main novelty of this paper.    \\

\noindent
$\bullet$ The linear Cosserat equations are {\it exactly} described by the formal adjoint $ad(D_1)$ of the {\it linear} first Spencer operator $D_1: {\hat{C}}_0 \rightarrow {\hat{C}}_1$ which is a first order operator ([33]). Accordingly, there is no conceptual difference between these equations and the second set $ad(d)$ of Maxwell equations where $d:T^* \rightarrow {\wedge}^2T^*$. This result explains why the {\it Cosserat} equations are quite different from the {\it Cauchy} equations which are described by the formal adjoint of the {\it Killing} operator in the Janet sequence used in classical elasticity, that is $Cauchy=ad(Killing)$ in the language of operators. It follows that the elastic and electromagnetic inductions are both specific sections of ${\wedge}^4T^* \otimes {\hat{C}}^*_1 \simeq {\wedge}^3T^*\otimes {\hat{R}}^*_2$, {\it independently of any constitutive relation}.  \\ 

\noindent
$\bullet$ Combining the two previous comments, respectively related to "{\it geometry} " and to " {\it physics} " according to H. Poincar\'{e} ([24]), there is no conceptual difference between the elastic constitutive constants of elasticity and the magnetic constant $\mu$ or rather $1/\mu$ of electromagnetism in the case of homogeneous isotropic materials on one side ({\it space}) or between the mass per unit volume and the dielectric constant $\epsilon$ on the other side ({\it time}), a result confirmed by the speeds of the various elastic or electromagnetic existing waves ([31],[46]). In general one has $\epsilon \mu c^2= n^2$ where $n$ is the index of refraction but in vacuum we have ${\epsilon}_0 {\mu}_0 c^2=1$ and we have thus only one electromagnetic constant involved in the corresponding Minkowski constitutive law of vacuum ([22]). \\

\noindent
$\bullet$ As for gravitation and the possibility to exhibit a conformal factor defined everywhere but at the origin, we may simply say that we needed $25$ years in order to correct the result we already obtained in 1994 ([28]). Such a possibility highly depends on the new mathematical tools involved in the construction of the Janet or Spencer nonlinear differential sequences for various groups, in particular for the conformal group of space-time because, {\it in this case},  the Spencer $\delta$-cohomology has very specific properties for the dimension $n=4$ {\it only}.  \\ 

We end this paper with the french proverb " AUTRES TEMPS, AUTRES MOEURS" as we do believe that a modern scientific translation could be " NEW MATHEMATICS, NEW PHYSICS".  \\  \\

\noindent
{\bf REFERENCES}  \\
  
\noindent
[1] P. APPELL: Trait\'{e} de M\'{e}canique Rationnelle, Gauthier-Villars, Paris (1909) Particularly t II concerned with Analytical Mechanics and t III with a Note by E. and F. Cosserat "Note sur la Th\'{e}orie de l'Action Euclidienne", 557-629.\\
\noindent
[2] V. ARNOLD: M\'{e}thodes Math\'{e}matiques de la M\'{e}canique Classique, Appendice 2 (G\'{e}od\'{e}siques des M\'{e}triques Invariantes \`{a} Gauche sur des Groupes de Lie et Hydrodynamique des Fluides Parfaits), MIR, moscow, 1974,1976. (For more details, see also: J.-F. POMMARET: Arnold's Hydrodynamics Revisited, AJSE-mathematics, 1, 1 (2009) 157-174).  \\
\noindent
[3] Bourbaki, N.: Alg\`{e}bre, Ch. 10, Alg\`{e}bre Homologique, Masson, Paris (1980). \\
\noindent
[4] Cartan, E.: Sur la Structure des Groupes Infinis de Transformations, Ann. Ec. Norm. Sup., 21 (1904) 153-206; 22 (1905) 219-308.  \\
\noindent
[5] Cartan, E.: Les Espaces \`{a} Connexion Conforme, Ann. Soc. Polonaise Math., 2 (1923) 171-221.  \\
\noindent
[6] Cartan, E.: Sur les Vari\'{e}t\'{e}s  \`{a} Connexion Affine et la Th\'{e}orie de la Relativit\'{e} G\'{e}n\'{e}ralis\'{e}e, Ann. Ec. Norm. Sup., 40 (1923) 325-412; 41 (1924) 1-25; 42 (1925) 17-88. Also "Colected Works", part III, Differential geometry, vol 1-2, Gauthiers-Villars, Paris (1955) 63-70, 619-624, 633-658, 747-799. \\
\noindent
[7] O. Chwolson, O.: Trait\'{e} de Physique (See III, 2, 537 + III, 3, 994 + V, 209), Hermann (1914).\\
\noindent
[8] Cosserat, E., Cosserat, F.: Th\'{e}orie des Corps D\'{e}formables, Hermann, Paris, (1909).\\
\noindent
[9] Einstein, A.: Zur Elektrodynamik Bewegter K\"{o}rper, Annalen der Physik, 17 (1905) 891-921.  \\
Sur l'Electrodynamique des Corps en mouvement, Gauthier-Villars, Paris (1955).   \\
\noindent
[10] Einstein, A., Fokker A.D.: Nordstr\^{o}m's Theory of Gravitation from the Point of View of the Absolute Differential Calculus, Annalen der Physik, 44 (1914) 321-328.  \\
\noindent
[11] Eisenhart, L.P.: Riemannian Geometry, Princeton University Press, Princeton (1926).\\
\noindent
[12] Goldschmidt, H.: Sur la Structure des Equations de Lie, J. Differential Geometry, 6 (1972) 357-373 and 7 (1972) 67-95.\\
\noindent
[13] Goldschmidt, H., Spencer,D.C.: On the Nonlinear Cohomology of Lie Equations, I+II, Acta. Math., 136 (1973) 103-239.\\
\noindent
[14] Gutt, S.: Invariance of Maxwell's equations, In: Cahen M., Lemaire L., Vanhecke L. (eds) Differential geometry and Mathematical Physics. Mathematical Physics Studies (A Supplementary Series to Letters in Mathematical Physics), vol 3 (1983) 27-29, Springer, Dordrecht.  \\
https://link.springer.com/chapter/10.1007/978-94-009-7022-9\_3  \\
\noindent
[15] Isaksson, E.: Gunar Nordstr\"{o}m (1881-1923) on Gravitation and Relativity (1985), XVIIth International Congress of History of Science, Berkeley (With a complete list of publications).  \\
\noindent
[16] Janet, M.: Sur les Syst\`{e}mes aux D\'{e}riv\'{e}es Partielles, Journal de Math., 8 (1920) 65-151. \\
\noindent 
[17] S. Kobayashi, S., Nomizu, K.: Foundations of Differential Geometry, Vol I, J. Wiley, New York (1963, 1969).\\
\noindent
[18] Koenig, G.: Le\c{c}ons de Cin\'{e}matique (The Note "Sur la Cin\'{e}matique d'un Milieu Continu" by E. Cosserat and F. Cosserat has rarely been quoted), Hermann, Paris (1897) 391-417.\\
\noindent
[19] Kumpera, A., Spencer, D.C.: Lie Equations, Ann. Math. Studies 73, Princeton University Press, Princeton (1972).\\
\noindent
[20] Nordstr\"{o}m, G.: Einstein Theory of Gravitation and Herglotz's Mechanics of Continua, Proc. Kon. Ned. Akad. Wet., 19 (1917) 884-891.  \\
\noindent
[21] Northcott, D.G.: An Introduction to Homological Algebra, Cambridge university Press (1966).  \\
\noindent
[22] Ougarov, V.: Th\'{e}orie de la Relativit\'{e} Restreinte, MIR, Moscow (1969), Paris (1979).\\
\noindent
[23] Pais, A.: Subtle is the Lord: The Science and the Life of Albert Einstein, Oxford University Press (1982) (Chapter 13).  \\
\noindent
[24] Poincar\'{e}, H.: Sur une Forme Nouvelle des Equations de la M\'{e}canique, C. R. Acad\'{e}mie des Sciences Paris, 132 (7) (1901) 369-371.  \\
\noindent
[25] Pommaret, J.-F.: Systems of Partial Differential Equations and Lie Pseudogroups, Gordon and Breach, New York (1978); Russian translation: MIR, Moscow (1983).\\
\noindent
[26] Pommaret, J.-F.: Differential Galois Theory, Gordon and Breach, New York (1983).\\
\noindent
[27] Pommaret, J.-F.: Lie Pseudogroups and Mechanics, Gordon and Breach, New York (1988).\\
\noindent
[28] Pommaret, J.-F.: Partial Differential Equations and Group Theory, Kluwer (1994).\\
https://doi.org/10.1007/978-94-017-2539-2    \\
\noindent
[29] Pommaret, J.-F.: Fran\c{c}ois Cosserat and the Secret of the Mathematical Theory of Elasticity, Annales des Ponts et Chauss\'ees, 82 (1997) 59-66 (Translation by D.H. Delphenich).  \\
\noindent
[30] Pommaret, J.-F.: Partial Differential Control Theory, Kluwer, Dordrecht (2001).\\
\noindent
[31] Pommaret, J.-F.: Group Interpretation of Coupling Phenomena, Acta Mech.,149 (2001) 23-39.\\
\noindent
[32] Pommaret, J.-F.: Algebraic Analysis of Control Systems Defined by Partial Differential Equations, in "Advanced Topics in Control Systems Theory", Springer, Lecture Notes in Control and Information Sciences 311 (2005) Chapter 5, pp. 155-223.\\
\noindent
[33] Pommaret, J.-F.: Parametrization of Cosserat Equations, Acta Mechanica, 215 (2010) 43-55.\\
https://doi.org/10.1007/s00707-010-0292-y  \\
\noindent
[34] Pommaret, J.-F.: Spencer Operator and Applications: From Continuum Mechanics to Mathematical Physics, in "Continuum Mechanics-Progress in Fundamentals and Engineering Applications", Dr. Yong Gan (Ed.), ISBN: 978-953-51-0447--6, InTech (2012) Available from: \\
DOI: 10.5772/35607   \\
\noindent
[35] Pommaret, J.-F.: The Mathematical Foundations of General Relativity Revisited, Journal of Modern Physics, 4 (2013) 223-239. \\
 https://doi.org/10.4236/jmp.2013.48A022   \\
  \noindent
[36] Pommaret, J.-F.: The Mathematical Foundations of Gauge Theory Revisited, Journal of Modern Physics, 5 (2014) 157-170.  \\
https://doi.org/10.4236/jmp.2014.55026  \\
 \noindent
[37] Pommaret, J.-F.: Relative Parametrization of Linear Multidimensional Systems, Multidim. Syst. Sign. Process., 26 (2015) 405-437.  \\
DOI 10.1007/s11045-013-0265-0   \\
\noindent
[38] Pommaret,J.-F.:From Thermodynamics to Gauge Theory: the Virial Theorem Revisited, pp. 1-46 in "Gauge Theories and Differential geometry,", NOVA Science Publisher (2015).  \\
\noindent
[39] Pommaret, J.-F.: Airy, Beltrami, Maxwell, Einstein and Lanczos Potentials revisited, Journal of Modern Physics, 7 (2016) 699-728. \\
\noindent
https://doi.org/10.4236/jmp.2016.77068   \\
\noindent
[40] Pommaret, J.-F.: Deformation Theory of Algebraic and Geometric Structures, Lambert Academic Publisher (LAP), Saarbrucken, Germany (2016). A short summary can be found in "Topics in Invariant Theory ", S\'{e}minaire P. Dubreil/M.-P. Malliavin, Springer 
Lecture Notes in Mathematics, 1478 (1990) 244-254.\\
https://arxiv.org/abs/1207.1964  \\
\noindent
[41] Pommaret, J.-F.:Why Gravitational Waves Cannot Exist, J. of Modern Physics, 8 (2017) 2122-2158.  \\
https://doi.org/104236/jmp.2017.813130    \\
\noindent
[42] Pommaret, J.-F.: New Mathematical Methods for Physics, Mathematical Physics Books, Nova Science Publishers, New York (2018) 150 pp. \\
\noindent
[43] Pommaret, J.-F.: Minkowski, Schwarrzschild and Kerr Metric Revisited, J. of Modern Physics, 9 (2018) 1970-2007.  \\
https://doi.org/10.4236/jmp.2018.910125   \\
\noindent
[44] Pommaret, J.-F.: Generating Compatibility Conditions and General Relativity, J. of Modern Physics, 10, 3 (2019) 371-401.  \\
https://doi.org/10.4236/jmp.2019.103025   \\
\noindent
[45] Pommaret, J.-F.: Differential Homological Algebra and General Relativity, J. of Modern Physics, 10 (2019) 1454-1486. \\
https://doi.org/10.4236/jmp.2019.1012097   \\
\noindent
[46] Pommaret, J.-F.: The Mathematical Foundations of Elasticity and Electromagnetism Revisited, J. of Modern Physics, 10 (2019) 1566-1595.  \\
https://doi.org/10.4236/jmp.2019.1013104    \\
\noindent
[47] Pommaret, J.-F.: The Conformal Group Revisited (2020).   \\
https://arxiv.org/abs/2006.03449   \\
\noindent
[48] Rotman, J.J.: An Introduction to Homological Algebra, Academic Press (1979).   \\
\noindent
[49] Spencer, D.C.: Overdetermined Systems of Partial Differential Equations, Bull. Am. Math. Soc., 75 (1965) 1-114.\\
\noindent
[50] Teodorescu, P.P.: Dynamics of Linear Elastic Bodies,  Abacus Press, Tunbridge, Wells (1975) (Editura Academiei, Bucuresti, Romania).\\
\noindent
[51] Vessiot, E.: Sur la Th\'{e}orie des Groupes Infinis, Ann. Ec. Norm. Sup., 20 (1903) 411-451.   \\ 
\noindent
[52] Vessiot, E.: Sur la Th\'{e}orie de Galois et ses Diverses G\'{e}n\'{e}ralisations, Ann. Ec. Norm. Sup., 21 (1904) 9-85.\\
\noindent
[53] Weinstein, G.: The Einstein-Nordstr\"{o}m Theory (2012), arXiv:1205.5966v1. \\
\noindent
[54] Weyl, H.: Space, Time, Matter, Springer (1918,1958); Dover (1952). \\
\noindent
[55] Yang, C.N., Mills, R.L.: Conservation of Isotopic Gauge Invariance, Phys. Rev., 96 (1954) 191-195.\\
\noindent
[56] Zou, Z., Huang, P., Zang ,Y., Li, G.: Some Researches on Gauge Theories of Gravitation, Scientia Sinica, XXII, 6 (1979) 628-636.\\

\end{document}